\documentclass[prd,amsfonts,showpacs,preprintnumbers,nofootinbib,amsmath,amssymb,twocolumn]{revtex4-2}

\usepackage{bm,graphics,graphicx,epsfig,soul,latexsym,hyperref,float,mathrsfs,multirow,comment,xcolor}
\usepackage[justification=raggedright,singlelinecheck=true,tableposition=top]{caption}
\usepackage{dblfloatfix}
\usepackage{enumitem}
\usepackage{orcidlink}
\usepackage{ragged2e}

\newcommand{\blue}{\color{black}}

\graphicspath{{figure/}{./}}
\usepackage[english]{babel}
\usepackage[normalem]{ulem}
\usepackage{setspace}
\definecolor{darkgreen}{rgb}{0,0.5,0}

\hypersetup{
    bookmarks=true,         
    unicode=false,          
    pdftoolbar=true,        
    pdfmenubar=true,        
    pdffitwindow=false,     
    pdfstartview={FitH},    
    pdftitle={Aligned_spins_phasing},    
    pdfauthor={Author},     
    pdfsubject={Subject},   
    pdfcreator={Creator},   
    pdfproducer={Producer}, 
    pdfkeywords={keyword1} {key2} {key3}, 
    pdfnewwindow=true,      
    colorlinks=true,       
    linkcolor=red,          
    citecolor=blue,        
    filecolor=magenta,      
    urlcolor=darkgreen,           
    linktocpage=true
}

\newcommand{\TFtwoE}{\texttt{TaylorF2Ecc}}

\newcommand{\TTtwo}{\texttt{TaylorT2}}
\newcommand{\TFtwo}{\texttt{TaylorF2}}
\newcommand{\TTthree}{\texttt{TaylorT3}}
\newcommand{\TTfour}{\texttt{TaylorT4}}
\newcommand{\TTone}{\texttt{TaylorT1}}

\allowdisplaybreaks
\begin{document}
\title{Spin effects in the phasing formula of eccentric compact binary inspirals up to the third post-Newtonian order}
\author{Omkar Sridhar\orcidlink{0009-0008-3375-1077
}}\email{omkarnm1401@gmail.com}
\author{Soham Bhattacharyya\orcidlink{0000-0001-8753-7799}}\email{xeonese@gmail.com}
\author{Kaushik Paul\orcidlink{0000-0002-8406-6503}}\email{kpaul.gw@gmail.com}
\author{Chandra Kant Mishra\orcidlink{0000-0002-8115-8728}}\email{ckm@physics.iitm.ac.in}
\affiliation{Department of Physics, Indian Institute of Technology Madras, Chennai 600036, India}
\affiliation{Centre for Strings, Gravitation and Cosmology, Department of Physics, Indian Institute of Technology Madras, Chennai 600036, India}

\begin{abstract}
Compact binary sources that emit gravitational waves (GW) are expected to be both spinning and {\blue on} eccentric orbits. {\blue N}o closed-form expression for the phasing of GWs {\blue are available to date} that contain information from both spin and eccentricity. The introduction of eccentricity can slow waveform generation, often requiring slower numerical methods {\blue governing its evolution}. However, closed-form expressions for the waveform phase can be obtained when eccentricity is treated as a small parameter, enabling quick waveform generation. In this paper, closed-form expressions for the GW phasing in the form of Taylor approximants up to the eighth power in initial eccentricity $(e_0)$ are obtained while also including aligned spins up to the third post-Newtonian order. The phasing is obtained in both time and frequency domains. The {\blue fully analytical approximant} (\TTtwo{}) is also resummed for usage in scenarios where initial eccentricities are as high as 0.5. {\blue The frequency domain approximant (\TFtwo{}) based on Stationary Phase approximation is compared with an existing model (\TFtwoE{}) to assess the importance of the newly computed eccentric/spinning terms.} The findings indicate that for eccentricities $\gtrsim 0.15$ (defined at 10 Hz) and small spins $(\sim 0.2 )$, the mismatches can be higher than 1\%. This leads to an overall loss in signal-to-noise ratio and lower detection efficiency of GWs coming from eccentric spinning compact binary inspirals if the combined effects of eccentricity and aligned spins are neglected in the waveforms.
\end{abstract}
\date{\today}
\maketitle
\section{Introduction} 
\label{sec:intro}
\noindent
Gravitational wave (GW) detections from binary black hole (BBH) ~\citep{LIGOScientific:2016aoc}, binary neutron star (BNS) ~\citep{LIGOScientific:2017vwq,LIGOScientific:2020aai}, and neutron-star-black hole~\citep{LIGOScientific:2021qlt} mergers by the ground-based network of LIGO~\citep{LIGOScientific:2014pky} and Virgo~\citep{VIRGO:2014yos} detectors have opened a new avenue for exploring astrophysical phenomena in the Universe. Nearly 90 compact binary coalescence (CBC) events~\citep{LIGOScientific:2018mvr,LIGOScientific:2020ibl,LIGOScientific:2021usb,KAGRA:2021vkt} have been detected by the LIGO-Virgo detector network until the end of the third observing run (O3). The detection of GWs relies on a technique known as matched filtering~\citep{Cutler:1994ys, Poisson:1995ef, Krolak:1995md}. In this method, the observed data is cross-correlated with simulated copies of pre-computed waveforms (also known as templates). Its accuracy heavily depends on how closely the templates match the signal hidden within the detector's noisy data. Typically, the disagreement between the templates and the signal should be no larger than $\sim3\%$ for detection purposes and $\sim 1\%$ for parameter estimation studies~\citep{LIGOScientific:2016vbw}.\\

The evolution of CBC systems can be divided into three distinct phases: the low-frequency, weak-field inspiral phase can be modelled accurately using post-Newtonian (PN) theory (for a detailed review on this, see Ref.~\citep{Blanchet:2013haa}), while the high-frequency, strong-field merger phase can be described using Numerical relativity (NR)~\citep{SXS:catalog, Healy:2022wdn, Ferguson:2023vta} and the final ringdown using black hole perturbation theory (BHPT) (see Ref.~\citep{Pound:2021qin} for a review). The inspiralling compact binaries shed eccentricity due to the emission of GWs, and thus their orbits are expected to circularize with time~\citep{Peters:1963ux,Peters:1964zz}. However, compact binaries formed via dynamical interactions in dense stellar environments or through Kozai-Lidov processes~\citep{Kozai:1962zz,LIDOV1962719} can have ``detectable" residual eccentricities $(e_{0} \gtrsim 0.1)$ while entering the frequency band of the current ground-based detectors,
which in turn can tell us 
about the binary formation channels and various astrophysical processes, such as the evolution of binaries in globular clusters and in galactic nuclei~\citep{Samsing:2013kua,Rodriguez:2016kxx,Antonini:2015zsa,Chomiuk:2013qya,Strader:2012wj,Osburn:2015duj,VanLandingham:2016ccd,Hoang:2017fvh,Gondan:2017hbp,Gondan:2020svr,Kumamoto:2018gdg,Fragione:2019hqt,OLeary:2005vqo}. Besides, a small eccentricity does not mean that its exclusion in waveform modelling will go unpunished while doing matched filtering~\citep{Allen:2005fk}, as was found in Ref.~\citep{Bhat:2022amc,Chattaraj:2022tay}. Ref.~\citep{Favata2013} demonstrates that even small initial eccentricities (e.g., $6\times10^{-3}$ at 10 Hz) can bias source parameter estimates if GW models used for matched filtering neglect eccentricity.

Current pipelines may have reduced sensitivity to eccentric binaries{\blue, as they use circularized templates,} potentially leading to missed detections, particularly for systems with larger eccentricities \citep{Divyajyoti:2023rht, phukon:2024amh}. The orbits of CBC systems are expected to have eccentricities, and the components themselves have spins. Electromagnetic observations like Ref.~\citep{Cui:2023uyb} have revealed that black holes have dimensionless spins close to 0.9.  Spins of individual compact objects significantly change the shape and length of the GW signal {\blue including those in eccentric orbits~\citep{Taracchini:2012ig}. 
Ignoring either of the two effects}
can lead to biases in parameter estimation or missed event detections~\citep{Ramos-Buades:2019uvh,OShea:2021faf,Divyajyoti:2023rht}.  

Though progress in waveform modelling including effects due to eccentricity and spins in inspiral-only~\citep{Klein:2013qda,Huerta:2014eca,Moore:2018kvz,Klein:2018ybm,Moore:2018kvz,Tanay:2019knc,Liu:2019jpg,Tiwari:2020hsu,Klein:2021jtd, Paul:2022xfy} and full inspiral-merger-ringdown (IMR) ~\citep{Huerta:2016rwp,Hinderer:2017jcs,Hinder:2017sxy,Huerta:2017kez,Chen:2020lzc,Chiaramello:2020ehz, Chattaraj:2022tay, Manna:2024ycx, Paul:2024ujx} models have been made, {\blue fully analytical  prescriptions useful for data analysis purposes involving the effect of both spin and eccentricity at high PN orders are rare.} 
Over the past few decades, significant progress has been made in computing {\blue separately the effect of spin and eccentricity} in GW amplitude and phase at high PN orders. We highlight those here. Earlier work on GW amplitude for non-spinning but eccentric binaries was performed in Ref.~\citep{Mishra:2015bqa}, where the instantaneous (the part of GW radiation that depends on the
state of the source at a given retarded time) contributions to the spherical harmonic modes till the $3$PN order was presented. Spin effects were introduced for quasi-circular orbits in modes at the $1.5$PN and $2$PN order in Refs.~\citep{Arun:2008kb} and \citep{Buonanno:2012rv} respectively. This was extended to $3.5$PN, including both instantaneous and hereditary (the part of GW radiation which depends on
the entire past history of the source) contributions in modes by Ref.~\citep{Henry:2022dzx}. The combined effect of spin and eccentricity in energy loss rate and GW radiation were calculated in Refs.~\citep{Kidder:1992fr, Kidder:1995zr} up to $2$PN and $2.5$PN order, respectively. Additionally, Refs.~\citep{Vasuth:2007gx,Majar:2008zz} and Refs.~\citep{Khalil:2021txt,Paul:2022xfy} calculated the GW polarizations and modes till $1.5$PN and $2$PN order respectively. The modes computation was pushed to $3$PN by Ref.~\citep{Henry:2023tka}. Earlier works on GW phasing in quasi-circular binaries can be found in Ref.~\citep{Blanchet:1995fg} where they calculated the 2PN phasing from the energy loss of the binary. However, the $3$PN extension was a formidable task completed almost a decade later in Ref.~\citep{Blanchet:2004ek}. Recently, the same group calculated the phasing for quasi-circular binaries till $4.5$PN in Ref.~\citep{Blanchet:2023bwj}. Previous work on the phasing of GWs in eccentric systems can be found in Ref.~\citep{Krolak:1995md}, where they computed the Newtonian part of the phase. The work was then extended to $2$PN in Ref.~\citep{Favata2013}. Finally, Ref.~\citep{Moore:2016qxz} extended the computation of non-spinning eccentric binaries to $3$PN. {\blue Our work here can be considered as an update to Ref.~\citep{Moore:2016qxz}. We utilize the inputs of Ref.~\citep{Henry:2023tka} to provide fully analytical, eccentricity expanded phasing formulae for eccentric, spinning compact binary systems to the 3PN order. We summarize our strategies and findings below.}

\subsection{Summary of the current work}\label{subsec:summary}
{\blue Reference~\citep{Moore:2016qxz} lists what are commonly referred to as Taylor approximants that describe the phase of the signal in either the time- or frequency domain. We discuss each of these approximants and how they may be computed in Sec.~\ref{subsec:TaylorApproximants}. 
One of the time-domain Taylor approximants, that takes a fully analytical form, referred as \TTtwo{}. We wish to briefly discuss its structure so as to clarify some of our notations before summarizing our important findings. Equation~\eqref{eq:basic-TT2-expr} below displays structural form of the \TTtwo{} approximant and is analogous to Eq.~(1.1) of Ref.~\citep{Moore:2016qxz}, albeit for a spinning system. Expressed as a PN series in terms of a frequency dependent parameter, $v(\propto \omega^{1/3})$, where $\omega$ is orbit averaged orbital frequency, it takes the following form   
}
\begin{widetext}
\begin{align}\label{eq:basic-TT2-expr}
     \phi &=   \phi_{\rm c}  - \frac{1}{32 \nu v^5} \biggl\{ 1  + \left(\cdots \right)v^2 +  \frac{565}{24}\biggl[\delta  \chi_a+\left(\cdots\right) \chi_s\biggr]v^3 + \cdots + \mathcal{O}(v^7)  -\frac{785}{272} e_0^2 \biggl(\frac{v_0}{v}\biggr)^{19/3} \biggl[1 + 
   \left(\cdots \right) v^2 \nonumber \\ 
   &+ \left(\cdots \right) v_0^{2} -\frac{157}{54} 
   \biggl[\delta \chi_a + \chi_s \left(\cdots\right)\biggr] v_0^{3}+ \frac{208012}{21195}\biggl[\delta  \chi_a+\chi_s \left(\cdots\right)\biggr] v^3 + \cdots + \mathcal{O}(v^6) \biggr]+\cdots + \mathcal{O}(e_0^8)\biggr\}\,.
\end{align}
\end{widetext}
{\blue Here, $\phi_c$ represents the orbital phase ($\phi$) at coalescence and the parameters, ($e_0, v_0$), refer to a reference (say at the start of the waveform) value of time-eccentricity ($e_t$; see Sec.~\ref{app:QK} for details) and the PN parameter $v$. The mass and spin of the binary enter the formula through parameters, ($\nu=m_1m_2/(m_1+m_2)^2, \chi_{s,a}=(\chi_1\pm\chi_2)/2$), where $m_{1,2}$ and $\chi_{1,2}$ represent mass and spin of the binary constituents. Finally, symbols $\mathcal{O}(v^6)$ and $\mathcal{O}(e_0^8)$ indicate that terms beyond 6th power in $v$ (3PN) and 8th power in $e_0$ in the expansion have been ignored.\footnote{Note that, a series with highest power as $v^{2n}$ is referred to as $n$-PN accurate.} Note that this is the order to which we provide all the results in this work, although, if provided in full, these results will run over multiple pages and thus we explicitly list the results only to leading order in $e_0$ ($\mathcal{O}(e^2)$); see for instance Eq.~\ref{TaylorT2 phase}. Full ($\mathcal{O}(e^8)$) expressions can be accessed through the supplemental file~\citep{SBPM:suppl}.

\begin{table}[t]
\vspace{2mm}
    \centering
    \begin{tabular}{|l|r|}
    \hline
        \multicolumn{1}{|c|}{PN order} & \multicolumn{1}{|c|}{$\Delta N_{\rm cyc}$} \\ \hline 
        0PN (circ) & 16031  \\ 
        0PN (ecc) & -463 \\ 
        1PN (circ) & 439 \\ 
        1PN (ecc) & -15.8 \\
        1.5PN (circ) & -208 (-156)
        \\ 
        1.5PN (ecc) & 1.67 (1.25)
        \\ 
        2PN (circ) & 9.54 (7.89)
        \\ 
        2PN (ecc) & -0.215 (-0.207)
        \\ 
        2.5PN (circ) & -10.6 (-3.9)
        \\ 
        2.5PN (ecc) & 0.0443 (0.0167)
        \\ 
        3PN (circ) & 2.02 (-0.21)
        \\ 
        3PN (ecc) & 0.00200 (0.00369)
        \\ 
        3.5PN (circ) & -0.662 \\ \hline
        Total & 15785 (15840)
        \\ \hline
    \end{tabular}
\caption{\justifying Post-Newtonian contributions to the number of GW cycles by the non-spinning (non-spinning+spinning) sections of the \TTtwo{} phase for $\chi_1=\chi_2=0.4$ assuming an initial eccentricity $e_0 = e(f_0) =0.1$ for an equal mass $m_1 = m_2 = 1.4 M_\odot$ system. This is computed using $\Delta N_{\rm cyc} = [\phi(f_2) - \phi(f_1)]/\pi$. In the first column, the term ``circ" corresponds to the contribution of the circular term at that PN order, while ``ecc" implies the corresponding eccentric term. Since there are no spinning contributions at 0PN, 1PN and 3.5PN, the parentheses at these orders have been omitted. The binary enters the LIGO band at $f_1$ = 10 Hz. Since this is a low-mass system, the innermost stable circular orbit (ISCO) frequency ($f_{\rm isco}$) extends beyond the LIGO band ($f_{\rm max}$ = 1000 Hz), and hence we curtail the signal at $f_2$ = 1000 Hz. The total cycles in the last row are the sum of all non-spinning (spinning) contributions. Note that these orbit calculations were done using \TTtwo{} phasing expressed in the PN parameter $x$, while the $y$ parameterization is used in the detailed expressions listed in Sec. \ref{sec:TaylorT2}. Also to be noted is that, in the spirit of Table I of Ref.~\citep{Moore:2016qxz}, we use the phasing expression curtailed to {\blue $\mathcal{O}(e_0^2)$} order, while the expressions listed in Sec. \ref{sec:TaylorT2} are extended till {\blue $\mathcal{O}(e_0^8)$} order.
}
\label{Intro cycles table}
\end{table}

The accuracy of these approximants can be accessed through estimates of number of GW cycles ($\Delta N_{\rm cyc}$) in the sensitivity band of a detector using the formula, $\Delta N_{\rm cyc} = [\phi(f_2) - \phi(f_1)]/\pi$; here ($f_1, f_2$) denote end-frequencies of a detector's sensitivity band. Table~\ref{Intro cycles table} displays these estimates by first setting spins to zero (and thus recovering estimates of \citep{Moore:2016qxz}) and then with spin terms included (corrected estimates shown in round brackets) assuming a binary neutron star system and advanced LIGO sensitivity band~\citep{Shoemaker:T0900288-v3}. Note that, for a fair comparison we restrict to $\mathcal{O}(e_0^2)$ as in \citep{Moore:2016qxz}. The estimates using full $\mathcal{O}(e_0^8)$ expression are displayed in Table~\ref{table_cycles_freq_bands} which also shows these estimates for a few representative binary black hole configurations and detector bands corresponding to a 3rd generation detector~\cite{Dwyer:2014fpa,Punturo:2010zz} and space based detectors such as DECIGO~\cite{Sato_2017} and LISA~\cite{amaro2017laser}. Estimates shown in Table~\ref{Intro cycles table}-\ref{table_cycles_freq_bands} assume small value of reference eccentricity ($e_0=0.2$) for which contributions from even leading spinning, eccentric terms (1.5PN order) seems negligible ($\mathcal{O}$(1) cycle). With eccentric corrections up to $\mathcal{O}(e^8)$ our expressions may be useful for larger values of eccentricities as well. (See for instance Fig.~\ref{fig:comparison} that compares our analytical estimates with a numerical result as a function of eccentricity.) For completeness we repeat the computations of Table~\ref{table_cycles_freq_bands} and provide these estimates for a moderate value of eccentricity ($e_0=0.5$) in Table~\ref{table_cycles_freq_bands_higher_ecc}. As expected the contributions at 1.5PN spinning, eccentric terms to number of GW cycles estimates is significant ($\mathcal{O}$(10) cycle.
}

{\blue Further, a fully analytical prescription for a frequency domain version of the \TTtwo{} approximant, referred as \TFtwo{}, can be obtained under the stationary phase approximation~\citep{VanDenBroeck:2006qu,VanDenBroeck:2006ar}. We follow the prescription of Ref.~\citep{Moore:2016qxz} to obtain the spinning version of the \TFtwo{} for eccentric systems; see Sec.~\ref{sec:TaylorF2} for details. Additionally, we also perform suitable resummation (via a trail and error method) of the \TTtwo{} phase to provide a form of the phase (again fully analytical) that helps us push the accuracy of the model to higher values of eccentricity; see for instance Fig.~\ref{fig:comparison}. Other approximants that are considered here are -- \TTone{}, \TTthree{}, and \TTfour{} and are listed in Appendix~\ref{sec: app-Taylor} and provide a semi-analytical prescription for phase.}

We also assess the importance of the newly computed spinning, eccentric corrections
in \TFtwo{} phase 
by comparing it with \TFtwoE{}~\citep{Moore:2016qxz}. 
{\blue Note that, while the implementation of \TFtwoE{} is based on Ref.~\citep{Moore:2016qxz}, which lists only non-spinning terms, it also includes 
spinning quasi-circular corrections.}
We compute mismatches (defined in Eq.~\eqref{eq:match}) between the two models and plot them as a function of chirp mass $(M_{\rm{chirp}})$, mass ratio $(q)$, eccentricity at 10 Hz $(e_{10})$, {\blue and} effective spin parameter $(\chi_{\rm{eff}})$ in Figs.~\ref{fig:mismatch_pn}-\ref{fig:mismatch_pn_2}.
{\blue Additionally, 96.5\% contours denote the detection threshold for current ground-based detectors that help us to distinguish the parameter space where these newly computed terms become important over \TFtwoE{}.} 
As it can be observed from Figs.~\ref{fig:mismatch_pn}-\ref{fig:mismatch_pn_2}, the mismatches for $e_{10}$ vs. $q$, $M_{\rm{chirp}}$, and $\chi_{\rm{eff}}$ achieves a maximum value of $13$\%, it is highest in $\chi_{\rm{eff}} - M_{\rm{chirp}}$ plane reaching up to $\sim15$\%. {\blue This  clearly highlights the need to include the effect of both the spin and eccentricity in waveform templates employed for searches and parameter estimation studies}.

{\blue The rest of the paper is organized as follows. Section~\ref{sec:methods} presents the method employed in computing various Taylor approximants briefly mentioned above in Sec.~\ref{subsec:summary} and presents explicit expressions for the \TTtwo{} approximant and those derived from it -- \TFtwo{} and the resummed version. (Remaining approximants, \TTone{}, \TTthree{} and \TTfour{} are listed in Appendix~\ref{sec: app-Taylor}.) Section~\ref{sec:noprec} presents our investigations assessing the accuracy and  importance of the newly computed phasing prescriptions. Finally, in Sec.~\ref{sec:discussions} we summarize our results and observations.} 
Throughout the paper we closely follow the notations of Ref.~\citep{Moore:2016qxz} unless otherwise mentioned and set $G = c = 1$. 

{\blue \section{Post-Newtonian phasing for spinning,  eccentric compact binaries}\label{sec:methods}}
{\blue As indicated in the previous paragraph, this section presents methods that are employed here for computing various phasing approximants and also lists our most important results obtained using evolution equations for parameters that describe the orbital phase (or equivalently the phase of the GW signal).}  

\subsection{Evolution of orbital elements}\label{subsec:ecc-evol}
Evolution of the orbital phase for binaries in quasi-circular orbits under the adiabatic approximation (i.e., the binary's orbital timescale is much shorter than the radiation reaction timescale) is given by the following differential equations,

\begin{subequations}
\label{eq:approx_timedom}
\begin{align}
\label{eq:approx_timedom_phi}
 \frac{d\phi}{dt} &=\frac{x^{3/2}}{M},\\
 \label{eq:approx_timedom_v}
 \frac{dx}{dt} &=-\frac{{\mathcal F}(x)}{dE(x)/dx},
 \end{align}
\end{subequations}
where $x$ is a gauge invariant PN parameter\footnote{\blue Note that, a series with highest power as $x^{n}$ is referred to as $n$-PN accurate.} {\blue like $v$; in fact, $x(=v^2)\propto\omega^{2/3}$, where $\omega$ is orbit averaged orbital frequency.} 
${\mathcal F}(x)$ is the GW flux and $E(x)$ is the orbital energy. 

For eccentric orbits, the the orbital phase ($\phi$) includes oscillatory contributions {\blue (see Appendix~\ref{app:QK} for details)} and 
hence Eq.~\eqref{eq:approx_timedom} need to be modified. However, it was shown (in the case of binaries with compact components without spins) that these oscillatory terms contribute negligibly to the orbital phase in Ref.~\citep{Moore:2016qxz} as long as {\blue the eccentricity is treated as a small parameter} which is also the case with the present study. {We find that this is true even for spinning case (see Appendix~\ref{sec:osc}) and thus one can still work with  Eq.~\eqref{eq:approx_timedom} in order to find orbital phase for spinning, eccentric binaries assuming neglecting the corrections due to the oscillatory piece may only lead to small errors.} 

Different approaches for solving these differential equations are referred to as {\blue Taylor} approximants (see, for instance, Refs.~\citep{buonanno-iyer-oshsner-pan-sathya-templatecomparison-PRD2009, Moore:2016qxz}) {\blue discussed earlier in Sec.~\ref{subsec:summary} and we discuss these in detail in Sec.~\ref{subsec:TaylorApproximants} below. It should also be clear from Eq.~\eqref{eq:approx_timedom} that all we need are expressions for the flux (${\mathcal F}(x)$) and energy ($E(x)$) and we can then solve it. In fact, Ref.~\citep{Henry:2023tka} explicitly provides the expression for the evolution of $x$ (\textit{i.e} for $dx/dt$) in terms of the parameter $x$ and time-eccentricity ($e_t$; represented as $e$ here); see Eqs. (3.18)-(3.20) there. Note that, the evolution of the $e_t$ (\textit{i.e} for $de_t/dt$) is also expressed as function of the parameter $x$ and $e_t$ itself and thus prior to solving the Eq.~\eqref{eq:approx_timedom} we should solve the evolution equation for $e_t(x)$. Further, we also wish to work with a new frequency dependent variable\footnote{{\blue Note that, a series with highest power as $y^{2n}$ is referred to as $n$-PN accurate.}} $y$ which was introduced in Ref.~\citep{Boetzel:2017zza} and is related to the parameter $x$ and $e_t\equiv e$ as}\footnote{For our purposes (\emph{i.e.} while working with small eccentricities), the parameter $x$ could also have been used to express the results; however, it was observed in previous studies (for instance, {\blue in Sec. II C of} Ref.~\citep{Boetzel:2017zza}) that the use of the parameter $y$ over $x$ helps with the convergence of PN series when dealing with cases with relatively high eccentricities and thus the new parametrization is adopted here.} 
\begin{eqnarray}
    y &=& \left(\frac{x}{1 - e^2}\right)^{1/2}\,. \label{eq:ydef}
\end{eqnarray}

{\blue With these in mind we next provide explicit expressions for the desired time-evolution equations that are to be solved in order formally write the phase.}
We indicated above that the evolution equations for $x$ and $e$ 
can be found in 
Ref.~\citep{Henry:2023tka}, although they are written in a \textsc{resummed} form. For our purpose, we first Taylor expand these and then re-express them in terms of the new parameter $y$. {\blue However, we only provide a structural form of these equations here both because they could be easily derived starting from the expressions for $x$, and $e$ and also because we need their content in slightly different forms which we explicitly list. Nevertheless, these are included in the supplemental file \citep{SBPM:suppl} for completeness.\footnote{\blue Note that 2PN (generic) spinning versions for evolution equations, $dy/dt$ and  $de^{2}/dt$, in terms of the parameter $y$ are listed in Appendix C of Ref.~\citep{Klein:2018ybm}.} Structure of the} 
evolution equations of $y$ and $e$ takes the following form:

\begin{subequations}
\label{eq:ydotedot-structure}
\begin{align}
\label{eq:ydot-structure}
    \frac{dy}{dt}&=\left( 1 - e^2\right)^{3/2} \nu y^9  \left[a_{\rm NS} + a_{\rm SO} + a_{\rm SS}\right] \,,\\ 
    \nonumber\\
    \label{eq:e2dot-structure}
    \frac{de^2}{dt}&=-\left( 1 - e^2\right)^{3/2} \nu y^8 \left[b_{\rm NS} + b_{\rm SO} + b_{\rm SS}\right]\,. 
\end{align}
\end{subequations}

The coefficients $a$ and $b$ are PN series, such that

\begin{subequations}\label{eq:acc}
\begin{align}
    a_{\rm NS} &= \sum_{i = 0}^{6} a^i_{\rm NS} y^i , \label{eq:a-series} \\
    a_{\rm SO} &= \sum_{i = 3}^{6} a^i_{\rm SO} y^i , \\
    a_{\rm SS} &= \sum_{i = 4}^{6} a^i_{\rm SS} y^i, 
\end{align}
\end{subequations}
and similarly for $b$. 

\subsubsection{Eccentricity evolution}\label{sec:ecc-evol}

We can now use the chain rule to find the rate of change of eccentricity with respect to the PN parameter $y$.  


\begin{align}\label{eq:deBdy-structure}
    \frac{de^2}{dy} &= -\frac{19 e^2}{3 y} \, \left(  
    1 + c_{\rm SO} + c_{\rm SS} \right)  
    + \cdots + \mathcal{O}\left(e^{8}\right)\,,
\end{align}
{\blue here, we present only the Newtonian non-spinning contribution and the complete 3PN non-spinning expression} can be found in {\blue Eq.~(4.15) of} Ref.~\citep{Moore:2016qxz}. The spin-orbit (SO) and the spin-spin (SS) coefficients of the $e^2$ part {\blue above} are given by
\begin{widetext}
    \begin{subequations}
    \begin{align}
     \label{eq:c_coeff}
        c^{\rm 1.5PN}_{\rm SO} &= \left(\frac{55 \nu }{57}-\frac{157}{114}\right) \chi_s-\frac{157 \delta  \chi_a}{114}\,, \\
        c^{\rm 2.5PN}_{\rm SO} &= \left(\frac{1631 \delta  \nu }{228}-\frac{373867 \delta }{19152}\right)\chi_a  +\left(-\frac{1195 \nu ^2}{228}+\frac{586343 \nu }{19152}-\frac{373867}{19152}\right)\chi_s \,,\\
        c^{\rm 3PN}_{\rm SO} &= \frac{8857 \pi  \delta  \chi_a}{1824}+\left(\frac{8857 \pi }{1824}-\frac{141 \pi  \nu }{152}\right) \chi_s \,,\\
        c^{\rm 2PN}_{\rm SS} &= \chi_a^2 \left[\frac{89 \delta  \kappa_a}{304}+\kappa_s \left(\frac{89}{304}-\frac{89 \nu }{152}\right)-\frac{623 \nu }{152}+\frac{293}{304}\right]+\chi_a \chi_s \left[\frac{89 \delta 
   \kappa_s}{152}+\frac{293 \delta }{152}+\kappa_a \left(\frac{89}{152}-\frac{89 \nu }{76}\right)\right]+\chi_s^2 \left[\frac{89 \delta  \kappa_a}{304} \right.\nonumber\\
   &\left.+ \kappa_s \left(\frac{89}{304}-\frac{89
   \nu }{152}\right)+\frac{37 \nu }{152}+\frac{293}{304}\right]\,, \\
   c^{\rm 3PN}_{\rm SS} &= \chi_a^2 \left[\kappa_a \left(\frac{326983 \delta }{102144}-\frac{10549 \delta  \nu }{3648}\right)+\kappa_s \left(\frac{1123 \nu ^2}{608}-\frac{158223 \nu }{17024}+\frac{326983}{102144}\right)+\frac{7861 \nu
   ^2}{608}-\frac{1373687 \nu }{21888}+\frac{4320187}{306432}\right] \nonumber\\
   &+ \chi_a \chi_s \left[\kappa_s \left(\frac{326983 \delta }{51072}-\frac{10549 \delta  \nu }{1824}\right)-\frac{189737 \delta  \nu
   }{5472}+\frac{4320187 \delta }{153216}+\kappa_a \left(\frac{1123 \nu ^2}{304}-\frac{158223 \nu }{8512}+\frac{326983}{51072}\right)\right] \nonumber\\
   &+ \chi_s^2 \left[\kappa_a \left(\frac{326983 \delta
   }{102144}-\frac{10549 \delta  \nu }{3648}\right)+\kappa_s \left(\frac{1123 \nu ^2}{608}-\frac{158223 \nu }{17024}+\frac{326983}{102144}\right)+\frac{24269 \nu ^2}{5472}-\frac{4337201 \nu
   }{153216}+\frac{4320187}{306432}\right]\,. \nonumber\\
    \end{align}
    \end{subequations}
\end{widetext}

{\blue Here, we define $\kappa_s = \left(\kappa_1 + \kappa_2\right)/2$ and $\kappa_a = \left(\kappa_1 - \kappa_2\right)/2$, where $\kappa_1$ and $\kappa_2$ are the spin-induced quadrupole moment (SIQM) parameters~\citep{Krishnendu:2017shb,Divyajyoti:2023izl}. Both $\kappa_1$ and $\kappa_2$ are equal to one for black holes.} Eq.~\eqref{eq:deBdy-structure} is a first-order differential equation and can be solved assuming $y$ and $e$ as small parameters. Note that, since we are extending our calculations to higher orders in eccentricity (the justification and benefit of which will be described in subsequent sections), Eq.~\eqref{eq:deBdy-structure} is not separable in nature, and a perturbative method needs to be devised to solve it. Here, we solve Eq.~\eqref{eq:deBdy-structure} by successively increasing the order of eccentricity we consider and using the previous order solution to maintain a quasi-separable form of the equation. 

The structure of the solution in terms of {\blue the parameter $(y)$}, an initial eccentricity $e_0$ at some initial frequency $y_0 = y(t_0)$ (where $t_0$ is some initial reference time) is as follows,
\begin{widetext}
\begin{eqnarray}\label{eq:e2-y}
    e^2(y, y_0{\blue ,} e_0) &=& e_0^2 \, \left(\frac{y_0}{y}\right)^{19/3} \biggl[ 
    1+ d_{\rm SO} + d_{\rm SS}\biggr] 
    + \cdots + \mathcal{O}(e_0^{8})\,.
\end{eqnarray}
{\blue Here, we provide only the Newtonian non-spinning contribution and the complete 3PN non-spinning expression} can be found in {\blue Eq.~(4.17) of} Ref.~\citep{Moore:2016qxz}. The SO and SS coefficients are given by
\begin{subequations}\label{eq:e2-y}
\begin{align}
   d_{\rm SO}^{\rm 1.5PN} &= y^3 \left[\frac{157 \delta  \chi_a}{54}+\left(\frac{157}{54}-\frac{55 \nu }{27}\right) \chi_s\right] + y_0^3 \left[\left(\frac{55 \nu }{27}-\frac{157}{54}\right) \chi_s-\frac{157 \delta  \chi_a}{54}\right]\,, \\
   d_{\rm SO}^{\rm 2.5PN} &= y^5 \left[\chi_a \left(\frac{66571 \delta  \nu }{9720}+\frac{4505701 \delta }{272160}\right) + \left(-\frac{2191 \nu ^2}{486}-\frac{1166491 \nu
   }{68040}+\frac{4505701}{272160}\right) \chi_s\right] + y_0^5 \left[\chi_a \left(\frac{242719 \delta  \nu }{9720} \right.\right.\nonumber\\
   &\left.\left.- \frac{1279073 \delta }{38880}\right)+\left(-\frac{4322 \nu ^2}{243}+\frac{146807 \nu
   }{2430} - \frac{1279073}{38880}\right) \chi_s\right] + y^2 y_0^3 \left[\chi_a \left(\frac{444781 \delta }{54432}-\frac{30929 \delta  \nu }{1944}\right)+\left(\frac{10835 \nu ^2}{972}\right.\right.\nonumber\\
   &\left.\left.- \frac{588821 \nu }{27216}+\frac{444781}{54432}\right)
   \chi_s\right] + y^3 y_0^2 \left[\chi_a \left(\frac{444781 \delta }{54432}-\frac{30929 \delta  \nu }{1944}\right)+\left(\frac{10835 \nu ^2}{972}-\frac{588821 \nu }{27216}+\frac{444781}{54432}\right)
   \chi_s\right]\,, \nonumber\\
   \\
   d_{\rm SO}^{\rm 3PN} &= y^6 \left[\left(\frac{45277 \pi  \nu }{3888}-\frac{316469 \pi }{15552}\right) \chi_s  - \frac{316469 \pi  \delta  \chi_a}{15552}\right] + y_0^6 \left[\left(\frac{37663 \pi  \nu }{3888}-\frac{157043 \pi }{15552}\right) \chi_s-\frac{157043 \pi  \delta  \chi_a}{15552}\right] \nonumber\\
   &+ y^3 y_0^3 \left[\frac{59189 \pi  \delta  \chi_a}{1944}+\left(\frac{59189 \pi }{1944} - \frac{20735 \pi  \nu }{972}\right) \chi_s\right]\,, \\
   d^{\rm 2PN}_{\rm SS} &= y^4\left\{\chi_a^2 \left[-\frac{89 \delta  \kappa_a}{192}+\kappa_s \left(\frac{89 \nu }{96}-\frac{89}{192}\right)+\frac{623 \nu }{96}-\frac{293}{192}\right]+\chi_a \chi_s \left[-\frac{89 \delta 
   \kappa_s}{96}-\frac{293 \delta }{96}+\kappa_a \left(\frac{89 \nu }{48}-\frac{89}{96}\right)\right] \right.\nonumber\\
   &\left. + \chi_s^2 \left[-\frac{89 \delta  \kappa_a}{192}+\kappa_s \left(\frac{89 \nu}{96}-\frac{89}{192}\right)-\frac{37 \nu }{96}-\frac{293}{192}\right]\right\} + y_0^4 \left\{ \chi_a^2 \left[\frac{89 \delta  \kappa_a}{192}+\kappa_s \left(\frac{89}{192}-\frac{89 \nu }{96}\right)-\frac{623 \nu }{96}+\frac{293}{192}\right] \right.\nonumber\\
   &\left.+ \chi_a \chi_s \left[\frac{89 \delta 
   \kappa_s}{96}+\frac{293 \delta }{96}+\kappa_a \left(\frac{89}{96}-\frac{89 \nu }{48}\right)\right]+\chi_s^2 \left[\frac{89 \delta  \kappa_a}{192}+\kappa_s \left(\frac{89}{192}-\frac{89 \nu
   }{96}\right)+\frac{37 \nu }{96}+\frac{293}{192}\right] \right\}\,, \\
   d_{\rm SS}^{\rm 3PN} &= y^6 \left\{\chi_a^2 \left[\kappa_a \left(\frac{3565 \delta  \nu }{6912}-\frac{133943 \delta }{64512}\right)+\kappa_s \left(\frac{10795 \nu ^2}{3456}+\frac{451739 \nu }{96768}-\frac{133943}{64512}\right)+\frac{75565 \nu
   ^2}{3456}+\frac{8491589 \nu }{373248} \right.\right.\nonumber\\
   &\left.\left.- \frac{33265999}{5225472}\right]+\chi_a \chi_s \left[\kappa_s \left(\frac{3565 \delta  \nu }{3456}-\frac{133943 \delta }{32256}\right)+\frac{1304159 \delta  \nu
   }{93312}-\frac{33265999 \delta }{2612736}+\kappa_a \left(\frac{10795 \nu ^2}{1728}+\frac{451739 \nu }{48384} \right.\right.\right.\nonumber\\
   &\left.\left.\left.- \frac{133943}{32256}\right)\right]+\chi_s^2 \left[\kappa_a \left(\frac{3565 \delta  \nu
   }{6912}-\frac{133943 \delta }{64512}\right)+\kappa_s \left(\frac{10795 \nu ^2}{3456}+\frac{451739 \nu }{96768}-\frac{133943}{64512}\right)-\frac{440045 \nu ^2}{93312} \right.\right.\nonumber\\
   &\left.\left.+ \frac{43607327 \nu
   }{2612736}-\frac{33265999}{5225472}\right]\right\} + y_0^6 \left\{ \chi_a^2 \left[\kappa_a \left(\frac{906103 \delta }{193536}-\frac{12877 \delta  \nu }{2304}\right)+\kappa_s \left(\frac{24271 \nu ^2}{3456}-\frac{1446937 \nu }{96768}+\frac{906103}{193536}\right) \right.\right.\nonumber\\
   &\left.+ \frac{169897
   \nu ^2}{3456}-\frac{40961143 \nu }{373248}+\frac{17465819}{746496}\right]+\chi_a \chi_s \left[\kappa_s \left(\frac{906103 \delta }{96768}-\frac{12877 \delta  \nu }{1152}\right)-\frac{5526373 \delta  \nu
   }{93312}+\frac{17465819 \delta }{373248} \right.\nonumber\\
   &\left.+ \kappa_a \left(\frac{24271 \nu ^2}{1728}-\frac{1446937 \nu }{48384}+\frac{906103}{96768}\right)\right]+\chi_s^2 \left[\kappa_a \left(\frac{906103 \delta
   }{193536}-\frac{12877 \delta  \nu }{2304}\right)+\kappa_s \left(\frac{24271 \nu ^2}{3456}-\frac{1446937 \nu }{96768} \right.\right.\nonumber\\
   &\left.\left.\left.+ \frac{906103}{193536}\right)+\frac{433639 \nu ^2}{93312}-\frac{16075987 \nu
   }{373248}+\frac{17465819}{746496}\right]\right\} + y^2 y_0^4 \left\{\chi_a^2 \left[\kappa_a \left(\frac{17533 \delta  \nu }{6912}-\frac{252137 \delta }{193536}\right) \right.\right.\nonumber\\
   &\left.+ \kappa_s \left(-\frac{17533 \nu ^2}{3456}+\frac{497599 \nu }{96768}-\frac{252137}{193536}\right)-\frac{122731
   \nu ^2}{3456}+\frac{367579 \nu }{13824}-\frac{830069}{193536}\right]+\chi_a \chi_s \left[\kappa_s \left(\frac{17533 \delta  \nu }{3456} \right.\right.\nonumber\\
   &\left.\left.- \frac{252137 \delta }{96768}\right)+\frac{57721 \delta  \nu
   }{3456}-\frac{830069 \delta }{96768}+\kappa_a \left(-\frac{17533 \nu ^2}{1728}+\frac{497599 \nu }{48384}-\frac{252137}{96768}\right)\right]+\chi_s^2 \left[\kappa_a \left(\frac{17533 \delta  \nu
   }{6912} \right.\right.\nonumber\\
   & \left.\left.\left.- \frac{252137 \delta }{193536}\right)+\kappa_s \left(-\frac{17533 \nu ^2}{3456}+\frac{497599 \nu }{96768}-\frac{252137}{193536}\right)+\frac{7289 \nu ^2}{3456}+\frac{703273 \nu }{96768}-\frac{830069}{193536}\right]\right\} \nonumber\\
   & + y^3 y_0^3 \left\{\chi_a \chi_s \left(\frac{8635 \delta  \nu }{729}-\frac{24649 \delta }{1458}\right)+\left(-\frac{3025 \nu ^2}{729}+\frac{8635 \nu }{729}-\frac{24649}{2916}\right) \chi_s^2+\left(\frac{24649 \nu
   }{729}-\frac{24649}{2916}\right) \chi_a^2\right\} \nonumber\\
   & + y^4 y_0^2 \left\{\chi_a^2 \left[\kappa_a \left(\frac{17533 \delta  \nu }{6912}-\frac{252137 \delta }{193536}\right)+\kappa_s \left(-\frac{17533 \nu ^2}{3456}+\frac{497599 \nu }{96768}-\frac{252137}{193536}\right)-\frac{122731
   \nu ^2}{3456}+\frac{367579 \nu }{13824} \right.\right.\nonumber\\
   &\left.- \frac{830069}{193536}\right]+\chi_a \chi_s \left[\kappa_s \left(\frac{17533 \delta  \nu }{3456}-\frac{252137 \delta }{96768}\right)+\frac{57721 \delta  \nu
   }{3456}-\frac{830069 \delta }{96768}+\kappa_a \left(-\frac{17533 \nu ^2}{1728}+\frac{497599 \nu }{48384} \right.\right.\nonumber\\
   &\left.\left.- \frac{252137}{96768}\right)\right]+\chi_s^2 \left[\kappa_a \left(\frac{17533 \delta  \nu
   }{6912}-\frac{252137 \delta }{193536}\right)+\kappa_s \left(-\frac{17533 \nu ^2}{3456}+\frac{497599 \nu }{96768}-\frac{252137}{193536}\right)+\frac{7289 \nu ^2}{3456} \right.\nonumber\\
   &\left.\left.+ \frac{703273 \nu }{96768}-\frac{830069}{193536}\right]\right\}\,.
   \end{align}
   \end{subequations}
\end{widetext}
\subsubsection{Evolution of time to coalescence as a function of $y$}
Equation~\eqref{eq:ydot-structure} can now be inverted to obtain the evolution of time with respect to the PN parameter $y$. The resulting expression is $\frac{dt}{dy}$ expanded in terms of the PN parameter $y$ and eccentricity $e$.
Further, Eq.~\eqref{eq:e2-y} is used to re-express $\frac{dt}{dy}$ explicitly in terms of parameters $y$, $y_0$, and $e_0$, up to 3PN and $\mathcal{O}(e_0^8)$. One then obtains
\begin{eqnarray}\label{eq:dtBdystructure}
    \frac{dt}{dy} &=& \frac{5 M}{32 y^9 \nu}\left(T_{\rm NS} + T_{\rm SO} + T_{\rm SS}\right)\,,
\end{eqnarray}
where, 
\begin{widetext}
    \begin{subequations}
    \begin{align}
        T_{\rm NS} &= 1  +  \frac{5 e_0^2}{8} \left(\frac{y_0}{y}\right)^{19/3} + \mathcal{O}\left(y^2 , e_0^4\right)\,, \\
        T_{\rm SO}^{\rm 1.5PN} &= y^3 \left[\frac{113 \delta  \chi_a}{12}+\left(\frac{113}{12}-\frac{19 \nu }{3}\right) \chi_s\right] + e_0^2 \left(\frac{y_0}{y}\right)^{19/3} \left\{ y^3 \left[\frac{611 \delta  \chi_a}{27}+\left(\frac{611}{27}-\frac{1319 \nu }{216}\right) \chi_s\right] + y_0^3 \left[\left(\frac{275 \nu }{216} \right.\right.\right. \nonumber\\
        & \left.\left.\left.- \frac{785}{432}\right) \chi_s-\frac{785 \delta  \chi_a}{432}\right] \right\} + \mathcal{O}\left(e_0^4\right)\,, \\
        T_{\rm SS}^{2 PN} &= y^4 \left[\chi_a^2 \left\{-5 \delta  \kappa_a +(10 \nu -5) \kappa_s +10 \nu -\frac{1}{16}\right\}+\chi_a \chi_s \left\{(20 \nu -10) \kappa_a-10 \delta  \kappa
   _S-\frac{\delta }{8}\right\} \right.\nonumber\\
   &\left.+\chi_s^2 \left\{-5 \delta  \kappa_a+(10 \nu -5) \kappa_s-\frac{39 \nu }{4}-\frac{1}{16}\right\}\right] + e_0^2 \left(\frac{y_0}{y}\right)^{19/3} \left[y^4 \left\{\chi_a^2 \left[-\frac{5149 \delta  \kappa_a}{384}+\frac{5149 \nu }{192}+\left(\frac{5149 \nu }{192} \right.\right.\right.\right.\nonumber\\
   &\left.\left. \left. - \frac{5149}{384}\right) \kappa_s -\frac{197}{768}\right]+\chi_A \chi_s \left[\left(\frac{5149 \nu }{96}-\frac{5149}{192}\right) \kappa_a-\frac{197 \delta }{384}-\frac{5149 \delta  \kappa_s}{192}\right]+\chi_s^2 \left[-\frac{5149
   \delta  \kappa_a}{384}-\frac{619 \nu }{24}+\left(\frac{5149 \nu }{192} \right.\right.\right. \nonumber\\
   &\left.\left.\left. -\frac{5149}{384}\right) \kappa_s-\frac{197}{768}\right]\right\} + y_0^4 \left\{\chi_a^2
   \left[\frac{445 \delta  \kappa_a}{384}-\frac{445 \nu }{192}+\left(\frac{445}{384}-\frac{445 \nu }{192}\right) \kappa_s+\frac{65}{768}\right]+\chi_a \chi_s
   \left[\left(\frac{445}{192}-\frac{445 \nu }{96}\right) \kappa_a \right.\right.\nonumber\\
   &\left.\left.\left.+ \frac{65 \delta }{384}+\frac{445 \delta  \kappa_s}{192}\right] +\chi_s^2 \left[\frac{445 \delta  \kappa
   _A}{384}+\frac{95 \nu }{48}+\left(\frac{445}{384}-\frac{445 \nu }{192}\right) \kappa_s+\frac{65}{768}\right]\right\}\right] + \mathcal{O}\left(e_0^4\right)\,, \\
   T_{\rm SO}^{\rm 2.5PN} &= y^5 \left[\chi_a \left(\frac{7 \delta  \nu }{2}+\frac{146597 \delta }{2016}\right)+\left(-\frac{17 \nu ^2}{2}-\frac{1213 \nu }{18}+\frac{146597}{2016}\right) \chi_s\right] + e_0^2 \left(\frac{y_0}{y}\right)^{19/3} \left[ y^5 \left\{\chi_a \left(\frac{1203689 \delta  \nu }{5184}\right. \right.\right.\nonumber\\
   &\left.\left. -\frac{13842365 \delta }{435456}\right) +\left(-\frac{39331 \nu ^2}{324}+\frac{19696429 \nu
   }{108864}-\frac{13842365}{435456}\right) \chi_s\right\}+y_0^2 y^3 \left\{\chi_a \left(\frac{1730963 \delta }{27216}-\frac{120367 \delta  \nu
   }{972}\right) +  \right.\nonumber\\
   &\left. \left(\frac{259843 \nu ^2}{7776}-\frac{30698935 \nu }{217728} + \frac{1730963}{27216}\right) \chi_s\right\}+y_0^3 y^2 \left\{\chi_a \left(\frac{26089475 \delta
   }{435456}-\frac{316355 \delta  \nu }{5184}\right)+\left(\frac{110825 \nu ^2}{2592} \right.\right. \nonumber\\
   &\left.\left.\left. -\frac{22426535 \nu }{217728} +\frac{26089475}{435456}\right) \chi_s\right\} + y_0^5
   \left\{\chi_a \left(\frac{242719 \delta  \nu }{15552}-\frac{1279073 \delta }{62208}\right)+\left(-\frac{10805 \nu ^2}{972}+\frac{146807 \nu
   }{3888} \right.\right.\right.\nonumber \\
   &\left.\left.\left. -\frac{1279073}{62208}\right) \chi_s\right\} \right]+ \mathcal{O}\left(e_0^4\right) \, ,\\
   T_{\rm SO}^{\rm 3PN} &= y^6 \left[ \left(26 \pi  \nu -\frac{227 \pi }{6}\right) \chi_s-\frac{227}{6} \pi  \delta  \chi_a \right] + e_0^2 \left(\frac{y_0}{y}\right)^{19/3} \left[ y^6 \left\{\left(\frac{4052537 \pi  \nu }{31104}-\frac{31241125 \pi }{124416}\right) \chi_s-\frac{31241125 \pi  \delta  \chi_a}{124416}\right\} \right.\nonumber\\
   &\left.+ y_0^3 y^3 \left\{\frac{4987553
   \pi  \delta  \chi_a}{31104}+\left(\frac{4987553 \pi }{31104}-\frac{476689 \pi  \nu }{7776}\right) \chi_s\right\}+y_0^6 \left\{\left(\frac{188315 \pi  \nu
   }{31104}-\frac{785215 \pi }{124416}\right) \chi_s-\frac{785215 \pi  \delta  \chi_a}{124416}\right\} \right] \, \nonumber \\
   &+ \mathcal{O}\left(e_0^4\right)\,, \nonumber\\
   \\
   T_{\rm SS}^{\rm 3PN} &= y^6 \left[ \chi_a^2 \left\{\kappa_a \left(\frac{299 \delta  \nu }{24}-\frac{5203 \delta }{112}\right)+12 \nu ^2-\frac{53563 \nu }{1008}+\left(12 \nu ^2+\frac{8851 \nu
}{84}-\frac{5203}{112}\right) \kappa_s+\frac{268895}{8064}\right\} \right.\nonumber\\
&+ \chi_a \chi_s \left\{\left(24 \nu ^2+\frac{8851 \nu }{42}-\frac{5203}{56}\right) \kappa_a-\frac{149
	\delta  \nu }{72}+\frac{268895 \delta }{4032}+\left(\frac{299 \delta  \nu }{12}-\frac{5203 \delta }{56}\right) \kappa_s\right\}+\chi_s^2 \left\{\kappa_a \left(\frac{299
	\delta  \nu }{24} \right. \right.\nonumber\\
&\left.\left.\left. -\frac{5203 \delta }{112}\right) - \frac{683 \nu ^2}{36}-\frac{165941 \nu }{2016}+\left(12 \nu ^2+\frac{8851 \nu }{84}-\frac{5203}{112}\right) \kappa_s+\frac{268895}{8064}\right\} \right] \nonumber\\
&+ e_0^2 \left(\frac{y_0}{y}\right)^{19/3} \left[ y^6 \left\{\chi_a^2 \left[\kappa_a \left(-\frac{836147 \delta  \nu }{13824}-\frac{4689653 \delta }{129024}\right) + \frac{197635 \nu ^2}{768}-\frac{819409469 \nu
}{1306368}+\left(\frac{197635 \nu ^2}{768}+\frac{2362901 \nu }{193536} \right.\right. \right.\right.\nonumber\\
& \left.\left. -\frac{4689653}{129024}\right) \kappa_s+\frac{3345863285}{20901888}\right]  +\chi_a \chi_s
\left[\left(\frac{197635 \nu ^2}{384}+\frac{2362901 \nu }{96768} - \frac{4689653}{64512}\right) \kappa_a-\frac{13785319 \delta  \nu }{373248}  \right.\nonumber\\
&\left. +\frac{3345863285 \delta
}{10450944}+\left(-\frac{836147 \delta  \nu }{6912}-\frac{4689653 \delta }{64512}\right) \kappa_s\right] +\chi_s^2 \left[\kappa_a \left(-\frac{836147 \delta  \nu
}{13824}-\frac{4689653 \delta }{129024}\right) \right. \left.\left.- \frac{25616525 \nu ^2}{93312}   \right. \right.\nonumber\\
& \left. \left. -\frac{37317125 \nu }{746496}+\left(\frac{197635 \nu ^2}{768}+\frac{2362901 \nu
}{193536}-\frac{4689653}{129024}\right) \kappa_s +\frac{3345863285}{20901888}\right]\right\} + y_0^2 y^4 \left\{\chi_a^2 \left[\kappa_a \left(\frac{1014353 \delta  \nu
}{13824} \right. \right.\right. \nonumber\\
& \left.\left. - \frac{14587117 \delta }{387072}\right)-\frac{1014353 \nu ^2}{6912}+\frac{3714695 \nu }{48384}+\left(-\frac{1014353 \nu ^2}{6912} +\frac{28788059 \nu
}{193536} -\frac{14587117}{387072}\right) \kappa_s-\frac{558101}{774144}\right]  \nonumber\\
& + \chi_a \chi_s \left[\left(-\frac{1014353 \nu ^2}{3456}+\frac{28788059 \nu
}{96768}-\frac{14587117}{193536}\right) \kappa_a+\frac{38809 \delta  \nu }{13824} \left.  -\frac{558101 \delta }{387072}+\left(\frac{1014353 \delta  \nu }{6912}-\frac{14587117
	\delta }{193536}\right) \kappa_s\right] \right.  \nonumber \\ 
& \left. + \chi_s^2 \left[\kappa_a \left(\frac{1014353 \delta  \nu }{13824}-\frac{14587117 \delta }{387072}\right)+\frac{121943 \nu
	^2}{864}  -\frac{13757353 \nu }{193536} +\left(-\frac{1014353 \nu ^2}{6912}+\frac{28788059 \nu }{193536}-\frac{14587117}{387072}\right) \kappa_s \right. \right. \nonumber\\
& \left. \left. -\frac{558101}{774144}\right]\right\} + y_0^3 y^3 \left\{\left(\frac{191854 \nu }{729}-\frac{95927}{1458}\right) \chi_a^2  +\chi_a \left(\frac{744763 \delta  \nu
}{11664}  -\frac{95927 \delta }{729}\right) \chi_s+\left(-\frac{72545 \nu ^2}{5832}+\frac{744763 \nu }{11664} \right. \right. \nonumber\\
& \left. \left. -\frac{95927}{1458}\right) \chi_s^2\right\} + y_0^4 y^2
\left\{\chi_a^2 \left[\kappa_a \left(\frac{179335 \delta  \nu }{4608}-\frac{14789575 \delta }{387072}\right)-\frac{179335 \nu ^2}{2304} +\frac{7669835 \nu
}{96768}+\left(-\frac{179335 \nu ^2}{2304} \right.\right. \right. \nonumber\\
& \left. \left.\left. +\frac{22321645 \nu }{193536}-\frac{14789575}{387072}\right) \kappa_s \right.\right.\left.- \frac{2160275}{774144}\right]+\chi_a \chi_s
\left[\left(-\frac{179335 \nu ^2}{1152}+\frac{22321645 \nu }{96768} -\frac{14789575}{193536}\right) \kappa_a   \right.  \nonumber\\
& \left. +\frac{26195 \delta  \nu }{4608}-\frac{2160275 \delta
}{387072}+\left(\frac{179335 \delta  \nu }{2304}-\frac{14789575 \delta }{193536}\right) \kappa_s\right] \left.+ \chi_s^2 \left[\kappa_a \left(\frac{179335 \delta  \nu
}{4608}-\frac{14789575 \delta }{387072}\right) +\frac{38285 \nu ^2}{576} \right. \right. \nonumber\\
&  \left. \left. -\frac{12079205 \nu }{193536}+\left(-\frac{179335 \nu ^2}{2304}+\frac{22321645 \nu
}{193536}-\frac{14789575}{387072}\right) \kappa_s-\frac{2160275}{774144}\right]\right\} + y_0^6 \left\{\chi_a^2 \left[\kappa_a \left(\frac{4530515 \delta
}{387072} \right.\right. \right.\nonumber\\
&\left.\left. -\frac{64385 \delta  \nu }{4608}\right)+\frac{121355 \nu ^2}{6912}-\frac{105953815 \nu }{2612736}+\left(\frac{121355 \nu ^2}{6912}-\frac{7234685 \nu
}{193536}+\frac{4530515}{387072}\right) \kappa_s+\frac{122165975}{20901888}\right] \nonumber\\
& + \chi_a \chi_s \left[\left(\frac{121355 \nu ^2}{3456}-\frac{7234685 \nu
}{96768}+\frac{4530515}{193536}\right) \kappa_a-\frac{5993155 \delta  \nu }{373248}+\frac{122165975 \delta }{10450944}+\left(\frac{4530515 \delta }{193536}-\frac{64385
	\delta  \nu }{2304}\right) \kappa_s\right] \nonumber\\
&+ \chi_s^2 \left[\kappa_a \left(\frac{4530515 \delta }{387072}-\frac{64385 \delta  \nu }{4608}\right)-\frac{957695 \nu
	^2}{93312}+\frac{5837485 \nu }{5225472}+\left(\frac{121355 \nu ^2}{6912}-\frac{7234685 \nu }{193536}+\frac{4530515}{387072}\right) \kappa_s \right.  \nonumber\\
& \left.\left.\left.+\frac{122165975}{20901888}\right]\right\} \right] + \mathcal{O}\left(e_0^4\right)\,.
   \end{align}
    \end{subequations}
\end{widetext}
The coalescence time $t$ can be obtained by integrating $dt = dy/\left(dy/dt\right)$. Now that we have already inverted Eq.~\eqref{eq:ydot-structure}, we can obtain the time to coalescence as a function of initial eccentricity and PN parameter $y$ by directly integrating Eq.~\eqref{eq:dtBdystructure}.
\subsection{Taylor approximants for eccentric, spin-aligned compact binaries}\label{subsec:TaylorApproximants}

{\blue As indicated above, different ways of solving Eq.~\eqref{eq:approx_timedom} (or equivalently for parameter $y$) are referred to as Taylor approximants. More specifically, the expressions for time ($t$) and phase ($\phi$) as a function of frequency ($x$ or $y$) and also in terms of a reference eccentricity ($e_0$) computed at a reference frequency ($x_0$ or $y_0$). We first give a quick overview of each of the phasing approximants computed here and then present explicit results for specific approximants which are fully analytical -- \TTtwo{}, \TFtwo{} and a resummed \TTtwo{}. Expressions for the semi-analytical approximants -- \TTone{}, \TTthree{} and \TTfour{} are explicitly listed in Appendix~\ref{sec: app-Taylor}.     
}

To approximate the \TTone{} series up to a specific PN order, we require the energy and energy flux expressed in terms of the PN parameter up to that PN order. In one of the supplementary files of Ref.~\citep{Henry:2023tka}, we can find the mean motion $n$, periastron advance $k$, and time eccentricity $e_t$ in terms of the binary's conserved energy and angular momentum. We observe that the orbital frequency $\omega$ is defined in Eq.~\eqref{eq:omega-def}, and the corresponding dimensionless PN parameter is defined in Eq.~\eqref{eq:x-def}. By multiplying the mean motion series with the periastron advance series in Ref.~\citep{Henry:2023tka} and truncating it to 3PN, we can find $\omega$ in terms of energy and angular momentum. Transforming from $\omega$ to $x$ using Eq.~\eqref{eq:x-def} is straightforward. As a result, we now have $x$ and $e_t$ in terms of the energy and angular momentum. Inverting the two series leads to the energy and angular momentum expressed in terms of $x$ and $e_t$. By transforming the energy to the PN parameter $y$, substituting the solution of $e_t$ from Eq.~\eqref{eq:e2-y}, and truncating it to 3PN, we have obtained the energy per unit mass $\mathcal{E}$ and the energy flux $\mathcal{F}$ in terms of $e_0$, $y_0$, and $y$ and have shown it in Appendix \ref{sec: app-Taylor}. Similarly, the \TTtwo{} approximant is obtained by performing a series expansion of the ratio on the RHS of Eq.~\eqref{eq:approx_timedom_v} and then solving the following set of equations by integration (see, for instance Sec. VIB of Ref.~\citep{Moore:2016qxz})

 The evolution equation of the \TTtwo{} phasing can be written as follows
\begin{subequations}
\label{eq:approx_timedom2}
\begin{align}
\label{eq:approx_timedom2_phi}
\frac{d\langle\phi\rangle}{dy} &= \frac{d\langle\phi\rangle}{dt}\frac{dt}{dy}=\left(1-e_t^2\right)^{3/2}{y^{3}\over M}\frac{dt}{dy}\,,
\\
\label{eq:approx_timedom2_t}
\frac{dt}{dy} &= -\frac{dE(y, e_t)/dy}{{\langle {\mathcal F }(y, e_t)\rangle}}\,,
\end{align}
\end{subequations}
where $\langle \cdot \rangle$ represent averaged quantities.\footnote{Note, $\phi\simeq \langle\phi\rangle$ over radiation reaction time-scales.}

Further, Eq.~\eqref{eq:dtBdystructure} can directly be used in Eq.~\eqref{eq:approx_timedom2_phi}, which can be integrated to obtain the \TTtwo{} phasing as a function of $e_0$ and $y$. This then gives us the  desired parametric solution for \TTtwo{}, i.e. $\left\{t(y, y_0, e_0), \phi(y, y_0, e_0)\right\}$. They're given by,
\begin{subequations}
\begin{equation}
\begin{split}
    t =  &- \frac{5 M}{256 \nu y^8} \biggl\{t_{\rm circ} +\frac{15}{43}e_0^2 \biggl(\frac{y_0}{y}\biggr)^{19/3} \biggl[ t_{\rm SO,ecc} + t_{\rm SS,ecc}\biggr] \\
    & 
    + \cdots + \mathcal{O}(e_0^{8}) \biggr\}\,,
\end{split}
\end{equation}
\begin{equation}
\begin{split}
    \phi =  &- \frac{1}{32 \nu y^5} \biggl\{\phi_{\rm circ} -\frac{105}{272}e_0^2 \biggl(\frac{y_0}{y}\biggr)^{19/3} \biggl[ \phi_{\rm SO,ecc} + \phi_{\rm SS,ecc} 
    \biggr] \\
    & 
    + \cdots + \mathcal{O}(e_0^{8}) \biggr\}\,,
\end{split}
\end{equation}
\end{subequations}

The \TTthree{} approximant is derived following the exact process followed by Ref.~\citep{Moore:2016qxz} in Sec VI. C. To recapitulate, $y(t)$ is obtained by performing a series inversion of $t(y)$ using an ansatz. Then, there is a simultaneous transformation of variables from $y$ to $F=y^3 (1-e^2)/(\pi M)$ and $t$ to $\theta = \left[\eta(t_c - t)/5M\right]^{-1/8}$. This gives us an expression for $F(\theta)$, including spins up to 3PN of the following form

\begin{subequations}
\begin{equation}
    F =  \frac{\theta^3}{8 M \pi} \biggl\{F_{\rm circ} -\frac{471}{344}e_0^2 \biggl(\frac{\theta_0}{\theta}\biggr)^{19/3} \biggl[F_{\rm SO,ecc} 
    + F_{\rm SS,ecc}  \biggr] \biggr\}\,,
\end{equation}
\begin{align}
    \langle \phi \rangle - \phi_c &=  -\frac{1}{\nu \theta^5} \biggl\{\phi_{\rm circ} -\frac{7065}{11696}e_0^2 \biggl(\frac{\theta_0}{\theta}\biggr)^{19/3} \biggl[\phi_{\rm SO,ecc} \nonumber \\
    &+ \phi_{\rm SS,ecc} \biggr] \biggr\}\,.
\end{align}
\end{subequations}

It must be noted that, as has been described in Sec. VI C of Ref.~\citep{Moore:2016qxz} and further elaborated upon in Ref.~\citep{buonanno-iyer-oshsner-pan-sathya-templatecomparison-PRD2009}, the \TTthree{} approximant behaves in a non-monotonic manner at certain PN orders. This renders its usage in practical applications ill-advised, and we have provided their detailed expressions in Appendix \ref{sec: app-Taylor} for the sake of completeness.

The \TTfour{} approximant is computed by expanding Eq.~\eqref{eq:ydot-structure} in $y$ till 3PN and in eccentricity till {\blue $\mathcal{O}\left(e_0^8\right)$}, using the eccentricity solution from Eq.~\eqref{eq:e2-y} and then re-expanding till 3PN. The expression for $dy/dt$ with spinning terms up to $\mathcal{O}(y^6)$ is provided in Appendix~\ref{sec: app-Taylor} and is of the functional form
\begin{eqnarray}
    \frac{dy}{dt} &=&   \frac{32 y^9 \nu}{5 M} \biggl\{\rho_{\rm circ} -\frac{5}{8}e_0^2 \biggl(\frac{y_0}{y}\biggr)^{19/3} \biggl[ \rho_{\rm SO,ecc} + \rho_{\rm SS,ecc} \biggr] \nonumber\\
    && 
    + \cdots+ \mathcal{O}(e_0^8)\biggr\}\,.
\end{eqnarray}

We also calculate the spinning eccentric corrections in the Fourier domain \TFtwo{} phase by performing a stationary phase approximation (SPA) of a time domain waveform \citep{Moore:2016qxz}. The frequency domain SPA waveform is widely used for fast parameter estimation and template bank generation and, hence, is a crucial part of this analysis. We have put the exact form of the \TFtwo{} phasing in Sec.~\ref{sec:TaylorF2}. The \TFtwo{} phase can be written in terms of the coalescence time $t$ and the \TTtwo{} phase $\phi$ as was shown in {\blue Sec. VI E of} Ref.~\citep{Moore:2016qxz}
\begin{equation}
    \Psi = 2 \pi f t(f) - 2 \phi [t(f)] + 2 \Phi_0 - \frac{\pi}{4}\,,
\end{equation}
where $\Psi$ has a structure given by
\begin{eqnarray}
    \Psi &=& \frac{3}{128 \nu y^5} \biggl\{\Psi_{\rm circ} +\frac{650}{731}e_0^2 \biggl(\frac{y_0}{y}\biggr)^{19/3} \biggl[ \Psi_{\rm SO,ecc} + \Psi_{\rm SS,ecc}\biggr] \nonumber\\
    &&  
    + \cdots + \mathcal{O}(e_0^{8})\biggr\}. \nonumber\\
\end{eqnarray}
\subsubsection{\TTtwo{}}\label{sec:TaylorT2}

The coalescence time as a function of the modified orbital frequency $y$ can be found out by integrating both sides of Eq.~\eqref{eq:dtBdystructure} with respect to $y$, as described in Sec.~\ref{subsec:TaylorApproximants}. The structure of the coalescence time {\blue (ignoring all higher order circular and non-spinning, eccentric corrections)}, and the corresponding SO and SS coefficients read as
\begin{widetext}

\begin{equation} \label{TaylorT2 time}
\begin{split}
    t =  &- \frac{5 M}{256 \nu y^8} \biggl\{ 
    1
    +\frac{15}{43}e_0^2 \biggl(\frac{y_0}{y}\biggr)^{19/3} \biggl[ 
    1
    + t_{\rm SO,ecc} + t_{\rm SS,ecc}\biggr] 
    + \cdots + \mathcal{O}(e_0^{8}) \biggr\}\,,
\end{split}
\end{equation}
where,
\begin{subequations}
\begin{align}
   t^{\rm 1.5PN}_{\rm SO,ecc}&= y^3 \biggl[\frac{105092 \delta  \chi_a}{2295}+\frac{105092}{2295} \biggl(1-\frac{1319 \nu }{4888}\biggr) \chi_s\biggr] +y_0^3 \biggl[-\frac{157 \delta  \chi_a}{54}-\frac{157}{54} \biggl(1-\frac{110 \nu }{157}\biggr) \chi_s\biggr]\,, \\
   t^{\rm 2PN}_{\rm SS,ecc}&=y_0^4 \biggl\{\biggl[\frac{13}{96}+\frac{89 \delta  \kappa_a}{48}+\frac{89}{48} \kappa_s (1-2 \nu )-\frac{89 \nu }{24}\biggr] \chi_a^2+\biggl[\delta  \biggl(\frac{13}{48}+\frac{89 \kappa_s}{24}\biggr)+\frac{89}{24} \kappa_a (1-2 \nu )\biggr]
   \chi_a \chi_s\nonumber\\
   &+\biggl[\frac{13}{96}+\frac{89 \delta  \kappa_a}{48}+\frac{89}{48} \kappa_s (1-2 \nu )+\frac{19 \nu }{6}\biggr] \chi_s^2\biggr\}+y^4
   \biggl\{\chi_a \chi_s \biggl[\delta  \biggl(-\frac{8471}{7440}-\frac{221407 \kappa_s}{3720}\biggr)\nonumber\\
   &-\frac{221407}{3720} (1-2 \nu
   ) \kappa_a\biggr]+\chi_s^2 \biggl[-\frac{8471}{14880}-\frac{221407 \delta  \kappa_a}{7440}-\frac{26617 \nu }{465}-\frac{221407}{7440} (1-2
   \nu )\kappa_s\biggr]\nonumber\\
   &+\chi_a^2 \biggl[-\frac{8471}{14880}-\frac{221407 \delta  \kappa_a}{7440}-\frac{221407 \nu }{3720}-\frac{221407}{7440} (1-2
   \nu )\kappa_s\biggr]\biggr\}\,,\\
   t^{\rm 2.5PN}_{\rm SO,ecc}&=y^2 y_0^3 \biggl[ \frac{224369485}{2013984} \biggl(1-\frac{33852
   \nu }{33235}\biggr)\delta  \chi_a+ \frac{224369485}{2013984} \biggl(1-\frac{8970614 \nu }{5217895}+\frac{744744 \nu ^2}{1043579}\biggr)\chi_s\biggr]\nonumber\\
   &+y^3
   y_0^2 \biggl[\frac{74431409}{578340} \biggl(1-\frac{5516 \nu }{2833}\biggr)\delta  \chi_a+ \frac{74431409}{578340}
   \biggl(1-\frac{30698935 \nu }{13847704}+\frac{1818901 \nu ^2}{3461926}\biggr)\chi_s\biggr]\nonumber\\
   &+y_0^5 \biggl[-\frac{1279073}{38880}
   \biggl(1-\frac{2348912 \nu }{1279073}+\frac{691520 \nu ^2}{1279073}\biggr)\chi_s-\frac{1279073}{38880} \biggl(1-\frac{970876 \nu
   }{1279073}\biggr)\delta  \chi_a\biggr]\nonumber\\
   &+y^5 \biggl[-\frac{119044339}{1524096} \biggl(1-\frac{78785716 \nu }{13842365}+\frac{52860864 \nu
   ^2}{13842365}\biggr)\chi_s-\frac{119044339}{1524096} \biggl(1-\frac{101109876 \nu }{13842365}\biggr)\delta  \chi_a\biggr]\,, \\ 
    t^{\rm 3PN}_{\rm SO,ecc}&=y_0^6 \biggl[-\frac{157043 \pi  \delta  \chi_a}{15552}-\frac{157043 \pi }{15552} \biggl(1-\frac{150652 \nu
   }{157043}\biggr)\chi_s\biggr]+y^3 y_0^3 \biggl[\frac{214464779 \pi  \delta  \chi_a}{660960}\nonumber\\
   &+ \frac{214464779 \pi }{660960}
   \biggl(1-\frac{1906756 \nu }{4987553}\biggr)\chi_s\biggr]+y^6 \biggl[-\frac{10746947 \pi  \delta  \chi_a}{15552} -\frac{10746947 \pi }{15552}
   \biggl(1-\frac{16210148 \nu }{31241125}\biggr)\chi_s\biggr]\,, \\
     t^{\rm 3PN}_{\rm SS,ecc} &=y^3 y_0^3 \biggl\{-\frac{8249722}{61965} (1-4 \nu ) \chi_a^2-\frac{16499444 \delta
   }{61965} \biggl(1-\frac{744763 \nu }{1534832}\biggl) \chi_a \chi_s-\frac{8249722}{61965}
   \biggl(1-\frac{744763 \nu }{767416}\nonumber\\
   & +\frac{72545 \nu ^2}{383708}\biggl) \chi_s^2\biggl\} +y^4 y_0^2
   \biggl\{\biggl[-\frac{23998343}{14999040} \biggl(1-\frac{59435120 \nu }{558101}+\frac{576688 \nu
   ^2}{2833}\biggl)-\frac{627246031 \delta }{7499520} \biggl(1-\frac{5516 \nu }{2833}\biggl) \kappa_a\nonumber\\
   &-\frac{627246031}{7499520} \biggl(1-\frac{11182 \nu }{2833}+\frac{11032 \nu ^2}{2833}\biggl)
   \kappa_s\biggl] \chi_a^2+\biggl[-\frac{23998343 \delta }{7499520} \biggl(1-\frac{5516 \nu
   }{2833}\biggl)-\frac{627246031}{3749760} \biggl(1-\frac{11182 \nu }{2833}\nonumber\\
   & +\frac{11032 \nu
   ^2}{2833}\biggl) \kappa_a -\frac{627246031 \delta }{3749760} \biggl(1-\frac{5516 \nu
   }{2833}\biggl) \kappa_s\biggl] \chi_a \chi_s+\biggl[-\frac{23998343}{14999040} \biggl(1+\frac{55029412
   \nu }{558101}-\frac{554624 \nu ^2}{2833}\biggl) \nonumber\\
   & -\frac{627246031 \delta }{7499520} \biggl(1-\frac{5516 \nu
   }{2833}\biggl) \kappa_a -\frac{627246031}{7499520} \biggl(1-\frac{11182 \nu }{2833}+\frac{11032
   \nu ^2}{2833}\biggl) \kappa_s\biggl] \chi_s^2\biggl\}+y^2 y_0^4 \biggl\{\biggl[-\frac{18578365}{3580416}
   \biggl(1 \nonumber\\
   & -\frac{12271736 \nu }{432055}+\frac{927024 \nu ^2}{33235}\biggl) -\frac{127190345 \delta
   }{1790208} \biggl(1-\frac{33852 \nu }{33235}\biggl) \kappa_a-\frac{127190345}{1790208}
   \biggl(1-\frac{100322 \nu }{33235}+\frac{67704 \nu ^2}{33235}\biggl) \kappa_s\biggl] \chi_a^2 \nonumber\\
   & +\biggl[-\frac{18578365 \delta }{1790208} \biggl(1-\frac{33852 \nu
   }{33235}\biggl) -\frac{127190345}{895104} \biggl(1-\frac{100322 \nu }{33235}+\frac{67704 \nu
   ^2}{33235}\biggl) \kappa_a-\frac{127190345 \delta }{895104} \biggl(1\nonumber\\
   & -\frac{33852 \nu
   }{33235}\biggl) \kappa_s\biggl] \chi_a \chi_s +\biggl[-\frac{18578365}{3580416} \biggl(1+\frac{9663364
   \nu }{432055}-\frac{791616 \nu ^2}{33235}\biggl)-\frac{127190345 \delta }{1790208} \biggl(1-\frac{33852
   \nu }{33235}\biggl) \kappa_a \nonumber\\
   & -\frac{127190345}{1790208} \biggl(1-\frac{100322 \nu
   }{33235} +\frac{67704 \nu ^2}{33235}\biggl) \kappa_s\biggl] \chi_s^2\biggl\}+y_0^6
   \biggl\{\biggl[\frac{24433195}{2612736} \biggl(1-\frac{169526104 \nu }{24433195}+\frac{73395504 \nu
   ^2}{24433195}\biggl) \nonumber\\
   & +\frac{906103 \delta }{48384} \biggl(1-\frac{1081668 \nu }{906103}\biggl)
   \kappa_a +\frac{906103}{48384} \biggl(1-\frac{2893874 \nu }{906103}+\frac{1359176 \nu
   ^2}{906103}\biggl) \kappa_s\biggl] \chi_a^2+\biggl[\frac{24433195 \delta }{1306368}
   \biggl(1\nonumber\\
   & -\frac{33561668 \nu }{24433195}\biggl)+\frac{906103}{24192} \biggl(1-\frac{2893874 \nu
   }{906103} +\frac{1359176 \nu ^2}{906103}\biggl) \kappa_a+\frac{906103 \delta }{24192}
   \biggl(1-\frac{1081668 \nu }{906103}\biggl) \kappa_s\biggl] \chi_a \chi
   _S\nonumber\\
   & +\biggl[\frac{24433195}{2612736} \biggl(1+\frac{4669988 \nu }{24433195}-\frac{42904736 \nu
   ^2}{24433195}\biggl) +\frac{906103 \delta }{48384} \biggl(1-\frac{1081668 \nu }{906103}\biggl)
   \kappa_a+\frac{906103}{48384} \biggl(1-\frac{2893874 \nu }{906103} \nonumber\\
   & +\frac{1359176 \nu
   ^2}{906103}\biggl) \kappa_s\biggl] \chi_s^2\biggl\} +y^6 \biggl\{\biggl[\frac{28774424251}{65318400}
   \biggl(1-\frac{13110551504 \nu }{3345863285}+\frac{1075766832 \nu ^2}{669172657}\biggl)-\frac{201655079
   \delta }{2016000} \biggl(1\nonumber\\
   & +\frac{23412116 \nu }{14068959}\biggl) \kappa_a -\frac{201655079}{2016000} \biggl(1-\frac{4725802 \nu }{14068959}-\frac{33202680 \nu
   ^2}{4689653}\biggl) \kappa_s\biggl] \chi_a^2+\biggl[\frac{28774424251 \delta }{32659200}
   \biggl(1-\frac{385988932 \nu }{3345863285}\biggl)\nonumber\\
   & -\frac{201655079}{1008000} \biggl(1-\frac{4725802 \nu
   }{14068959} -\frac{33202680 \nu ^2}{4689653}\biggl) \kappa_a-\frac{201655079 \delta }{1008000}
   \biggl(1+\frac{23412116 \nu }{14068959}\biggl) \kappa_s\biggl] \chi_a \chi
   _s\nonumber\\
   & +\biggl[\frac{28774424251}{65318400} \biggl(1-\frac{208975900 \nu }{669172657}-\frac{1147620320 \nu
   ^2}{669172657}\biggl) -\frac{201655079 \delta }{2016000} \biggl(1+\frac{23412116 \nu
   }{14068959}\biggl) \kappa_a-\frac{201655079}{2016000} \biggl(1 \nonumber\\
   &-\frac{4725802 \nu
   }{14068959}-\frac{33202680 \nu ^2}{4689653}\biggl) \kappa_s\biggl] \chi_s^2\biggl\}\,.
\end{align}
\end{subequations}
\end{widetext}
The \TTtwo{} phase can then be obtained by integrating both sides of Eq.~\eqref{eq:approx_timedom2_phi} with respect to $y$, as described in Sec.~\ref{subsec:TaylorApproximants}. This is the time domain orbital phasing formula and is given by the following form, which is followed by the various coefficients of the eccentric SO and SS terms {\blue (ignoring all higher order circular and non-spinning, eccentric corrections)}, as they appear in the structure of the phase
\begin{widetext}
\begin{equation} \label{TaylorT2 phase}
\begin{split}  
     \phi =  &- \frac{1}{32 \nu y^5} \biggl\{ 
     1
    -\frac{105}{272}e_0^2 \biggl(\frac{y_0}{y}\biggr)^{19/3} \biggl[
    1+\phi_{\rm SO,ecc} + \phi_{\rm SS,ecc} 
    \biggr]  
    + \cdots + \mathcal{O}(e_0^{8}) \biggr\}\,,
\end{split}
\end{equation}
where,
\begin{subequations}
\begin{align}
    \phi^{\rm 1.5PN}_{\rm SO,ecc} &=y_0^3
   \biggl[-\frac{157 \delta  \chi_a}{54}- \frac{157 \chi_s}{54} \biggl(1-\frac{110 \nu }{157}\biggr)\biggr]+y^3 \biggl[-\frac{6086 \delta  \chi_a}{945}-\frac{6086 \chi_s}{945} \biggl(1+\frac{1393 \nu }{895}\biggr)\biggr] \,, \\
    \phi^{\rm 2PN}_{\rm SS,ecc} &=y_0^4 \biggl\{\biggl[\frac{293}{192} \biggl(1-\frac{1246 \nu }{293}\biggr)+\frac{89 \delta  \kappa_a}{192}+\frac{89}{192}
   (1-2 \nu ) \kappa_s\biggr] \chi_a^2+\biggl[\frac{89}{96} (1-2 \nu ) \kappa_a+\delta  \biggl(\frac{293}{96}+\frac{89 \kappa
   _S}{96}\biggr)\biggr] \chi_a \chi_s\nonumber\\
   &+\biggl[\frac{293}{192} \biggl(1+\frac{74 \nu }{293}\biggr)+\frac{89 \delta  \kappa_a}{192}+\frac{89}{192} (1-2 \nu
   ) \kappa_s\biggr] \chi_s^2\biggr\}+y^4 \biggl\{\biggl[\frac{60197}{14784} \biggl(1-\frac{16814 \nu }{3541}\biggr)+\frac{20417 \delta  \kappa_a}{14784}\nonumber\\
   &+\frac{20417}{14784} (1-2 \nu ) \kappa_s\biggr] \chi_a^2+\biggl[\frac{20417}{7392} (1-2 \nu ) \kappa_a+\delta 
   \biggl(\frac{60197}{7392}+\frac{20417 \kappa_s}{7392}\biggr)\biggr] \chi_a \chi_s+\biggl[\frac{60197}{14784} \biggl(1+\frac{2650 \nu }{3541}\biggr)\nonumber\\
   &+\frac{20417
   \delta  \kappa_a}{14784}+\frac{20417}{14784} (1-2 \nu ) \kappa_s\biggr] \chi_s^2\biggr\} \,,\\
    \phi^{\rm 2.5PN}_{\rm SO,ecc} &=y_0^5 \biggl[ 
   -\frac{1279073}{38880} \biggl(1-\frac{970876 \nu }{1279073}\biggr)\delta \chi_a-\frac{1279073}{38880} \biggl(1-\frac{2348912 \nu
   }{1279073}+\frac{691520 \nu ^2}{1279073}\biggr) \chi_s\biggr]\nonumber\\
   &+y^5 \biggl[  \frac{717229337}{2585520} \biggl(1-\frac{39620772 \nu
   }{42189961}\biggr)\delta \chi_a+\frac{717229337}{2585520} \biggl(1-\frac{432179708 \nu }{295329727}+\frac{12905280 \nu ^2}{42189961}\biggr) \chi_s\biggr]\nonumber\\
   &+y^3 y_0^2 \biggl[-\frac{8620819}{476280} \biggl(1-\frac{5516 \nu }{2833}\biggr)\delta \chi_a-\frac{8620819}{476280}
   \biggl(1-\frac{990451 \nu }{2535535}-\frac{7683788 \nu ^2}{2535535}\biggr) \chi_s\biggr]\nonumber\\
   &+y^2 y_0^3 \biggl[-\frac{424176163}{5334336}
   \biggl(1-\frac{69804 \nu }{158927}\biggr)\delta \chi_a-\frac{424176163}{5334336} \biggl(1-\frac{28441198 \nu }{24951539}+\frac{7678440 \nu
   ^2}{24951539}\biggr) \chi_s\biggr]\,,\\ 
    \phi^{\rm 3PN}_{\rm SO,ecc} &=y_0^6 \biggl[-\frac{157043 \pi  \delta  \chi_a}{15552}-\frac{157043 \pi }{15552} \biggl(1-\frac{150652 \nu }{157043}\biggr) \chi
   _S\biggr]-y^3 y_0^3 \biggl[\frac{12307507 \pi  \delta  \chi_a}{340200}\nonumber\\
   &+\frac{12307507 \pi }{340200} \biggl(1+\frac{1015892 \nu
   }{723971}\biggr) \chi_s\biggr]+y^6 \biggl[\frac{153181985 \pi  \delta  \chi_a}{870912}+\frac{153181985 \pi }{870912} \biggl(1-\frac{1475156 \nu
   }{9010705}\biggr) \chi_s\biggr]\,, \\
    \phi^{\rm 3PN}_{\rm SS,ecc} &= y^3 y_0^3 \biggl[\frac{477751}{25515} (1-4 \nu ) \chi_a^2+  \frac{955502}{25515} \biggl(1+\frac{120251 \nu }{281030}\biggr)\delta \chi_a \chi
   _S+\frac{477751}{25515} \biggl(1+\frac{120251 \nu }{140515}-\frac{30646 \nu ^2}{28103}\biggr)\chi_s^2\biggr]\nonumber\\
   &+y^2 y_0^4 \biggl\{\biggl[\frac{35122867}{9483264} \biggl(1-\frac{57485464 \nu }{2066051}+\frac{24850224
   \nu ^2}{2066051}\biggr)+  \frac{240456551}{4741632} \biggl(1-\frac{69804 \nu }{158927}\biggr)\delta
   \kappa_a\nonumber\\
   &+\frac{240456551}{4741632} \biggl(1-\frac{387658 \nu }{158927}+\frac{139608 \nu
   ^2}{158927}\biggr) \kappa_s\biggr] \chi_a^2+\biggl[\frac{240456551}{2370816} \biggl(1-\frac{387658
   \nu }{158927}+\frac{139608 \nu ^2}{158927}\biggr) \kappa_a\nonumber\\
   &+\delta  \biggl[\frac{35122867}{4741632}
   \biggl(1-\frac{69804 \nu }{158927}\biggr)+\frac{240456551}{2370816} \biggl(1-\frac{69804 \nu
   }{158927}\biggr) \kappa_s\biggr]\biggr] \chi_a \chi_s+\biggl[\frac{35122867}{9483264}
   \biggl(1+\frac{47406356 \nu }{2066051} \nonumber\\
   & -\frac{21220416 \nu ^2}{2066051}\biggr) + 
   \frac{240456551}{4741632} \biggl(1-\frac{69804 \nu }{158927}\biggr)\delta \kappa_a+\frac{240456551}{4741632} \biggl(1-\frac{387658 \nu }{158927}+\frac{139608 \nu
   ^2}{158927}\biggr) \kappa_s\biggr] \chi_s^2\biggr\}\nonumber\\
   &+y_0^6 \biggl\{\biggl[\frac{24433195}{2612736}
   \biggl(1-\frac{169526104 \nu }{24433195}+\frac{73395504 \nu ^2}{24433195}\biggr)+ 
   \frac{906103}{48384} \biggl(1-\frac{1081668 \nu }{906103}\biggr) \delta \kappa_a\nonumber\\
   &+\frac{906103}{48384} \biggl(1-\frac{2893874 \nu }{906103}+\frac{1359176 \nu ^2}{906103}\biggr)
   \kappa_s\biggr] \chi_a^2+\biggl[\frac{906103}{24192} \biggl(1-\frac{2893874 \nu }{906103}+\frac{1359176
   \nu ^2}{906103}\biggr) \kappa_a\nonumber\\
   &+\delta  \biggl[\frac{24433195}{1306368} \biggl(1-\frac{33561668 \nu
   }{24433195}\biggr)+\frac{906103}{24192} \biggl(1-\frac{1081668 \nu }{906103}\biggr) \kappa_s\biggr]\biggr] \chi_a \chi_s+\biggl[\frac{24433195}{2612736} \biggl(1+\frac{4669988 \nu
   }{24433195}\nonumber\\
   & -\frac{42904736 \nu ^2}{24433195}\biggr) +  \frac{906103}{48384} \biggl(1-\frac{1081668
   \nu }{906103}\biggr)\delta \kappa_a+\frac{906103}{48384} \biggl(1-\frac{2893874 \nu
   }{906103}+\frac{1359176 \nu ^2}{906103}\biggr)  \kappa_s\biggr] \chi_s^2\biggr\}\nonumber\\
   &+y^6
   \biggl\{\biggl[-\frac{24653523709}{146313216} \biggl(1-\frac{7638735632 \nu }{1450207277}+\frac{2575441008 \nu
   ^2}{1450207277}\biggr)  -\frac{67698947}{903168} \biggl(1-\frac{12996788 \nu
   }{11946873}\biggr)\delta \kappa_a\nonumber\\
   &-\frac{67698947}{903168} \biggl(1-\frac{36890534 \nu
   }{11946873}+\frac{15897784 \nu ^2}{3982291}\biggr) \kappa_s\biggr] \chi_a^2+\biggl[-\frac{67698947}{451584} \biggl(1-\frac{36890534 \nu }{11946873}+\frac{15897784 \nu
   ^2}{3982291}\biggr) \kappa_a\nonumber\\
   &+\delta  \biggl[-\frac{24653523709}{73156608} \biggl(1+\frac{87630620 \nu
   }{1450207277}\biggr)-\frac{67698947}{451584} \biggl(1-\frac{12996788 \nu }{11946873}\biggr) \kappa
   _S\biggr]\biggr] \chi_a \chi_s\nonumber\\
   &+\biggl[-\frac{24653523709}{146313216} \biggl(1+\frac{2013167764 \nu
   }{1450207277}-\frac{3160271968 \nu ^2}{1450207277}\biggr)-\frac{67698947}{903168}
   \biggl(1-\frac{12996788 \nu }{11946873}\biggr)\delta  \kappa_a\nonumber\\
   &-\frac{67698947}{903168}
   \biggl(1-\frac{36890534 \nu }{11946873}+\frac{15897784 \nu ^2}{3982291}\biggr) \kappa_s\biggr] \chi_s^2\biggr\}+y^4 y_0^2 \biggl\{\biggl[-\frac{1492991}{7451136} \biggl(1+\frac{13438736 \nu
   }{87823}-\frac{26498864 \nu ^2}{87823}\biggr)\nonumber\\
   &+  \frac{57841361}{3725568} \biggl(1-\frac{5516 \nu
   }{2833}\biggr) \delta \kappa_a+\frac{57841361}{3725568} \biggl(1-\frac{11182 \nu }{2833}+\frac{11032 \nu
   ^2}{2833}\biggr) \kappa_s\biggr] \chi_a^2+\biggl[\frac{57841361}{1862784} \biggl(1-\frac{11182 \nu
   }{2833}\nonumber\\
   & +\frac{11032 \nu ^2}{2833}\biggr)\kappa_a +\delta  \biggl[-\frac{1492991}{3725568}
   \biggl(1-\frac{5516 \nu }{2833}\biggr)+\frac{57841361}{1862784} \biggl(1-\frac{5516 \nu
   }{2833}\biggr) \kappa_s\biggr]\biggr] \chi_a \chi_s+\biggl[-\frac{1492991}{7451136}
   \biggl(1\nonumber\\
   & -\frac{14132020 \nu }{87823}+\frac{27182848 \nu ^2}{87823}\biggr) + \frac{57841361}{3725568}
   \biggl(1-\frac{5516 \nu }{2833}\biggr) \delta \kappa_a+\frac{57841361}{3725568} \biggl(1-\frac{11182 \nu
   }{2833}+\frac{11032 \nu ^2}{2833}\biggr) \kappa_s\biggr] \chi_s^2\biggr\}\,.
\end{align}
\end{subequations}
\end{widetext}

\subsubsection{\TFtwo{}}\label{sec:TaylorF2}

Using the parametric Eq.~\eqref{TaylorT2 time} and \eqref{TaylorT2 phase} given in terms of the PN parameter $y$, we can then write our final result for the eccentric spinning SPA phase {\blue (ignoring all higher order circular and non-spinning, eccentric corrections)} in the following form.

\begin{widetext}

\begin{equation}
\begin{split}
    \Psi =  & \frac{3}{128 \nu y^5} \biggl\{1 
    +\frac{650}{731}e_0^2 \biggl(\frac{y_0}{y}\biggr)^{19/3} \biggl[1 
    + \Psi_{\rm SO,ecc} + \Psi_{\rm SS,ecc}\biggr] 
    + \cdots + \mathcal{O}(e_0^{8})\biggr\}\,,
\end{split}
\end{equation}
where,
\begin{subequations}
\begin{align}
    \Psi^{\rm 1.5PN}_{\rm SO,ecc} &=y^3
   \biggl[\frac{1969099 \delta  \chi_a}{70200}+ \frac{1969099}{70200} \biggl(1-\frac{118636 \nu }{228965}\biggr)\chi_s\biggr]+
   y_0^3 \biggl[-\frac{157 \delta  \chi_a}{54}-\frac{157}{54} \biggl(1-\frac{110 \nu }{157}\biggr) \chi_s\biggr] \,, \\
    \Psi^{\rm 2PN}_{\rm SS,ecc} &=y_0^4 \biggl[\chi_a \chi_s \biggl[\delta  \biggl(\frac{13}{48}+\frac{89 \kappa_s}{24}\biggr)+ \frac{89}{24} (1-2 \nu )\kappa_a\biggr]+\chi_a^2 \biggl[\frac{13}{96}+\frac{89
   \delta  \kappa_a}{48}-\frac{89 \nu }{24}+ \frac{89}{48} (1-2 \nu )\kappa_s\biggr]\nonumber\\
   &+\chi_s^2
   \biggl[\frac{13}{96}+\frac{89
   \delta  \kappa_a}{48}+\frac{19 \nu }{6}+ \frac{89}{48} (1-2 \nu )\kappa_s\biggr]\biggr]+y^4 \biggl[\chi_a \chi_s \biggl[\delta  \biggl(-\frac{236113}{425568}-\frac{43122421
   \kappa_s}{1063920}\biggr)\nonumber\\
   &-\frac{43122421}{1063920} (1-2 \nu )\kappa_a\biggr]+\chi_s^2
   \biggl[-\frac{236113}{851136}-\frac{43122421 \delta  \kappa_a}{2127840}-\frac{238306 \nu }{6045}-\frac{43122421}{2127840} (1-2 \nu )\kappa_s\biggr]\nonumber\\
   &+\chi_a^2
   \biggl[-\frac{236113}{851136}-\frac{43122421 \delta  \kappa_a}{2127840}+\frac{43122421 \nu }{1063920}-\frac{43122421}{2127840} (1-2 \nu )\kappa_s\biggr]\biggr]\,,\\
    \Psi^{\rm 2.5PN}_{\rm SO,ecc} &=y_0^5 \biggl[
   -\frac{1279073}{38880} \biggl(1-\frac{2348912 \nu }{1279073}+\frac{691520 \nu
   ^2}{1279073}\biggr)\chi_s -\frac{1279073}{38880} \biggl(1-\frac{970876 \nu
   }{1279073}\biggr)\delta  \chi_a\biggr]\nonumber\\
   &+y^2 y_0^3 \biggl[ \frac{132601447499}{7330901760}
   \biggl(1-\frac{2615284 \nu }{1155397}\biggr)\delta  \chi_a+ \frac{132601447499}{7330901760}
   \biggl(1-\frac{537693258 \nu }{181397329}+\frac{287681240 \nu ^2}{181397329}\biggr)\chi_s\biggr]\nonumber\\
   &+y^3
   y_0^2 \biggl[ \frac{5578457467}{70761600} \biggl(1-\frac{5516 \nu
   }{2833}\biggr)\delta  \chi_a+ \frac{5578457467}{70761600} \biggl(1-\frac{1599066728 \nu
   }{648657845}+\frac{654396176 \nu ^2}{648657845}\biggr)\chi_s\biggr]\nonumber\\
   &+y^5 \biggl[
   \frac{4267187295851}{18822585600} \biggl(1+\frac{6072295628 \nu }{5837465521}\biggr)\delta  \chi_a+
   \frac{4267187295851}{18822585600} \biggl(1-\frac{1266783204 \nu }{5837465521}-\frac{4030104736 \nu
   ^2}{5837465521}\biggr)\chi_s\biggr]\,,\\  
    \Psi^{\rm 3PN}_{\rm SO,ecc} &=y_0^6 \biggl[-\frac{157043 \pi  \delta 
   \chi_a}{15552}-\frac{157043 \pi }{15552} \biggl(1-\frac{150652 \nu
   }{157043}\biggr)\chi_s\biggr]+y^3 y_0^3 \biggl[\frac{4787739497 \pi  \delta  \chi_a}{25272000}\nonumber\\
   &+
   \frac{4787739497 \pi }{25272000} \biggl(1-\frac{62257792 \nu }{111342779}\biggr)\chi_s\biggr]+y^6
   \biggl[-\frac{82262188063 \pi  \delta  \chi_a}{161740800}-\frac{82262188063 \pi }{161740800}
   \biggl(1-\frac{342805652 \nu }{562668865}\biggr)\chi_s\biggr]\,, \\
    \Psi^{\rm 3PN}_{\rm SS,ecc} &=y^4 y_0^2 \biggl\{\chi_a \chi_s \biggl[-\frac{122165818693}{1072431360} \biggl(1-\frac{11182 \nu }{2833}+\frac{11032 \nu
   ^2}{2833}\biggr)\kappa_a-\frac{668908129 \delta }{428972544} \biggl(1-\frac{5516 \nu }{2833}\biggr)\nonumber\\
   &-\frac{122165818693 \delta }{1072431360}
   \biggl(1-\frac{5516 \nu }{2833}\biggr)\kappa_s\biggr]+\chi_a^2 \biggl[-\frac{122165818693}{2144862720} \biggl(1-\frac{11182 \nu }{2833}+\frac{11032
   \nu ^2}{2833}\biggr)\kappa_s\nonumber\\
   &-\frac{122165818693 \delta }{2144862720} \biggl(1-\frac{5516 \nu }{2833}\biggr)\kappa_a-\frac{668908129}{857945088}
   \biggl(1-\frac{677394352 \nu }{4575295}+\frac{1301577424 \nu ^2}{4575295}\biggr)\biggr]\nonumber\\
   &+\chi_s^2 \biggl[-\frac{122165818693}{2144862720}
   \biggl(1-\frac{11182 \nu }{2833}+\frac{11032 \nu ^2}{2833}\biggr)\kappa_s-\frac{122165818693 \delta }{2144862720} \biggl(1-\frac{5516 \nu
   }{2833}\biggr)\kappa_a\nonumber\\
   &-\frac{668908129}{857945088} \biggl(1+\frac{641276492 \nu }{4575295}-\frac{1265944064 \nu ^2}{4575295}\biggr)\biggr]\biggr\}+y^2 y_0^4
   \biggl\{\chi_a \chi_s \biggl[-\frac{75168973423}{3258178560} \biggl(1-\frac{4926078 \nu }{1155397}\nonumber\\
   & +\frac{5230568 \nu
   ^2}{1155397}\biggr)\kappa_a -\frac{844595207 \delta }{501258240} \biggl(1-\frac{2615284 \nu }{1155397}\biggr)-\frac{75168973423 \delta
   }{3258178560} \biggl(1-\frac{2615284 \nu }{1155397}\biggr)\kappa_s\biggr]\nonumber\\
   & +\chi_a^2 \biggl[-\frac{75168973423}{6516357120} \biggl(1-\frac{4926078 \nu
   }{1155397} +\frac{5230568 \nu ^2}{1155397}\biggr)\kappa_s-\frac{75168973423 \delta }{6516357120} \biggl(1-\frac{2615284 \nu
   }{1155397}\biggr)\kappa_a \nonumber\\
   & -\frac{844595207}{1002516480} \biggl(1-\frac{445320024 \nu }{15020161}+\frac{931041104 \nu ^2}{15020161}\biggr)\biggr] +\chi_s^2
   \biggl[-\frac{75168973423}{6516357120} \biggl(1-\frac{4926078 \nu }{1155397}+\frac{5230568 \nu ^2}{1155397}\biggr)\kappa_s \nonumber\\
   & -\frac{75168973423 \delta }{6516357120} \biggl(1-\frac{2615284 \nu }{1155397}\biggr)\kappa_a -\frac{844595207}{1002516480} \biggl(1+\frac{317241996 \nu
   }{15020161}-\frac{795046336 \nu ^2}{15020161}\biggr)\biggr]\biggr\} \nonumber\\
   & +y_0^6 \biggl\{\chi_a^2 \biggl[\frac{906103}{48384} \biggl(1-\frac{2893874 \nu
   }{906103}+\frac{1359176 \nu ^2}{906103}\biggr)\kappa_s + \frac{906103 \delta }{48384} \biggl(1-\frac{1081668 \nu
   }{906103}\biggr)\kappa_a+\frac{24433195}{2612736} \biggl(1 \nonumber\\
   & -\frac{169526104 \nu }{24433195}+\frac{73395504 \nu ^2}{24433195}\biggr)\biggr]+\chi_a \chi_s
   \biggl(\frac{906103}{24192} \biggl(1-\frac{2893874 \nu }{906103} +\frac{1359176 \nu ^2}{906103}\biggr)\kappa_a+ \frac{906103 \delta
   }{24192} \biggl(1 \nonumber\\
   & -\frac{1081668 \nu }{906103}\biggr)\kappa_s+\frac{24433195 \delta }{1306368} \biggl(1-\frac{33561668 \nu }{24433195}\biggr)\biggr)+\chi_s^2
   \biggl[\frac{906103}{48384} \biggl(1-\frac{2893874 \nu }{906103} +\frac{1359176 \nu ^2}{906103}\biggr)\kappa_s \nonumber\\
   & + \frac{906103 \delta
   }{48384} \biggl(1-\frac{1081668 \nu }{906103}\biggr)\kappa_a+\frac{24433195}{2612736} \biggl(1+\frac{4669988 \nu }{24433195}-\frac{42904736 \nu
   ^2}{24433195}\biggr)\biggr]\biggr\} \nonumber\\
   & +y^6 \biggl\{\chi_a^2 \biggl[-\frac{830182833661 \delta }{2515968000} \biggl(1+\frac{43459444 \nu
   }{1135681031}\biggr)\kappa_a-\frac{830182833661}{2515968000} \biggl(1-\frac{2227902618 \nu }{1135681031}-\frac{1374645160 \nu
   ^2}{1135681031}\biggr)\kappa_s \nonumber\\
   & +\frac{10996032441971}{27172454400} \biggl(1-\frac{203534409296 \nu }{75212260205}+\frac{14846167728 \nu
   ^2}{15042452041}\biggr)\biggr]+\chi_s^2 \biggl[-\frac{830182833661 \delta }{2515968000} \biggl(1+\frac{43459444 \nu
   }{1135681031}\biggr)\kappa_a \nonumber\\
   & -\frac{830182833661}{2515968000} \biggl(1-\frac{2227902618 \nu }{1135681031}-\frac{1374645160 \nu
   ^2}{1135681031}\biggr)\kappa_s+\frac{10996032441971}{27172454400} \biggl(1-\frac{22377479356 \nu }{15042452041} \nonumber\\
   & -\frac{17050930400 \nu
   ^2}{15042452041}\biggr)\biggr] +\chi_a \chi_s \biggl[-\frac{830182833661 \delta }{1257984000} \biggl(1+\frac{43459444 \nu
   }{1135681031}\biggr)\kappa_s-\frac{830182833661}{1257984000} \biggl(1-\frac{2227902618 \nu }{1135681031}  \nonumber\\
   & -\frac{1374645160 \nu
   ^2}{1135681031}\biggr)\kappa_a +\frac{10996032441971 \delta }{13586227200} \biggl(1-\frac{7286382628 \nu }{75212260205}\biggr)\biggr]\biggr\}+y^3 y_0^3
   \biggl[-\frac{309148543 \delta }{1895400}\biggl(1 \nonumber\\
   & -\frac{21906001 \nu }{35947505}\biggr) \chi_a \chi_s -\frac{309148543}{3790800}
   \biggl(1-\frac{43812002 \nu }{35947505}+\frac{237272 \nu ^2}{653591}\biggr)\chi_s^2-\frac{309148543}{3790800} (1-4 \nu )\chi_a^2\biggr]\,.
\end{align}
\end{subequations}
\end{widetext}

Note that we only list the leading (Newtonian) non-spinning term here. The full 3PN expression for the non-spinning part of the phase is listed as Eq.~(6.26) of Ref.~\citep{Moore:2016qxz}. Also, while we haven't explicitly listed it here, we provide full expressions for all the Taylor approximants, including the spinning part up to $\mathcal{O}(e_0^8)$ till 3PN, which we compute here as additional material.
\subsubsection{Resummed \TTtwo{} Phase}\label{sec:comparison}
Since the current study deals with an extension of Ref.~\citep{Moore:2016qxz} both in terms of an increase in eccentricity order and spins, it is natural to ask about the validity (in terms of the eccentricity) of the current model. Since Ref.~\citep{Moore:2016qxz} deals with $\mathcal{O}(e_0^2)$ in eccentricity, and the current model deals with $\mathcal{O}(e_0^8)$ the extent in eccentricity to which the current model is applicable is naturally higher than the previous works. To quantify that, we compare the $\mathcal{O}(e_0^2)$, $\mathcal{O}(e_0^4)$, $\mathcal{O}(e_0^6)$, $\mathcal{O}(e_0^8)$,  and a resummed version of the $\mathcal{O}(e_0^4)$ and $\mathcal{O}(e_0^8)$ \TTtwo{} phase with the numerically calculated \TTtwo{} that is valid for arbitrary initial eccentricities ($e_0 < 1$). We use only the secularly increasing parts of the phase, ignoring the oscillatory pieces (which will be discussed in Appendix \ref{sec:osc}). 

Since an eccentricity expanded phase will naturally only be valid for a small range of initial eccentricities, we attempt to increase its accuracy by performing a simple resummation, as highlighted in Sec. III D of Ref.~\citep{Henry:2023tka}. To do this, we choose a resummation ansatz of the form

\begin{equation}\label{eq: resum-ans}
    \phi^{(\rm resum)}_{ans} = {\blue y^{-5}}(1-e_0^2)^3 \left(d + f ~e_0^2 + g ~e_0^4 + h ~e_0^6 + l ~e_0^8\right) \,.
\end{equation}
Note that in Eq.~\eqref{eq: resum-ans}, the exponent of the $(1-e_0^2)$ segment was arrived at by experimenting with various powers, and we finally settled with the cubic power since it proved to accumulate the smallest error in the number of cycles vis-a-vis the numerical solution. 

This resummation ansatz is then expanded once again in $e_0$ and compared with the PN version of the phase to determine the ansatz coefficients $d$, $f$, $g$, $h$ and $l$. Fig.~\ref{fig:comparison} includes comparing the resummed analytical expression with numerical phase results. The resummed phase was found to have the following structure
\begin{align}
    \phi^{(\rm resum)} &=  - \frac{(1-e_0^2)^3}{32 \nu y^5} \biggl\{\phi^{(\rm resum)}_{\rm NS} + \phi^{(\rm resum)}_{\rm SO} + \phi^{(\rm resum)}_{\rm SS} \nonumber  \\
    &+3 e_0^2 \biggl[\phi^{(\rm resum)}_{\rm SO,ecc} + \phi^{(\rm resum)}_{\rm SS,ecc} \biggr] 
    + \cdots + \mathcal{O}(e_0^{8}) \biggr\}\,,
\end{align}
where the various pieces were found to be as follows
\begin{widetext}
\begin{subequations} \label{TaylorT2 resummed phase}
\begin{align}
    \phi^{\rm (resum)}_{\rm NS} &= 1+3 e_0^2 \biggl[1  -\frac{35}{272}  \biggl(\frac{y_0}{y}\biggr)^{19/3}\biggr] 
 + \cdots +  \mathcal{O}(e_0^{8})\,, \\
    \phi^{\rm (resum)1.5PN}_{\rm SO} &= y^3 \biggl[\frac{565 \delta  \chi_a}{24}+\frac{565}{24} \biggl(1-\frac{76 \nu }{113}\biggr) \chi_s\biggr]\,,\\
    \phi^{\rm (resum)1.5PN}_{\rm SO,ecc} &=y^3 \biggl[\frac{565 \delta  \chi_a}{24}+\frac{565}{24} \biggl(1-\frac{76 \nu }{113}\biggr) \chi_s\biggr]-\frac{35}{272} \biggl(\frac{y_0}{y}\biggr)^{19/3} \biggl\{y_0^3 \biggl[-\frac{157 \delta 
   \chi_a}{54}-\frac{157}{54} \biggl(1-\frac{110 \nu }{157}\biggr)\chi_s\biggr] \nonumber \\
   &+y^3 \biggl[-\frac{6086 \delta  \chi_a}{945}-\frac{6086}{945} \biggl(1+\frac{1393 \nu }{895}\biggr)\chi_s\biggr]\biggr\}\,,  \\
    \phi^{\rm (resum)2PN}_{\rm SS} &= y^4 \biggl\{\biggl[-\frac{5}{16} (1-160 \nu )-25 \delta  \kappa_a-25 (1-2 \nu ) \kappa_s\biggr] \chi_a^2+\biggl[-50 (1-2 \nu ) \kappa_a+\delta 
   \biggl(-\frac{5}{8}-50 \kappa_s\biggr)\biggr] \chi_a \chi_s \nonumber \\
   &+\biggl[-\frac{5}{16} (1+156 \nu )-25 \delta  \kappa_a-25 (1-2 \nu ) \kappa_s\biggr] \chi_s^2\biggr\}\,, \\
    \phi^{\rm (resum)2PN}_{\rm SS,ecc} &=y^4 \biggl\{\biggl[-\frac{5}{16} (1-160 \nu )-25 \delta  \kappa_a-25 (1-2 \nu ) \kappa_s\biggr] \chi_a^2+\biggl[-50 (1-2 \nu )
   \kappa_a+\delta  \biggl(-\frac{5}{8}-50 \kappa_s\biggr)\biggr] \chi_a \chi_s \nonumber \\
   &+\biggl[-\frac{5}{16} (1+156 \nu )-25 \delta  \kappa_a-25 (1-2 \nu ) \kappa_s\biggr] \chi_s^2\biggr\}-\frac{35}{272} \biggl(\frac{y_0}{y}\biggr)^{19/3} \biggl\{y_0^4 \biggl[\biggl[\frac{13}{96} \biggl(1-\frac{356 \nu }{13}\biggr)+\frac{89
   \delta  \kappa_a}{48} \nonumber \\
   &+\frac{89}{48} (1-2 \nu ) \kappa_s\biggr] \chi_a^2+\biggl[\frac{89}{24} (1-2 \nu ) \kappa_a+\delta 
   \biggl(\frac{13}{48}+\frac{89 \kappa_s}{24}\biggr)\biggr] \chi_a \chi_s+\biggl[\frac{13}{96} \biggl(1+\frac{304 \nu }{13}\biggr) +\frac{89 \delta  \kappa_a}{48}\nonumber \\
   &+\frac{89}{48} (1-2 \nu ) \kappa_s\biggr] \chi_s^2\biggr]+y^4 \biggl[\biggl[-\frac{527}{7392} \biggl(1+\frac{4804 \nu }{31}\biggr)+\frac{20417
   \delta  \kappa_a}{3696}+\frac{20417}{3696} (1-2 \nu ) \kappa_s\biggr] \chi_a^2  \nonumber \\
   & +\biggl[\frac{20417}{1848} (1-2 \nu ) \kappa_a +\delta 
   \biggl(-\frac{527}{3696}+\frac{20417 \kappa_s}{1848}\biggr)\biggr] \chi_a \chi_s+\biggl[-\frac{527}{7392} \biggl(1-\frac{4928 \nu }{31}\biggr)+\frac{20417 \delta
    \kappa_a}{3696} \nonumber\\
    & +\frac{20417}{3696} (1-2 \nu ) \kappa_s\biggr] \chi_s^2\biggr]\biggr\}\,,\\
    \phi^{\rm (resum)2.5PN}_{\rm SO} &= -\frac{732985}{2016} y^5  \log (y) \biggl[\biggl(1+\frac{7056 \nu }{146597}\biggr) \delta \chi_a+\biggl(1-\frac{135856 \nu }{146597}-\frac{17136 \nu
   ^2}{146597}\biggr) \chi_s\biggr]\,,\\
    \phi^{\rm (resum)2.5PN}_{\rm SO,ecc} &= -\frac{732985}{2016} y^5  \log (y) \biggl[\biggl(1+\frac{7056 \nu }{146597}\biggr) \delta \chi_a+\biggl(1-\frac{135856 \nu }{146597}-\frac{17136 \nu
   ^2}{146597}\biggr) \chi_s\biggr] \nonumber \\
   &-\frac{35}{272} \biggl(\frac{y_0}{y}\biggr)^{19/3} \biggl\{y_0^5 \biggl[-\frac{1279073}{38880}
   \biggl(1-\frac{970876 \nu }{1279073}\biggr)\delta \chi_a-\frac{1279073}{38880} \biggl(1-\frac{2348912 \nu }{1279073}+\frac{691520 \nu
   ^2}{1279073}\biggr) \chi_s\biggr] \nonumber \\
   &+y^5 \biggl[  \frac{717229337}{2585520} \biggl(1-\frac{39620772 \nu }{42189961}\biggr)\delta \chi_a+\frac{717229337}{2585520} \biggl(1-\frac{432179708 \nu }{295329727}+\frac{12905280 \nu ^2}{42189961}\biggr) \chi_s\biggr] \nonumber \\
   &+y^3 y_0^2 \biggl[ 
   -\frac{8620819}{476280} \biggl(1-\frac{5516 \nu }{2833}\biggr)\delta \chi_a-\frac{8620819}{476280} \biggl(1-\frac{990451 \nu
   }{2535535}-\frac{7683788 \nu ^2}{2535535}\biggr) \chi_s\biggr] \nonumber \\
   &+y^2 y_0^3 \biggl[-\frac{424176163}{5334336} \biggl(1-\frac{69804 \nu
   }{158927}\biggr)\delta \chi_a-\frac{424176163}{5334336} \biggl(1-\frac{28441198 \nu }{24951539}+\frac{7678440 \nu ^2}{24951539}\biggr) \chi_s\biggr]\biggr\}\,,\\ 
    \phi^{\rm (resum)3PN}_{\rm SO} &= y^6 \biggl[\frac{1135}{6} \pi  \delta  \chi_a+\frac{1135 \pi }{6} \biggl(1-\frac{156 \nu }{227}\biggr) \chi_s\biggr]\,,\\ 
    \phi^{\rm (resum)3PN}_{\rm SO,ecc} &=y^6
   \biggl[\frac{1135}{6} \pi  \delta  \chi_a+\frac{1135 \pi }{6} \biggl(1-\frac{156 \nu }{227}\biggr) \chi_s\biggr]+\frac{35}{272}
   \biggl(\frac{y_0}{y}\biggr)^{19/3} \biggl\{y_0^6 \biggl[\frac{157043 \pi  \delta  \chi_a}{15552}+\frac{157043 \pi }{15552} \biggl(1 \nonumber \\
   & -\frac{150652 \nu
   }{157043}\biggr) \chi_s\biggr] +y^3 y_0^3 \biggl[\frac{12307507 \pi  \delta  \chi_a}{340200}+\frac{12307507 \pi }{340200} \biggl(1+\frac{1015892
   \nu }{723971}\biggr) \chi_s\biggr] \nonumber \\
   &-y^6 \biggl[\frac{153181985 \pi  \delta  \chi_a}{870912}+\frac{153181985 \pi }{870912} \biggl(1-\frac{1475156 \nu
   }{9010705}\biggr) \chi_s\biggr]\biggr\}\,, \\
    \phi^{\rm (resum)3PN}_{\rm SS} &= y^6 \biggl\{\biggl[-\frac{1344475}{8064} \biggl(1-\frac{428504 \nu }{268895}+\frac{96768 \nu ^2}{268895}\biggr)+  \frac{26015}{112} \biggl[\biggl(1-\frac{4186 \nu
   }{15609}\biggr)\delta \kappa_a + \biggl(1-\frac{35404 \nu }{15609}  \nonumber \\
   & -\frac{1344 \nu ^2}{5203}\biggr) \kappa_s \biggr]\biggr] \chi_a^2 +\biggl[\frac{26015}{56} \biggl(1-\frac{35404 \nu }{15609}-\frac{1344 \nu ^2}{5203}\biggr) \kappa_a+\delta  \biggl[-\frac{1344475}{4032}
   \biggl(1-\frac{8344 \nu }{268895}\biggr) \nonumber \\
   & +\frac{26015}{56} \biggl(1-\frac{4186 \nu }{15609}\biggr) \kappa_s\biggr]\biggr] \chi_a \chi_s +\biggl[-\frac{1344475}{8064} \biggl(1-\frac{663764 \nu }{268895}-\frac{152992 \nu ^2}{268895}\biggr)+  \frac{26015}{112} \biggl[\biggl(1 \nonumber\\
   &-\frac{4186 \nu
   }{15609}\biggr)\delta \kappa_a + \biggl(1-\frac{35404 \nu }{15609}-\frac{1344 \nu ^2}{5203}\biggr) \kappa_s \biggr]\biggr] \chi_s^2\biggr\}\,,\\
    \phi^{\rm (resum)3PN}_{\rm SS,ecc} &= y^6 \biggl\{\biggl[-\frac{1344475}{8064} \biggl(1-\frac{428504 \nu }{268895}+\frac{96768 \nu ^2}{268895}\biggr)+  \frac{26015}{112} \biggl[\biggl(1-\frac{4186 \nu
   }{15609}\biggr)\delta \kappa_a + \biggl(1-\frac{35404 \nu }{15609} \nonumber \\
   & -\frac{1344 \nu ^2}{5203}\biggr) \kappa_s \biggr]\biggr] \chi_a^2 +\biggl[\frac{26015}{56} \biggl(1-\frac{35404 \nu }{15609}-\frac{1344 \nu ^2}{5203}\biggr) \kappa_a+\delta  \biggl[-\frac{1344475}{4032} \biggl(1-\frac{8344 \nu }{268895}\biggr) +\frac{26015}{56} \biggl(1 \nonumber \\
   & -\frac{4186 \nu }{15609}\biggr) \kappa_s\biggr]\biggr] \chi_a \chi_s +\biggl[-\frac{1344475}{8064} \biggl(1-\frac{663764 \nu }{268895}-\frac{152992 \nu ^2}{268895}\biggr)+ 
   \frac{26015}{112} \biggl[\biggl(1-\frac{4186 \nu
   }{15609}\biggr)\delta \kappa_a + \biggl(1 \nonumber \\
   & -\frac{35404 \nu }{15609}-\frac{1344 \nu ^2}{5203}\biggr) \kappa_s \biggr]\biggr] \chi_s^2\biggr\} -\frac{35}{272} \biggl(\frac{y_0}{y}\biggr)^{19/3} \biggl\{y^3 y_0^3 \biggl[\frac{477751}{25515}
   (1-4 \nu ) \chi_a^2+ \frac{955502}{25515} \biggl(1 \nonumber \\
   & +\frac{120251 \nu }{281030}\biggr)\delta \chi_a \chi_s +\frac{477751}{25515}
   \biggl(1+\frac{120251 \nu }{140515}-\frac{30646 \nu ^2}{28103}\biggr) \chi_s^2\biggr]+y^4 y_0^2 \biggl[\biggl[-\frac{1492991}{7451136}
   \biggl(1+\frac{13438736 \nu }{87823} \nonumber \\
   & -\frac{26498864 \nu ^2}{87823}\biggr) +  \frac{57841361}{3725568} \biggl(1-\frac{5516 \nu }{2833}\biggr)\delta
   \kappa_a+\frac{57841361}{3725568} \biggl(1-\frac{11182 \nu }{2833}+\frac{11032 \nu ^2}{2833}\biggr) \kappa_s\biggr] \chi_a^2\nonumber \\
   &+\biggl[\frac{57841361}{1862784} \biggl(1-\frac{11182 \nu }{2833}+\frac{11032 \nu ^2}{2833}\biggr)\kappa_a+\delta 
   \biggl[-\frac{1492991}{3725568} \biggl(1-\frac{5516 \nu }{2833}\biggr)+\frac{57841361}{1862784} \biggl(1 \nonumber \\
   & -\frac{5516 \nu }{2833}\biggr) \kappa_s\biggr]\biggr] \chi_a \chi_s +\biggl[-\frac{1492991}{7451136} \biggl(1-\frac{14132020 \nu }{87823}+\frac{27182848 \nu ^2}{87823}\biggr)+ 
   \frac{57841361}{3725568} \biggl(1-\frac{5516 \nu }{2833}\biggr)\delta \kappa_a\nonumber \\
   &+\frac{57841361}{3725568} \biggl(1-\frac{11182 \nu }{2833}+\frac{11032
   \nu ^2}{2833}\biggr) \kappa_s\biggr] \chi_s^2\biggr]+y^2 y_0^4 \biggl[\biggl[\frac{35122867}{9483264} \biggl(1-\frac{57485464 \nu }{2066051}+\frac{24850224
   \nu ^2}{2066051}\biggr)\nonumber \\
   &+ \frac{240456551}{4741632} \biggl(1-\frac{69804 \nu }{158927}\biggr)\delta \kappa_a+\frac{240456551}{4741632}
   \biggl(1-\frac{387658 \nu }{158927}+\frac{139608 \nu ^2}{158927}\biggr) \kappa_s\biggr] \chi_a^2\nonumber \\
   &+\biggl[\frac{240456551}{2370816}
   \biggl(1-\frac{387658 \nu }{158927}+\frac{139608 \nu ^2}{158927}\biggr) \kappa_a+\delta  \biggl[\frac{35122867}{4741632} \biggl(1-\frac{69804 \nu
   }{158927}\biggr)\nonumber \\
   &+\frac{240456551}{2370816} \biggl(1-\frac{69804 \nu }{158927}\biggr) \kappa_s\biggr]\biggr] \chi_a \chi_s+\biggl[\frac{35122867}{9483264} \biggl(1+\frac{47406356 \nu }{2066051}-\frac{21220416 \nu ^2}{2066051}\biggr)\nonumber \\
   &+  \frac{240456551}{4741632}
   \biggl(1-\frac{69804 \nu }{158927}\biggr)\delta \kappa_a+\frac{240456551}{4741632} \biggl(1-\frac{387658 \nu }{158927}+\frac{139608 \nu
   ^2}{158927}\biggr) \kappa_s\biggr] \chi_s^2\biggr]\nonumber \\
   &+y_0^6 \biggl[\biggl[\frac{24433195}{2612736} \biggl(1-\frac{169526104 \nu }{24433195}+\frac{73395504 \nu
   ^2}{24433195}\biggr)+\frac{906103}{48384} \biggl(1-\frac{1081668 \nu }{906103}\biggr)\delta  \kappa_a\nonumber \\
   &+\frac{906103}{48384}
   \biggl(1-\frac{2893874 \nu }{906103}+\frac{1359176 \nu ^2}{906103}\biggr) \kappa_s\biggr] \chi_a^2+\biggl[\frac{906103}{24192} \biggl(1-\frac{2893874
   \nu }{906103}+\frac{1359176 \nu ^2}{906103}\biggr) \kappa_a\nonumber \\
   &+\delta  \biggl[\frac{24433195}{1306368} \biggl(1-\frac{33561668 \nu
   }{24433195}\biggr)+\frac{906103}{24192} \biggl(1-\frac{1081668 \nu }{906103}\biggr) \kappa_s\biggr]\biggr] \chi_a \chi_s\nonumber \\
   &+\biggl[\frac{24433195}{2612736} \biggl(1+\frac{4669988 \nu }{24433195}-\frac{42904736 \nu ^2}{24433195}\biggr)+  \frac{906103}{48384}
   \biggl(1-\frac{1081668 \nu }{906103}\biggr)\delta \kappa_a\nonumber \\
   &+\frac{906103}{48384} \biggl(1-\frac{2893874 \nu }{906103}+\frac{1359176 \nu
   ^2}{906103}\biggr) \kappa_s\biggr] \chi_s^2\biggr]+y^6 \biggl[\biggl[-\frac{24653523709}{146313216} \biggl(1-\frac{7638735632 \nu
   }{1450207277}\nonumber \\
   & +\frac{2575441008 \nu ^2}{1450207277}\biggr) -\frac{67698947}{903168} \biggl(1-\frac{12996788 \nu }{11946873}\biggr)\delta \kappa_a+\biggl(-\frac{67698947}{903168} \biggl(1-\frac{36890534 \nu }{11946873} \nonumber \\
   & +\frac{15897784 \nu ^2}{3982291}\biggr)\biggr) \kappa_s\biggr] \chi_a^2 +\biggl[-\frac{67698947}{451584} \biggl(1-\frac{36890534 \nu }{11946873}+\frac{15897784 \nu ^2}{3982291}\biggr) \kappa_a+\delta 
   \biggl[-\frac{24653523709}{73156608} \biggl(1 \nonumber \\
   & +\frac{87630620 \nu }{1450207277}\biggr) -\frac{67698947}{451584} \biggl(1-\frac{12996788 \nu
   }{11946873}\biggr) \kappa_s\biggr]\biggr] \chi_a \chi_s+\biggl[-\frac{24653523709}{146313216} \biggl(1+\frac{2013167764 \nu
   }{1450207277} \nonumber \\
   & -\frac{3160271968 \nu ^2}{1450207277}\biggr) -\frac{67698947}{903168} \biggl(1-\frac{12996788 \nu }{11946873}\biggr)\delta \kappa_a-\frac{67698947}{903168} \biggl(1-\frac{36890534 \nu }{11946873} \nonumber\\
   &+\frac{15897784 \nu ^2}{3982291}\biggr) \kappa_s\biggr] \chi_s^2\biggr]\biggr\} \,.
   \end{align}
\end{subequations}
\end{widetext}
{\blue \section{Significance of eccentric, spinning terms in the phasing formula}
\label{sec:noprec}}
In this section, we display the eccentricity evolution and phase of the orbit of a spin-aligned eccentric binary system using the PN approximants in the time domain (\TTtwo{}) and in the frequency domain (\TFtwo{}), respectively. Additionally, we also quantify the effect of newly computed spinning eccentric corrections expanded in eccentricity up to $\mathcal{O}\left(e_0^2\right)$ in \TFtwo{} phase by performing a mismatch study between \TFtwoE{}~\citep{Moore:2016qxz} and all these terms added to \TFtwoE{}. We also calculate a resummed version of \TTtwo{} in this section. We give the Taylor approximants  \TTone{}, \TTthree{}, and \TTfour{} in Appendix \ref{sec: app-Taylor}.


\subsection{PN contribution to the number of GW cycles}\label{sec:num_cycles}
Now that \TTtwo{} has been constructed, we can compute the number of GW cycles the time-domain waveform sweeps from an initial frequency $f_1$ to a final frequency $f_2$. The final frequency is taken to be the minimum of the upper limit of a particular detector band (for example, $f_2 = 1000\,\, Hz $ for the LIGO band) and $f_{\rm ISCO}$ (where the PN formalism is supposed to break down), given by the following
\begin{equation}\label{eq:f_ss_isco}
    f_{\rm ISCO} = \frac{c^3}{6^{3/2} \pi G M}\,.
\end{equation}
We will take $f_1 = 10 \,$ Hz as the lower end of the LIGO sensitivity bucket. We also quote the number of cycles accumulated in 3G, DECIGO, and LISA bands by utilizing different ranges of $f_1$ and $f_2$. The number of accumulated GW cycles between $f_1$ and $f_2$ can be defined as
\begin{equation}\label{eq:num-cyc}
    N_{\rm cyc} = \frac{1}{\pi} \left[\phi(f_2) - \phi(f_1)\right]\,.
\end{equation}

Table \ref{table_cycles_freq_bands} shows the PN breakup of $N_{\rm cyc}$ for a fixed initial eccentricity and spins for various detector and mass configurations. 
\begin{widetext}
\begin{center}
\begin{table}[H]
	\centering
	\begin{tabular}{|l||c|c||c|c|c||c|c||c|c|}
		\hline
		Detector & \multicolumn{2}{c||}{LIGO/Virgo} & \multicolumn{3}{c||}{3G}& \multicolumn{2}{c||}{DECIGO} & \multicolumn{2}{c|}{LISA}  \\
		\hline
		Masses ($M_\odot$) & $1.4 + 1.4$ & $ 10 + 10$ &  $1.4 + 1.4$ & $50 + 50$ & $500 + 500$ & $500 + 500$ & $5000 + 5000$ & $10^5 + 10^5$ & $10^7 + 10^7$  \\
		\hline
		PN order &\multicolumn{9}{|c|}{cumulative number of cycles} \\
		\hline
		\hline
		1.5PN (circ) & $51.9072$ & $28.8143$ & $250.7890$ & $48.2115$ & $7.0857$ & $232.0150$ & $48.2115$ & $149.0670$ & $2.9067$ \\
		\hline
		1.5PN (ecc) & $-1.7260$ & $-1.0468$ & $-8.0115$ & $-1.6621$ & $-0.3493$ & $-7.7149$ & $-1.6621$ & $-4.8601$ & $-0.1916$\\
		\hline
		2PN (circ)& $-1.6502$ & $-3.5553$ & $-4.1416$ & $-4.9924$ &  $-1.2597$ & $-11.7745$ & $-4.9924$ & $-9.9381$ & $-0.5927$ \\
		\hline
		2PN (ecc) & $0.0334$ & $0.0878$ & $0.0720$ & 
$0.1107$ & $0.0491$ & $0.2385$ & $0.1107$ & $0.1893$ & $0.0321$ \\
		\hline
		2.5PN (circ) & $6.7599$ & $10.2069$ & $10.8024$ & $12.4962$ & $4.8913$ & $15.2098$ & $12.4962$ & $17.8117$ & $2.6020$\\
		\hline	
		2.5PN (ecc)& $-0.1148$ & $-0.2577$ & $-0.1148$ & $-0.2581$ & $-0.2418$ & $-0.2582$ & $-0.2581$ & $-0.2583$ & $-0.1921$ \\
		\hline 
		3PN (circ)& $-2.2268$ & $-4.9561$ & $-3.0152$ & $-5.5237$ & $-3.0028$ & $-3.6909$ & $-5.5237$ & $-6.4303$ & $-1.7800$ \\
		\hline	
		3PN (ecc)& $0.0070$ & $0.0385$ & $0.0033$ & $0.0309$ & $0.0515$ & $0.0144$ & $0.0309$ & $0.0182$ & $0.0330$ \\
		\hline
            \textbf{Total} & $52.9898$ & $29.3315$ & $246.384$ & $48.4129$ & $7.2239$ & $224.0390$ & $48.4129$ & $145.5990$ & $2.8173$ \\
		\hline
	\end{tabular}
	\caption{\justifying Contribution of each PN order to the total number of accumulated cycles by the spinning section of the {\blue $e^8_0$ expanded} phase inside the detector's frequency band, for typical eccentric spinning ($e_0 = 0.2$; $\chi_1 = \chi_2 = 0.4$ for BNS and $\chi_1 = \chi_2 = 0.9$ for BBH) compact binaries observed by current and future detectors. {\blue (See Table~\ref{table_cycles_freq_bands_higher_ecc} for estimates with a higher $e_0=0.5$.)} We {\blue approximate} the frequency bands of LIGO/Virgo A+, Einstein Telescope (ET/CE/3G), DECIGO, and LISA with step functions, respectively between $\bigl[10\,\text{Hz},10^3\,\text{Hz}\bigr]$, $\bigl[1\,\text{Hz},10^4\,\text{Hz}\bigr]$, $\bigl[10^{-2}\,\text{Hz},1\,\text{Hz}\bigr]$ and $\bigl[10^{-4}\,\text{Hz},10^{-1}\,\text{Hz}\bigr]$.
    }\label{table_cycles_freq_bands}
\end{table}
\end{center}
\end{widetext}
From Table \ref{table_cycles_freq_bands} we can now draw the following conclusions
\begin{enumerate}[label=(\alph*)]
    \item The number of accumulated GW cycles contributed by the circular part of the phasing always has a relative negative sign with respect to the number of cycles contributed by the eccentric part. This is evident from the \TTtwo{} phasing formula in Eq.~\eqref{TaylorT2 phase} that the eccentric part of the phasing has a relative negative sign with respect to the circular part. Moreover, it is expected that the presence of eccentricity will take away several cycles from an inspiral in general, even though for certain PN orders, it does not hold, as can be seen in the LIGO/Virgo column in Table \ref{table_cycles_freq_bands} for the 2PN (as well as 3PN) circular and eccentric cases.
    \item From Table \ref{Intro cycles table} and \ref{table_cycles_freq_bands} it is to be noted that the spin-orbit effect for aligned spinning binaries contributes positively to the number of accumulated GW cycles (they make the waveform longer/elongates the inspiral process). It is also to be noted that anti-alignment of spins with respect to the orbital angular momentum works reversely, i.e. they take away the number of cycles (makes the binary merge faster compared to the non-spinning case).
    \item While the spin-orbit coupling positively contributes to the number of GW cycles for aligned spins, the spin-spin coupling, however, owing to the relative sign difference between the spin-orbit and spin-spin sections of Eq.~\eqref{TaylorT2 phase}, contributes negatively to the number of cycles. This can also be seen in Table \ref{table_cycles_freq_bands} for the 2PN (circular) row. However, since the spin-spin coupling is a subdominant effect (owing to its quadratic nature), the number of cycles contributed by the spin-orbit case (owing to its linear nature) trumps the number of cycles taken away by the spin-spin case. 
    \item The combined presence of eccentricity and spins has a subdominant effect on the number of GW cycles. For the leading order spin-orbit spin-aligned case, as seen from Table \ref{table_cycles_freq_bands}, some cycles, ranging from $-0.1$ to $-8$ for various total mass ranges, are taken away. The same holds for the leading order spin-spin spin-aligned case where a negligible number of cycles gets added to the net waveform.
    \item For the mixed spin case (where one spin is parallel to the orbital angular momentum and the other is anti-parallel to the orbital angular momentum), the numbers entirely depend on the relative magnitude of positive and negative spins. If the aligned-spin magnitude is greater than that of the anti-aligned spin, or equivalently if $\chi_{\rm{eff}} > 0$, it follows the trends of both spin-aligned cases. But for the opposite scenario, when $\chi_{\rm{eff}} < 0$, it follows the anti-aligned case. However, the magnitude of cycles accumulated is (predictably) lower than the corresponding spin-aligned and spin-anti-aligned cases.
    \item The importance of an effect while using the GW cycles metric can be judged by whether or not the corresponding $ N_{\rm{cyc}} > 1$. A general analysis of the total cycles in each table indicates that the eccentric spinning contributions (the new addition of our study to \TTtwo{}) have become increasingly important for larger spinning systems, especially those with aligned/anti-aligned spin configurations.
\end{enumerate}

{\blue Note that the estimates displayed in Table~\ref{Intro cycles table}-\ref{table_cycles_freq_bands} do not account for the contributions from the oscillatory part of the phase which contributes negligibly at small eccentricities; see for instance, Table~\ref{table_cycles_freq_bands_osc_phase} and Fig.~\ref{fig:spin-osscillatory} and discussions around them.}
\subsection{Difference in number of GW cycles using resummed \TTtwo{} phase}
\label{subsub:NGW_resum_T2}
\begin{figure}[H]
    \centering
    \includegraphics[width=\linewidth]{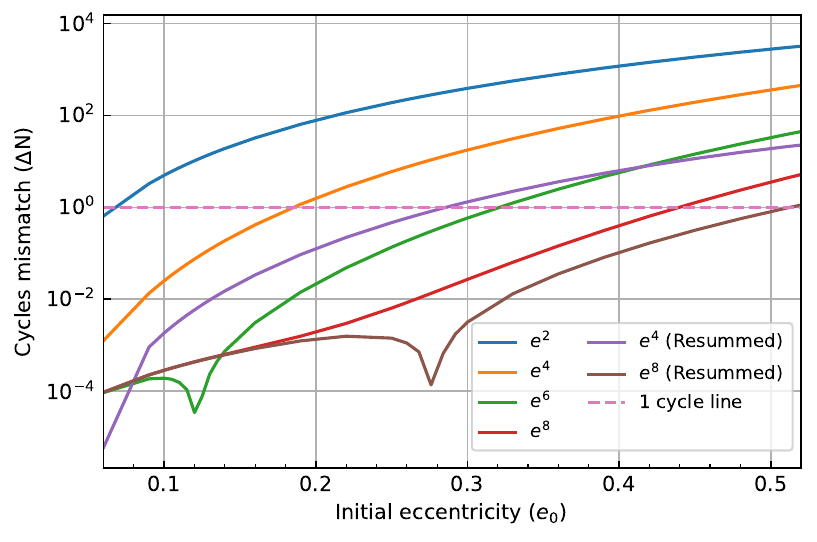}
    \caption{\justifying{\blue Absolute value of the} difference in the number of GW cycles between a numerical evolution of $\phi$ (valid for any initial eccentricity $e_0$) and the analytical $\phi$ derived above, plotted using Eq. (\ref{eq:DeltaN}). The 1-cycle line roughly represents where the phase error between the exact eccentricity solution and the analytical low eccentricity solution becomes significant and our approximation breaks down. Further, the analytical resummed expressions (as in Eq. \ref{TaylorT2 resummed phase}) for the $e_0^4$, $e_0^6$ and $e_0^8$ expanded phase are also shown. The plot has been made for a system with spins $\chi_1 = 0.7$, $\chi_2 = 0.8$ and with component masses $m_1 = m_2 = 1.4\,M_\odot$.
    }
    \label{fig:comparison}
\end{figure}

We plot in Fig.~\ref{fig:comparison} the difference between the number of cycles, that is
\begin{equation}\label{eq:DeltaN}
    \Delta N_{\rm cyc} = N_{\rm cyc,numerical} - N_{\rm cyc,analytical},
\end{equation}
as a function of the eccentricity. The figure indicates that as the initial eccentricity increases, the difference in the number of cycles between the numerical and the analytical increases. It shows that the $\mathcal{O}(e_0^2)$ solution is valid only for initial eccentricities around 0.07, whereas for the $\mathcal{O}(e_0^8)$ solution, the validity increases to about 0.45. The validity is further improved up to close to 0.55 for the $\mathcal{O}(e_0^8)$ resummed version.

\subsection{Mismatch computations}\label{subsubsec:WFsys}
To estimate the effects due to newly computed spinning eccentric corrections, we start by defining the scalar product between any two waveforms $h_1$ and $h_2$, as,
\begin{align}\label{eq:scalar_product}
    \left( h_1 | h_2\right) \equiv 4 \Re \left[\int_{f_{\rm{low}}}^{f_{\rm{high}}} \frac{\tilde{h}_1 (f) \tilde{h}^{*}_2 (f)}{S_n (f)} df\right]\,,
\end{align}
where, $\tilde{h}_1(f)$, $\tilde{h}_2(f)$ are the Fourier transforms of $h_1$, $h_2$, $f_{\rm{low}}$, $f_{\rm{high}}$, $S_n (f)$ are lower and upper cutoff frequencies, and one-sided power-spectral density (PSD) of the detector noise, respectively. The $*$ denotes the complex conjugate of the quantity. One natural way of quantifying the agreement between two waveforms is to compute the \textit{match} (also known as \textit{overlap}), defined as~\citep{Purrer:2014fza},
\begin{align}\label{eq:match}
\mathcal{M}(h_1, h_2) \equiv \, \underset{\phi_c,  t_c}{\rm{max}}    \dfrac{(e^{\rm{i} (\phi_c + 2\pi f t_c)} h_1|h_2)}{\sqrt{(h_1|h_1) (h_2|h_2)}}\,.
\end{align}
The quantity $1-\mathcal{M}$ is referred to as the \textit{mismatch} between the two waveforms $h_1$ and $h_2$. The numerical value of mismatch is used extensively in GW modelling and parameter estimation studies to denote the disagreement between two waveforms. 

Following the discussion above, we, too, compute and quote mismatch values between \TFtwoE{}~\citep{Moore:2016qxz} and with the newly computed spinning eccentric corrections in \TFtwo{} phase up to 3PN and expanded in second power in eccentricity added in \TFtwoE{}. We perform the mismatch study in the parameter space of total mass $(M)= (5-100) M_{\odot}$, mass ratio $(q) = 1-4$, initial eccentricity at $10$Hz $(e_{10})  = 0.1-0.3$, and dimensionless spins of $(\chi_{1}, \chi_{2}) = 0.0-0.9$. {\blue Note that, following \TFtwoE{}, we too set $\kappa_1$ and $\kappa_2$ equal to one while performing the comparison study.} Instead of dealing with $M$ and individual dimensionless spins $(\chi_1, \chi_2)$ of the binary, we define two parameters related to component masses $(m_1, m_2)$ and spins of the binary called the \textit{chirp mass} $(M_{\rm{chirp}})$ and \textit{effective spin parameter} $(\chi_{\rm{eff}})$, defined as,
\begin{equation}\label{eq:chi_eff}
    M_{\rm{chirp}} = \frac{(m_1 m_2)^{3/5}}{(m_1 + m_2 )^{1/5}}\,,\,\,
    \chi_{\rm{eff}} = \frac{m_1 \chi_1  + m_2 \chi_2 }{M}.
\end{equation}
 \begin{widetext}
 \begin{center}
\begin{figure}[H]
    \centering
    \includegraphics[width=\linewidth]{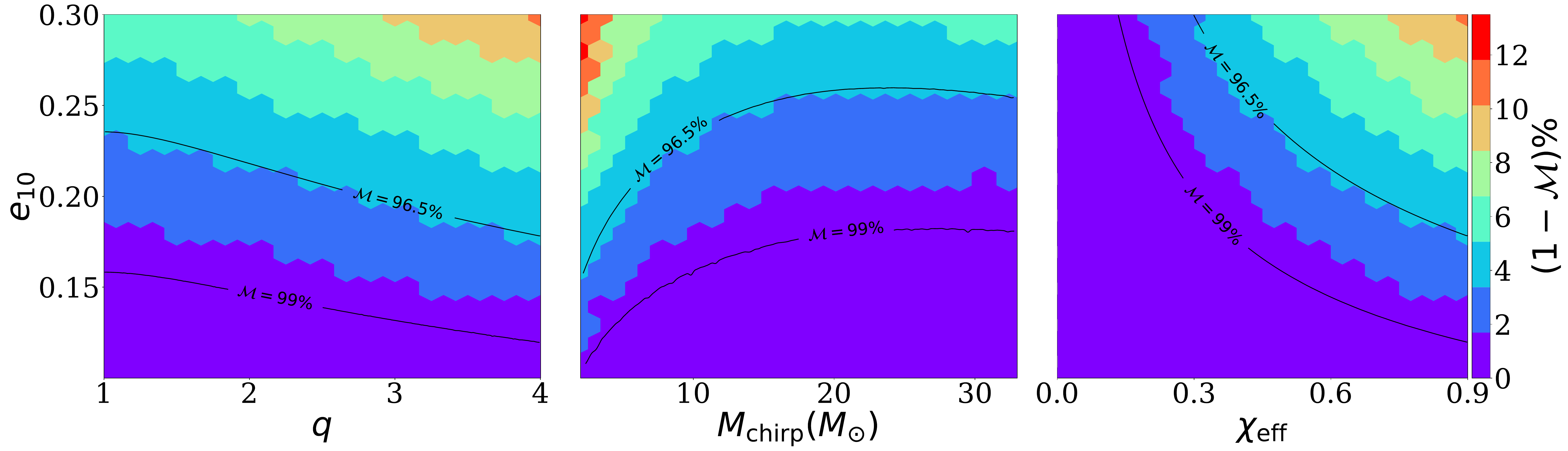}
    \caption{\justifying The comparison of \TFtwoE{}~\citep{Moore:2016qxz} with the spinning eccentric corrections in \TFtwo{} up to 3PN order, expanded in eccentricity up to second power in eccentricity added in \TFtwoE{} is shown here. We compute mismatch $(1-\mathcal{M})\%$ between the two waveforms in various parameter planes, such as (a) eccentricity at $10$Hz $(e_{10})$ - mass-ratio $(q)$ (\textit{left}), (b) $e_{10}$ - chirp mass $(M_{\rm{chirp}})$ (\textit{middle}), (c) $e_{10}$ - effective spin parameter $(\chi_{\rm{eff}})$ (\textit{right}), respectively. The total mass $(M)$ is fixed to $10M_{\odot}$ ({\textit{left and right}), $q$ is fixed to $4$ (\textit{middle and right}), the $z$-component of dimensionless spins $(\chi_{1_z},\chi_{2_z})$ are fixed to $(0.9, 0.9)$ (\textit{left and middle}). The $96.5\%$ and $99\%$ match contours (as indicated by the solid black curves) are shown in each figure. The mismatch is performed using a lower cutoff frequency $10$Hz, choosing Schwarzschild ISCO frequency (defined in Eq.~\eqref{eq:f_ss_isco}) as the upper cutoff frequency, and using advanced LIGO (aLIGO) zero-detuned high-power PSD.}
    }
    \label{fig:mismatch_pn}
\end{figure}
\end{center}
\begin{figure}[H]
    \centering
    \includegraphics[width=0.41\linewidth]{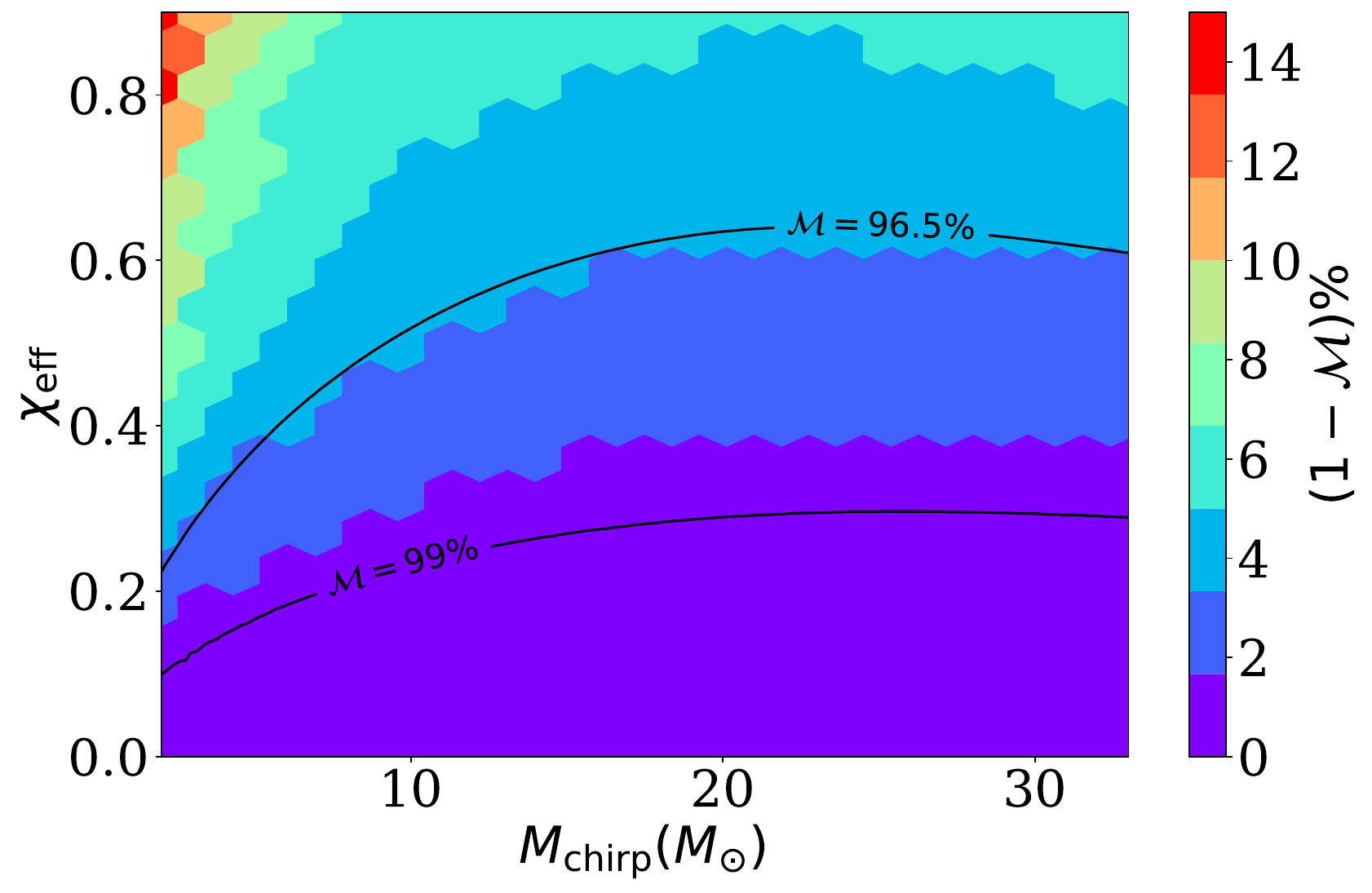}
    \caption{\justifying Similar to Fig.~\ref{fig:mismatch_pn}, here we show the mismatch $(1-\mathcal{M})\%$ as a function of $\chi_{\rm{eff}}$ and $M_{\rm{chirp}}$. $q$ and $e_{10}$ are fixed to $4$ and $0.3$, respectively. The $96.5\%$ and $99\%$ match contours (as indicated by the solid black curves) are shown in the figure. The mismatch is performed using a lower cutoff frequency $10$Hz, choosing Schwarzschild ISCO frequency (defined in Eq.~\eqref{eq:f_ss_isco}) as the upper cutoff frequency, and using advanced LIGO (aLIGO) zero-detuned high-power PSD.}
    \label{fig:mismatch_pn_2}
\end{figure}
\end{widetext}


In Fig.~\ref{fig:mismatch_pn}, we show the mismatch (denoted as color bar) in (a) $e_{10} - q$ (\textit{left}), (b) $e_{10} - M_{\rm{chirp}}$ (\textit{middle}), (c) $e_{10} - \chi_{\rm{eff}}$ (\textit{right}) panels, respectively. We show the mismatch as a function of $\chi_{\rm{eff}}$ and $M_{\rm{chirp}}$ separately in Fig.~\ref{fig:mismatch_pn_2}. We fix various parameters while performing the mismatch study in different parameter planes. For instance, total mass $(M)$ is fixed to $10M_{\odot}$ in the left and right figures of Fig.~\ref{fig:mismatch_pn}, $q$ is fixed to $4$ for middle, right 
 figures of Fig.~\ref{fig:mismatch_pn} and in Fig.~\ref{fig:mismatch_pn_2}. The dimensionless spins are fixed to $(0.9,0.9)$ for the left and middle figures of Fig.~\ref{fig:mismatch_pn}. Further, $e_{10}$ is fixed to $0.3$ in Fig.~\ref{fig:mismatch_pn_2}. We denote the $96.5\%$ and $99\%$ match values between the two waveforms using contours (denoted by solid black curves) to identify the region of parameter space where these spinning eccentric terms become important and can't be ignored {\blue for detection and parameter estimation.} In $e_{10} - q$ plane, though the mismatches are negligible $(\lesssim 1\%)$ up to $e_{10} \sim 0.15$ for all mass ratios, it becomes as large as $\sim 11\%$ for $e_{10} = 0.3$ and $q=4$. Next, in $e_{10}-M_{\rm{chirp}}$ plane, the effect of these corrections is mostly negligible up to $e_{10}\sim 0.18$ for $M_{\rm{chirp}}>10M_{\odot}$. The mismatch achieves a value of $13\%$ for lower chirp mass $(M_{\rm{chirp}}< 10M_{\odot})$ and higher eccentricity $(e_{10}\gtrsim 0.25)$ value. Further, the mismatch variation due to eccentricity and spins can be quantified by looking at the panel's right figure. The mismatches between the two can be as high as $\sim 11\%$ for high spin $(\sim (0.9, 0.9))$ and high $(\sim 0.3)$ eccentricity value. Finally, the mismatches attain an even larger value in $\chi_{\rm{eff}}-M_{\rm{chirp}}$ plane (shown in Fig.~\ref{fig:mismatch_pn_2}). For systems with $\chi_{\rm{eff}} \gtrsim 0.2$ shows a mismatch of $> 1\%$ for all $M_{\rm{chirp}}$, while for lighter $(M_{\rm{chirp}} \sim 2M_{\odot})$ and maximally spinning systems $(\chi_{\rm{eff}}\gtrsim 0.8)$ the mismatch can achieve a value $\sim15\%$. This observation is consistent as lower mass positive aligned systems will have more cycles than a high mass case in the frequency band of current ground-based detectors. This will lead to a larger accumulation of mismatches in the inspiral cycles for lighter mass cases compared to the heavier ones. All these observations point to the importance of including effects due to eccentricity and spins in the waveform templates while doing parameter estimation and searches for spinning eccentric GW signals.

\section{Conclusions and Discussions}\label{sec:discussions}
The main objective of our work has been to include spinning eccentric corrections, as well as higher-order eccentricity ($\mathcal{O}(e_0^8)$) corrections to the existing analytical inspiral-only Taylor approximants derived in Ref.~\citep{Moore:2016qxz}. We perform this task and give the reader a prescription to calculate the eccentricity evolution to any arbitrary eccentricity order of $e_0$. Although we only quote low eccentricity solution and Taylor approximants (till {\blue $\mathcal{O}(e_0^2)$}) in the main paper, we provide the full {\blue $\mathcal{O}(e_0^8)$} solution in a supplementary file~\citep{SBPM:suppl}. 

We estimate the importance of the newly computed spinning eccentric expressions reported in the \TFtwo{} phase. To do so, we compute mismatch (defined in Eq.~\eqref{eq:match}) between \TFtwoE{}~\citep{Moore:2016qxz} and spinning eccentric corrections in \TFtwo{} phase expanded in eccentricity up to $\mathcal{O}\left(e_{0}^{2}\right)$ added in \TFtwoE{} in Figs.~\ref{fig:mismatch_pn} ~and~\ref{fig:mismatch_pn_2}. We show the mismatch in eccentricity at 10 Hz $(e_{10})$, chirp mass $(M_{\rm{chirp}})$, effective spin parameter\, $(\chi_{\rm{eff}})$ (defined in Eq.~\eqref{eq:chi_eff}), mass ratio $(q)$ parameter space. {\blue The 96.5\% and 99\% match contours (denoted by solid black lines) indicate parts of the parameter space where these newly calculated terms involving both eccentricity and spin become important for detection and parameter estimation, respectively. Clearly, these contributions are most important for low $M_{\rm chirp}$, high $q$ and high $\chi_{\rm eff}$ values as the mismatch grows to values $>10\%$. Further, as can be seen from the (extreme) right panel of Fig.~\ref{fig:mismatch_pn}, the match drops below 96.5\% (theoretical detection threshold) for $e_{10}\gtrsim 0.18$ and $\chi_{\rm eff}\gtrsim 0.3$ assuming a $M=10M_{\odot}$ and $q=4$ binary. Naturally, these contributions become relevant for parameter estimation studies for even smaller values of eccentricity and spins, see Fig.~\ref{fig:mismatch_pn}.}

While our study primarily focuses on the secular contribution to the orbit and waveform in terms of the \TTtwo{} approximants, we also provide a derivation of the oscillatory part of the phase, $W(f)$, in Appendix \ref{sec:osc}. It has also been shown that the eccentric spinning section of the oscillatory terms contributes negligibly to the total GW cycles for $e_0 = 0.1$ and any spin setting; {\blue see Fig.~\ref{fig:spin-osscillatory}. Further, as can be verified from Table~\ref{table_cycles_freq_bands_osc_phase}, for $e_0=0.2$ the contribution to number of cycles from these terms is typically below one tenth of a GW cycle.}  

While we include eccentricity corrections to $\mathcal{O}(e_0^8)$ to the phasing approximants, we do not consider eccentricity corrections to the waveform amplitude. But, our low-eccentricity approximation has been compared in Sec.~\ref{sec:comparison} with a numerically calculated solution to the evolution equations (Eq.~\eqref{eq:ydotedot-structure}), which is exact in eccentricity. For most spin systems, our analytic expressions are valid up to $e_0 \lesssim 0.45$ (arrived at by setting the threshold of accumulated cycles difference to $\Delta N_{\rm cyc} \leq 1$), and the resummed expressions extend this validity up to $e_0 \lesssim 0.55$. This validity is subject to the specific settings of the system under consideration.

{\blue On at least three separate occasions we use cumulative number of cycles estimates to quantify the importance of the results presented here. Table~\ref{Intro cycles table} highlights corrections in number of GW cycles estimates due to spins in the phasing formula, obtained by performing comparisons with its non-spinning counterpart presented in Ref.~\citep{Moore:2016qxz}. Clearly, the inclusion of spins is critical; the leading (1.5 PN) spin-orbit terms correct non-spinning estimates by more than 50 cycles. On the other hand, the correction due to eccentric, spinning terms at the same order is much lower ($\sim0.5$ cycles). Note however that, Table~\ref{Intro cycles table} employs phasing results with only leading eccentricity corrections. As can be verified from Table~\ref{table_cycles_freq_bands}, which uses \TTtwo{} phasing with eccentric corrections to $\mathcal{O}(e^8)$, the eccentric, spinning contributions at this order are consistently larger than one GW cycle, which hints at the importance of including higher order eccentricity corrections. Moreover, these estimates correspond to a low value of eccentricity ($e_0=0.2$), while the phasing formulae presented here can handle eccentricities up to $\sim0.5$ as should be clear from Fig.~\ref{fig:comparison}. Table~\ref{table_cycles_freq_bands_higher_ecc} displays these estimates for $e_0$ value of $0.5$. As expected the contributions from eccentric, spinning terms grew significantly to few to $\mathcal{O}(10)$ GW cycles depending upon the detector configuration. Estimates are similar (within $\mathcal{O}(1)$ cycle) when resummed version of \TTtwo{} presentd in Sec.~\ref{sec:comparison} is used.}

In this study, we calculated both time and frequency domain phasing for spin-aligned and eccentric compact binary systems. An immediate consequence of this study will be the estimation of source parameters of GWs coming from spinning and eccentric binary systems. The results of this study can be used to create new inspiral waveform templates or augment existing waveform models. The inclusion of the combined effect of spins and orbital eccentricity in waveform models will lead to lesser systematic biases in the estimation of GW source properties and can possibly lead to more detections because of a better match between a template waveform and the data, leading to an increase in the signal-to-noise ratio. A possible extension of this study is to include the effects of misaligned spins on the spherical harmonic modes and the phasing of compact binary mergers with eccentric orbits. In such systems, there is a precession of the orbit due to spin misalignment, which leads to amplitude modulation of the GW signal since the plane of the orbit keeps changing with respect to the line connecting the source and the detector. In the absence of matter (for example, accretion disks), spin-misaligned eccentric systems form the most general situation a compact binary system can be in. Hence, we hope to include the combined effect of spin precession and eccentricity in GW waveforms in a future study.

\acknowledgments
We thank Prayush Kumar for insightful discussions on waveform related issues. We also thank Guillaume Faye for key inputs throughout the course of this study. C.K.M. acknowledges the support of SERB’s Core Research Grant No. CRG/2022/007959. K.P. acknowledges the hospitality of the International Centre for Theoretical Sciences (ICTS) during the final stages of writing the paper.  We thank Anuradha Gupta for useful comments and suggestions on our manuscript. This document has LIGO preprint number LIGO-P2400592.
\appendix
\section{Quasi-Keplerian (QK) parametrization}\label{app:QK}
The energy and angular momentum emitted by an inspiralling compact binary system can be written as functions of the distance of separation $r$, the orbital phase $\phi$, and their derivatives. Looking only at the conservative problem, the derivatives of $r$ and $\phi$ can be written instead as a function of the energy and angular momentum. Treating the energy and angular momentum as constants in the conservative problem, these two first-order differential equations can be used to find the following 3PN relations between the $r$ and the eccentric anomaly $u$ and the relationship between various other angles. In the following, we have reproduced the set of QK equations from Ref.~\citep{Moore:2016qxz} for the reader's advantage.
\begin{subequations}
\begin{align}
r &= S(l;n,e_t) = a_r(1-e_r \cos u) \, ,  \label{eq:req}\\
\label{eq:rdot}
\dot{r} &= n \frac{\partial S}{\partial l}(l;n,e_t) \, , \\
\label{eq:phi}
\phi &= \lambda + W(l;n,e_t) \, , \\
\label{eq:lambda}
\lambda &= (1+k) n (t-t_0)+c_{\lambda} \, , \\ 
\label{eq:W}
W(l;n,e_t)
&= ( 1 + k ) ( v - l ) 
+ \bigg( \frac{ f_{4\phi} }{c^4} + \frac{ f_{6\phi} }{c^6} \bigg) \sin 2 v \nonumber\\
& + \bigg( \frac{ g_{4\phi} }{c^4} + \frac{ g_{6\phi} }{c^6} \bigg) \sin 3 v
+ \frac{ i_{6\phi} }{c^6} \sin 4 v \nonumber\\
& + \frac{ h_{6\phi} }{c^6} \sin 5 v
\, , \\
\label{eq:phidot}
\dot{\phi} &= (1+k) n+n\frac{\partial W}{\partial l}(l;n,e_t) \, , \\
\label{eq:keplereq}
l &= n(t-t_0)+c_l = u - e_t \sin u \nonumber\\
&+ \bigg( \frac{g_{4t}}{c^4} + \frac{g_{6t}}{c^6} \bigg) (v - u)
+ \bigg( \frac{f_{4t}}{c^4} + \frac{f_{6t}}{c^6} \bigg) \sin v \nonumber\\
&+ \frac{i_{6t}}{c^6} \sin 2v + \frac{h_{6t}}{c^6} \sin 3v
\, , \\
\label{eq:veqn}
v &= V(u) \equiv 2 \arctan \left[ \left( \frac{1+e_{\phi}}{1-e_{\phi}}\right)^{1/2} \tan\Big(\frac{u}{2}\Big) \right] \, . \nonumber\\
\end{align}
\end{subequations}
In the above, the various symbols have a corresponding PN series, which are energy and angular momentum functions. A non-exhaustive list of the meanings of the symbols in Eqs. (\ref{eq:req})-(\ref{eq:veqn}) are given as follows
\begin{itemize}
    \item $a_r$: Semi-major axis
    \item $e_r$: Radial eccentricity
    \item  $u$: Eccentric anomaly
    \item $\lambda$: Secularly increasing part of the orbital phase
    \item $W$: Oscillatory part of the orbital phase
    \item $k$: Periastron precession under a single orbit
    \item $v$: True anomaly
    \item $l$: Mean anomaly
    \item $n = \frac{2 \pi}{P}$: Mean motion
    \item $P$: Radial orbital period
    \item $e_t$: time eccentricity
    \item $e_\phi$: phase eccentricity
\end{itemize}
It is to be noted that while all the quantities are gauge-dependent objects, the mean motion and the periastron advance are gauge-invariant. It is also to be noted that the orbital frequency $\omega$ is defined as follows
\begin{equation}\label{eq:omega-def}
    \omega = \left(1 + k\right) \, n\,,
\end{equation}
which is also a gauge-invariant quantity. Hence, one can define a dimensionless and gauge-invariant quantity $x$ that is given by
\begin{equation}\label{eq:x-def}
    x = \left(\frac{G\,M\,\omega}{c^3}\right)^{2/3}\,,
\end{equation}
which is a monotonic and increasing function of time and can be used as a substitute of time itself. {\blue Note that $e$ is used instead of $e_t$ in the rest of the paper.}
\section{Oscillatory part of the orbital phase till {\blue $\mathcal{O}(e_0^2)$}}\label{sec:osc}
One finds the oscillatory part of the phase from Eq.~\eqref{eq:W}. To obtain it as a function of the GW frequency, however, one needs to calculate quantities like the periastron precession ($k$), the true anomaly ($v$), the mean anomaly ($l$), and $f_\phi$, $g_\phi$, $i_\phi$ and $h_\phi$ as a function of the PN parameter or the GW frequency.

Considering a binary with no radiation reaction, differentiation of Eq. (\ref{eq:keplereq}) with respect to time yields the following equation for the mean anomaly evolution
\begin{equation}\label{eq:l-evol}
    \frac{dl}{dt} = n\,.
\end{equation}
Again, using the chain rule, we rewrite the $l$ evolution with respect to time as $l$ evolution with respect to the PN parameter $y$, that is
\begin{eqnarray}\label{eq:l-evol-y}
    \frac{dl}{dy} = \frac{n}{\frac{dy}{dt}}\,,
\end{eqnarray}
where $\frac{dy}{dt}$ was given in Eq.~\eqref{eq:ydot-structure}, and we obtain $n$ from the supplementary material of Ref.~\citep{Henry:2023tka}. Rewriting $n$ as $y$, and integrating both sides of Eq.~\eqref{eq:l-evol-y}, we obtain $l$ as follows
\begin{widetext}
    \begin{equation}
        l = -\frac{1}{32 y^5 \nu} \left(\mathcal{L}_{\rm NS} + \mathcal{L}_{\rm SO} + \mathcal{L}_{\rm SS} \right)\,,
    \end{equation}
    where,
    \begin{subequations}
        \begin{align}
            \mathcal{L}_{\rm NS} &= 1 - e_0^2 \frac{105}{272} \left(\frac{y_0}{y}\right)^{19/3} + \mathcal{O}\left(y^2, e_0^4\right)\,, \\
            \mathcal{L}_{\rm SO,circ}^{\rm 1.5PN} &= y^3 \left[\frac{805 \delta  \chi_a}{24}+\left(\frac{805}{24}-\frac{125 \nu }{6}\right) \chi_s\right]\,, \\
            \mathcal{L}_{\rm SO,ecc}^{\rm 1.5PN} &= - e_0^2 \frac{105}{272} \left(\frac{y_0}{y}\right)^{19/3} \left\{ 
y^3 \left[\left(-\frac{8602 \nu }{675}-\frac{4726}{4725}\right) \chi_s-\frac{4726 \delta  \chi_a}{4725}\right] + y_0^3 \left[\left(\frac{55 \nu }{27}-\frac{157}{54}\right) \chi_s-\frac{157 \delta  \chi_a}{54}\right] \right\}\,, \\
            \mathcal{L}_{\rm SS,circ}^{\rm 2PN} &= y^4 \left\{ \chi_a^2 \left[-\frac{65 \delta  \kappa_a}{2}+\kappa_s \left(65 \nu -\frac{65}{2}\right)+65 \nu -\frac{5}{16}\right]+\chi_a \chi_s \left[-65 \delta  \kappa_s-\frac{5 \delta
   }{8}+\kappa_a (130 \nu -65)\right] \right.\nonumber\\
   &\left.+ \chi_s^2 \left[-\frac{65 \delta  \kappa_a}{2}+\kappa_s \left(65 \nu -\frac{65}{2}\right)-\frac{255 \nu }{4}-\frac{5}{16}\right] \right\}\,, \\
            \mathcal{L}_{\rm SS,ecc}^{\rm 2PN} &= - e_0^2 \frac{105}{272} \left(\frac{y_0}{y}\right)^{19/3} \left( y^4 \left\{ \chi_a^2 \left[\frac{461873 \delta  \kappa_a}{14784}+\kappa_s \left(\frac{461873}{14784}-\frac{461873 \nu }{7392}\right)+\frac{33371 \nu }{672}-\frac{415531}{14784}\right] \right.\right.\nonumber\\
            &+ \chi_a \chi_s
   \left[\frac{461873 \delta  \kappa_s}{7392}-\frac{415531 \delta }{7392}+\kappa_a \left(\frac{461873}{7392}-\frac{461873 \nu }{3696}\right)\right]+\chi_s^2 \left[\frac{461873 \delta  \kappa_a}{14784}+\kappa_s \left(\frac{461873}{14784}-\frac{461873 \nu }{7392}\right) \right.\nonumber\\
   &\left.\left.+ \frac{66283 \nu }{1056}-\frac{415531}{14784}\right] \right\} + y_0^4 \left\{ \chi_a^2 \left[\frac{89 \delta  \kappa_a}{192}+\kappa_s \left(\frac{89}{192}-\frac{89 \nu }{96}\right)-\frac{623 \nu }{96}+\frac{293}{192}\right]+\chi_a \chi_s \left[\frac{89 \delta 
   \kappa_s}{96}+\frac{293 \delta }{96} \right.\right.\nonumber\\
            &\left.\left.\left.+ \kappa_a \left(\frac{89}{96}-\frac{89 \nu }{48}\right)\right]+\chi_s^2 \left[\frac{89 \delta  \kappa_a}{192}+\kappa_s \left(\frac{89}{192}-\frac{89
            \nu }{96}\right)+\frac{37 \nu }{96}+\frac{293}{192}\right] \right\} \right)\,, \\
            \mathcal{L}_{\rm SO,circ}^{\rm 2.5PN} &= y^5 \left[\chi_a \left(-30 \delta  \nu  \log (y)-\frac{638185 \delta  \log (y)}{2016}\right)+\chi_s \left(60 \nu ^2 \log (y)+\frac{177185}{504} \nu  \log (y)-\frac{638185 \log (y)}{2016}\right)\right]\,, \\
            \mathcal{L}_{\rm SO,ecc}^{\rm 2.5PN} &= - e_0^2 \frac{105}{272} \left(\frac{y_0}{y}\right)^{19/3} \left\{ y^5 \left[ \chi_a \left(\frac{7612064279 \delta }{18098640}-\frac{7610543 \delta  \nu }{23940}\right)+\left(\frac{56185 \nu ^2}{513}-\frac{576320281 \nu }{1131165}+\frac{7612064279}{18098640}\right) \chi_s \right] \right.\nonumber\\
            &+ y_0^5 \left[ \chi_a \left(\frac{242719 \delta  \nu }{9720}-\frac{1279073 \delta }{38880}\right)+\left(-\frac{4322 \nu ^2}{243}+\frac{146807 \nu }{2430}-\frac{1279073}{38880}\right) \chi_s \right] + y^2 y_0^3 \left[ \chi_a \left(\frac{739313 \delta  \nu }{21168} \right.\right.\nonumber\\
            &\left.\left.- \frac{367678771 \delta }{5334336}\right)+\left(-\frac{258995 \nu ^2}{10584}+\frac{221958103 \nu }{2667168}-\frac{367678771}{5334336}\right) \chi_s \right] + y^3 y_0^2 \left[ \chi_a \left(\frac{465511 \delta  \nu }{85050}-\frac{6694379 \delta }{2381400}\right) \right.\nonumber\\
            &\left.\left.+ \left(\frac{847297 \nu ^2}{12150}-\frac{10322689 \nu }{340200}-\frac{6694379}{2381400}\right) \chi_s \right] \right\}\,, \\
            \mathcal{L}_{\rm SO,circ}^{\rm 3PN} &= y^6 \left[\frac{1615 \pi  \delta  \chi_a}{6}+\left(\frac{1615 \pi }{6}-170 \pi  \nu \right) \chi_s\right] \,,\\
            \mathcal{L}_{\rm SO,ecc}^{\rm 3PN} &= - e_0^2 \frac{105}{272} \left(\frac{y_0}{y}\right)^{19/3} \left\{ y^6 \left[ \frac{157779329 \pi  \delta  \chi_a}{870912}+\left(\frac{157779329 \pi }{870912}-\frac{6844081 \pi  \nu }{217728}\right) \chi_s \right] + y_0^6 \left[ \left(\frac{37663 \pi  \nu }{3888} \right.\right.\right.\nonumber\\
            &\left.\left.\left.- \frac{157043 \pi }{15552}\right) \chi_s-\frac{157043 \pi  \delta  \chi_a}{15552} \right] + y^3 y_0^3 \left[\left(-\frac{2764421 \pi  \nu }{42525}-\frac{2617099 \pi }{340200}\right) \chi_s-\frac{2617099 \pi  \delta  \chi_a}{340200}\right] \right\}\,,\\
            \mathcal{L}_{\rm SS,circ}^{\rm 3PN} &= y^6 \left\{ \chi_a^2 \left[\kappa_a \left(\frac{50705 \delta }{224}-\frac{505 \delta  \nu }{6}\right)+\kappa_s \left(-\frac{305 \nu ^2}{4}-\frac{180395 \nu }{336}+\frac{50705}{224}\right)-\frac{305 \nu
   ^2}{4}+\frac{273635 \nu }{252}-\frac{2951395}{8064}\right]\right.\nonumber\\
   &+\chi_a \chi_s \left[\kappa_s \left(\frac{50705 \delta }{112}-\frac{505 \delta  \nu }{3}\right)+\frac{8725 \delta  \nu }{72}-\frac{2951395 \delta
   }{4032}+\kappa_a \left(-\frac{305 \nu ^2}{2}-\frac{180395 \nu }{168}+\frac{50705}{112}\right)\right] \nonumber\\
   &\left.+ \chi_s^2 \left[\kappa_a \left(\frac{50705 \delta }{224}-\frac{505 \delta  \nu }{6}\right)+\kappa_s \left(-\frac{305 \nu ^2}{4}-\frac{180395 \nu }{336}+\frac{50705}{224}\right)+\frac{880 \nu ^2}{9}+\frac{1006615 \nu }{2016}-\frac{2951395}{8064}\right] \right\}\,, \\
            \mathcal{L}_{\rm SS,ecc}^{\rm 3PN} &= - e_0^2 \frac{105}{272} \left(\frac{y_0}{y}\right)^{19/3} \left( y^6 \left\{ \chi_a^2 \left[\kappa_a \left(\frac{33341947 \delta  \nu }{387072}+\frac{398332253 \delta }{3612672}\right)+\kappa_s \left(-\frac{40026449 \nu ^2}{64512}-\frac{728209501 \nu
   }{5419008} \right.\right.\right.\right.\nonumber\\
   &\left.\left.+ \frac{398332253}{3612672}\right)-\frac{394111 \nu ^2}{7168}+\frac{240105088291 \nu }{146313216}-\frac{130039497587}{292626432}\right]+\chi_a \chi_s \left[\kappa_s \left(\frac{33341947 \delta 
   \nu }{193536}+\frac{398332253 \delta }{1806336}\right) \right.\nonumber\\
   &\left.+ \frac{68731561 \delta  \nu }{5225472}-\frac{130039497587 \delta }{146313216}+\kappa_a \left(-\frac{40026449 \nu ^2}{32256}-\frac{728209501 \nu
   }{2709504}+\frac{398332253}{1806336}\right)\right]  \nonumber\\
   &+\chi_s^2 \left[\kappa_a \left(\frac{33341947 \delta  \nu }{387072}+ \frac{398332253 \delta }{3612672}\right)+\kappa_s \left(-\frac{40026449 \nu
   ^2}{64512}-\frac{728209501 \nu }{5419008}+\frac{398332253}{3612672}\right) \right.\nonumber\\
   &\left.\left. +\frac{3649860101 \nu ^2}{5225472}+\frac{3128341513 \nu }{20901888}  - \frac{130039497587}{292626432}\right] \right\} + y_0^6 \left\{ 
\chi_a^2 \left[\kappa_a \left(\frac{906103 \delta }{193536}-\frac{12877 \delta  \nu }{2304}\right)+\kappa_s \left(\frac{24271 \nu ^2}{3456} \right. \right.\right.\nonumber\\
   &\left. \left. -\frac{1446937 \nu
   }{96768}+\frac{906103}{193536}\right) + \frac{169897 \nu ^2}{3456}-\frac{40961143 \nu }{373248}+\frac{17465819}{746496}\right]+\chi_a \chi_s \left[\kappa_s \left(\frac{906103 \delta }{96768}-\frac{12877
   \delta  \nu }{1152}\right)  \right.\nonumber\\
   &\left. -\frac{5526373 \delta  \nu }{93312}+\frac{17465819 \delta }{373248} + \kappa_a \left(\frac{24271 \nu ^2}{1728}-\frac{1446937 \nu }{48384}+\frac{906103}{96768}\right)\right]+\chi_s^2
   \left[\kappa_a \left(\frac{906103 \delta }{193536}-\frac{12877 \delta  \nu }{2304}\right)  \right.\nonumber\\
   &\left.\left. +\kappa_s \left(\frac{24271 \nu ^2}{3456}-\frac{1446937 \nu }{96768} + \frac{906103}{193536}\right)+\frac{433639 \nu
   ^2}{93312}-\frac{16075987 \nu }{373248}+\frac{17465819}{746496}\right] \right\}\nonumber\\
   &+ y^2 y_0^4 \left\{ \chi_a^2 \left[\kappa_a \left(\frac{208429367 \delta }{18966528}-\frac{419101 \delta  \nu }{75264}\right)+ \kappa_s \left(\frac{419101 \nu ^2}{37632}-\frac{261236093 \nu
   }{9483264}+\frac{208429367}{18966528}\right)+\frac{419101 \nu ^2}{5376} \right. \right. \nonumber\\
   & \left. -\frac{33323519 \nu }{193536}+\frac{686177579}{18966528}\right] +\chi_a \chi_s \left[\kappa_s \left(\frac{208429367 \delta
   }{9483264}-\frac{419101 \delta  \nu }{37632}\right)-\frac{1379737 \delta  \nu }{37632}+\frac{686177579 \delta }{9483264}  \right.\nonumber\\
   &\left. +\kappa_a \left(\frac{419101 \nu ^2}{18816}-\frac{261236093 \nu
   }{4741632} + \frac{208429367}{9483264}\right)\right]+\chi_s^2 \left[\kappa_a \left(\frac{208429367 \delta }{18966528}-\frac{419101 \delta  \nu }{75264}\right) \right.\nonumber\\
   &\left.\left. +\kappa_s \left(\frac{419101 \nu
   ^2}{37632}-\frac{261236093 \nu }{9483264}+\frac{208429367}{18966528}\right)  - \frac{174233 \nu ^2}{37632}-\frac{87196451 \nu }{9483264}+\frac{686177579}{18966528}\right] \right\} \nonumber\\
   &+ y^3 y_0^3 \left[ \chi_a \chi_s \left(\frac{4466869 \delta  \nu }{127575} + \frac{741982 \delta }{127575}\right)+\left(-\frac{94622 \nu ^2}{3645}+\frac{4466869 \nu }{127575} + \frac{370991}{127575}\right) \chi_s^2+\left(\frac{370991}{127575}  \right.\right.\nonumber\\
   &\left.\left. -\frac{1483964 \nu }{127575}\right) \chi_a^2 \right] + y^4 y_0^2 \left\{ \chi_a^2 \left[\kappa_a \left(\frac{1308486209 \delta }{14902272}-\frac{90988981 \delta  \nu }{532224}\right) + \kappa_s \left(\frac{90988981 \nu ^2}{266112}-\frac{2582331943 \nu
   }{7451136} \right.\right.\right.\nonumber\\
   & \left.\left.  +\frac{1308486209}{14902272}\right)-\frac{6574087 \nu ^2}{24192}+\frac{2185974971 \nu }{7451136}-\frac{1177199323}{14902272}\right] + \chi_a \chi_s \left[\kappa_s \left(\frac{1308486209 \delta
   }{7451136}-\frac{90988981 \delta  \nu }{266112}\right)  \right.\nonumber\\
   &\left. +\frac{81859607 \delta  \nu }{266112}-\frac{1177199323 \delta }{7451136}+\kappa_a \left(\frac{90988981 \nu ^2}{133056}-\frac{2582331943 \nu
   }{3725568} + \frac{1308486209}{7451136}\right)\right] \nonumber\\
   & +\chi_s^2 \left[\kappa_a \left(\frac{1308486209 \delta }{14902272}-\frac{90988981 \delta  \nu }{532224}\right)+\kappa_s \left(\frac{90988981 \nu
   ^2}{266112}-\frac{2582331943 \nu }{7451136}+\frac{1308486209}{14902272}\right)  - \frac{13057751 \nu ^2}{38016} \right. \nonumber \\
   & \left.\left.\left. +\frac{351498953 \nu }{1064448}-\frac{1177199323}{14902272}\right] \right\} \right)\,.
        \end{align} 
    \end{subequations}
\end{widetext}
Inverting Eq.~\eqref{eq:keplereq}, one obtains $u$ in terms of $l$. Similarly, once $u$ is obtained as a function of $l$, $v$ can also be obtained as a function of u through Eq.~\eqref{eq:veqn}. The expressions of $u$ as a function of $l$ and $v$ as a function of $u$ are rather large and have been given in a supplementary file~\citep{SBPM:suppl}. We also find that out of the set {$f_\phi, g_\phi, i_\phi, \text{ and } h_\phi$}, only $f_\phi$ contributes at 3PN for spinning systems. That is, the rest of the contribution to $g_\phi, i_\phi, \text{ and } h_\phi$ comes due to the non-spinning (and eccentric) parts of the system, so we ignore them and concentrate on the contribution of the spinning parts. Before calculating $W$, we find that we have one more quantity to calculate: $K = (1 + k)$. We find $K$ in terms of $y$ as follows
\begin{widetext}
    \begin{equation}
        K = \mathcal{K}_{\rm NS} + \mathcal{K}_{\rm SO} + \mathcal{K}_{\rm SS}\,,
    \end{equation}
    where,
    \begin{subequations}
        \begin{align}
            \mathcal{K}_{\rm NS} &= 1 + \mathcal{O}\left(y^2, e_0^2 y^4\right)\,,\\
            \mathcal{K}_{\rm SO,circ}^{\rm 1.5PN} &= y^3 \left[(2 \nu -4) \chi_s-4 \delta  \chi_a\right] \,,\\
            \mathcal{K}_{\rm SO,ecc}^{\rm 1.5PN} &= 0 \,,\\
            \mathcal{K}_{\rm SS,circ}^{\rm 2PN} &= y^4 \left\{ \chi_a^2 \left[\frac{3 \delta  \kappa_a}{2}+\kappa_s \left(\frac{3}{2}-3 \nu \right)-3 \nu \right]+\chi_a \chi_s \left[3 \delta  \kappa_s+\kappa_a (3-6 \nu )\right]+\chi_s^2
   \left[\frac{3 \delta  \kappa_a}{2}+\kappa_s \left(\frac{3}{2}-3 \nu \right)+3 \nu \right] \right\}\,,\\
            \mathcal{K}_{\rm SS,ecc}^{\rm 2PN} &= 0 \,,\\
            \mathcal{K}_{\rm SO,circ}^{\rm 2.5PN} &= y^5 \left[\chi_a \left(\frac{17 \delta  \nu }{2}-34 \delta \right)+\left(-2 \nu ^2+\frac{81 \nu }{2}-34\right) \chi_s\right] \,,\\
            \mathcal{K}_{\rm SO,ecc}^{\rm 2.5PN} &= e_0^2 \left( \frac{y_0}{y} \right)^{19/3} y^5 \left[\chi_a (14 \delta  \nu -30 \delta )+\left(-7 \nu ^2+29 \nu -30\right) \chi_s\right]\,,\\
            \mathcal{K}_{\rm SO,circ}^{\rm 3PN} &= 0 \,,\\
            \mathcal{K}_{\rm SO,ecc}^{\rm 3PN} &= 0 \,,\\
            \mathcal{K}_{\rm SS,circ}^{\rm 3PN} &= y^6 \left\{ \chi_a^2 \left[\kappa_a \left(\frac{39 \delta }{2}-\frac{17 \delta  \nu }{2}\right)+\kappa_s \left(5 \nu ^2-\frac{95 \nu }{2}+\frac{39}{2}\right)+5 \nu ^2-92 \nu +14\right]+\chi_a \chi_s
   \left[\kappa_s (39 \delta -17 \delta  \nu ) - 38 \delta  \nu \right.\right.\nonumber\\
   &\left.\left. +28 \delta +\kappa_a \left(10 \nu ^2-95 \nu +39\right)\right]+\chi_s^2 \left[\kappa_a \left(\frac{39 \delta }{2}-\frac{17 \delta  \nu
   }{2}\right)+\kappa_s \left(5 \nu ^2-\frac{95 \nu }{2}+\frac{39}{2}\right)+9 \nu ^2-2 \nu +14\right] \right\}\,, \\
            \mathcal{K}_{\rm SS,ecc}^{\rm 3PN} &= e_0^2 \left( \frac{y_0}{y} \right)^{19/3} y^6 \left\{ \chi_a^2 \left[\kappa_a \left(\frac{39 \delta }{2}-\frac{35 \delta  \nu }{4}\right)+\kappa_s \left(\frac{29 \nu ^2}{2}-\frac{191 \nu }{4}+\frac{39}{2}\right)+\frac{29 \nu ^2}{2}-\frac{249 \nu
   }{4}+6\right] \right.\nonumber\\
   &+ \chi_a \chi_s \left[\kappa_s \left(39 \delta -\frac{35 \delta  \nu }{2}\right)-\frac{15 \delta  \nu }{2}+12 \delta +\kappa_a \left(29 \nu ^2-\frac{191 \nu
   }{2}+39\right)\right]+\chi_s^2 \left[\kappa_a \left(\frac{39 \delta }{2}-\frac{35 \delta  \nu }{4}\right) \right.\nonumber\\
   &\left.\left.+ \kappa_s \left(\frac{29 \nu ^2}{2}-\frac{191 \nu }{4}+\frac{39}{2}\right)-\frac{29 \nu
   ^2}{2}+\frac{123 \nu }{4}+6\right] \right\}\,.
        \end{align}
    \end{subequations}
\end{widetext}
The oscillatory part of the phase $W$ as a function of $y$ is a rather lengthy expression, which we don't show in this text. However, we plot $W$ as a function of the GW frequency $f$ for the spinning case in Fig. \ref{fig:spin-osscillatory}. It is seen that the value of $W$ decreases rapidly as the binary inspiral progresses towards higher frequencies. 

When the spin is not zero, the oscillatory phase is similar to the non-spinning case. We have observed that the behaviour of the oscillatory phase does not depend much on the component spins. In Fig. \ref{fig:spin-osscillatory}, we have shown the oscillatory phase for different orientations and values of the spins. We noticed that the phase only changes while the amplitude remains constant.
\begin{widetext}
\begin{center}
\begin{figure}[h!]
    \centering
    \includegraphics[width=0.5\linewidth]{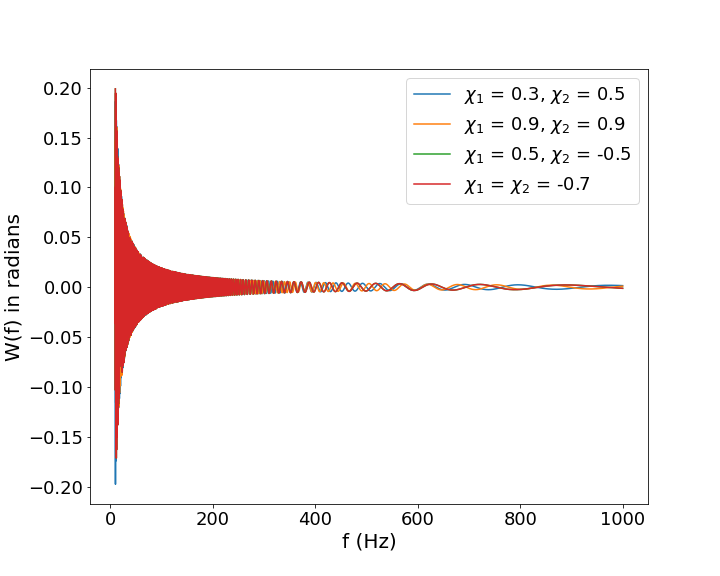}
    \caption{\justifying The oscillatory part of the phase $W(f)$ has been plotted as a function of the GW frequency $f$. We have limited the masses of the components to be $m_1 = m_2 = 1.4\,M_\odot$ in this figure and have varied the spins while keeping the initial eccentricity $e_0$ fixed at 0.1 at 10 Hz. We observed that there was very little difference from the non-spinning system. The amplitude of each scenario remained at 0.2. However, in the $W(f)$ phase, there were shifts when the spins and their orientations were altered.
    }
    \label{fig:spin-osscillatory}
\end{figure}

\begin{table}[H]
	\!\!\!\!\!\!\!\!\centering
 \resizebox{\columnwidth}{!}{
	\begin{tabular}{|l||c|c||c|c|c||c|c||c|c|}
		\hline
		Detector & \multicolumn{2}{c||}{LIGO/Virgo} & \multicolumn{3}{c||}{3G}& \multicolumn{2}{c||}{DECIGO} & \multicolumn{2}{c|}{LISA}  \\
		\hline
		Masses ($M_\odot$) & $1.4 + 1.4$ & $ 10 + 10$ &  $1.4 + 1.4$ & $50 + 50$ & $500 + 500$ & $500 + 500$ & $5000 + 5000$ & $10^5 + 10^5$ & $10^7 + 10^7$  \\
		\hline
		PN order &\multicolumn{9}{|c|}{cumulative number of cycles} \\
		\hline
		\hline
		0PN & $-0.005$ & $-0.121$ & $-0.025$ & $0.075$ & $0.131$ & $0.044$ & $0.075$ & $0.128$ & $-0.075$ \\
		\hline
		1PN &  & $-0.012$ &  & $-0.002$ & $0.034$ & $-0.001$ & $-0.002$ & $-0.003$ & $-0.034$\\
		\hline
		1.5PN &  & $0.003$ &  &  & $0.010$ &  &  &  & $0.042$ \\
		\hline
		2PN &  & $-0.002$ &  & $-0.001$ & $0.006$ &  & $-0.001$ &  & $0.015$ \\
		\hline
		2.5PN &  & $-0.002$ &  &  & $-0.007$ &  &  &  & $-0.011$\\
		\hline	
		3PN &  &  &  &  & $0.002$ &  &  &  & $0.001$ \\
		\hline
        \textbf{Total} & $-0.005$ & $-0.134$ & $-0.025$ & $0.072$ & $0.176$ & $0.043$ & $0.072$ & $0.125$ & $-0.062$ \\
		\hline
	\end{tabular} 
 }
	\caption{\justifying Similar to Table~\ref{table_cycles_freq_bands} except number of cycles estimates refer to contributions from the oscillatory phase (see Sec.~\ref{sec:osc}). {\blue Note that entries with value smaller than $10^{-3}$ are not displayed.}
    }\label{table_cycles_freq_bands_osc_phase}
\end{table}
\end{center}
\end{widetext}
\section{\TTone{}, \TTthree{} and \TTfour{}} \label{sec: app-Taylor}

\subsection{\TTone{}}
The orbital binding energy for a spin-aligned system in an eccentric orbit has the following structure in the $y$ parametrization
\begin{widetext}
    \begin{eqnarray}\label{eq:Energy-structure}
    \mathcal{E}(y) &=& -\frac{y^2}{2} \left(\mathcal{E}_{\rm NS}^{\rm {\blue Newt}} + \mathcal{E}_{\rm SO}^{\rm 1.5PN} + \mathcal{E}_{\rm SS}^{\rm 2PN} + \mathcal{E}_{\rm SO}^{\rm 2.5PN} + \mathcal{E}_{\rm SO}^{\rm 3PN} + \mathcal{E}_{\rm SS}^{\rm 3PN} \right) \,,
\end{eqnarray}
where,
\begin{subequations}
    \begin{align}
        \mathcal{E}_{\rm NS}^{\rm Newt} &= 1 - e_0^2 \left(\frac{y_0}{y}\right)^{19/3} + \mathcal{O}\left(y^2,e_0^4\right)\,,\\
        \mathcal{E}_{\rm SO,circ}^{\rm 1.5PN} &= y^3 \left[\frac{8 \delta  \chi_a}{3}+\left(\frac{8}{3}-\frac{4 \nu }{3}\right) \chi_s\right] \,,\\
        \mathcal{E}_{\rm SO,ecc}^{\rm 1.5PN} &= e_0^2 \left(\frac{y_0}{y}\right)^{19/3} \left\{y^3 \left[\left(\frac{91 \nu }{27}-\frac{301}{54}\right) \chi_s-\frac{301 \delta  \chi_a}{54}\right] + y_0^3 \left[\frac{157 \delta  \chi_a}{54}+\left(\frac{157}{54}-\frac{55 \nu }{27}\right) \chi_s\right]\right\} + \mathcal{O}\left(e_0^4\right) \,,
        \end{align}
    \begin{align}
     \mathcal{E}_{\rm SS,circ}^{\rm 2PN} &=  y^4 \left\{\chi_a^2 \left[-\frac{\delta  {\blue \kappa_{-}}}{2}+{\blue \kappa_{+}} \left(\nu -\frac{1}{2}\right)+2 \nu \right] + \chi_a \chi_s \left[{\blue \kappa_{-}} (2 \nu -1)-\delta  {\blue \kappa_{+}}\right] + \chi_s^2 \left[-\frac{\delta  {\blue \kappa_{-}}}{2}+{\blue \kappa_{+}} \left(\nu -\frac{1}{2}\right)-2 \nu \right]\right\}\,,\\
        \mathcal{E}_{\rm SS,ecc}^{\rm 2PN} &= e_0^2 \left(\frac{y_0}{y}\right)^{19/3} \left( y^4 \left\{\chi_a^2 \left[\frac{185 \delta  {\blue \kappa_{-}}}{192}+{\blue \kappa_{+}} \left(\frac{185}{192}-\frac{185 \nu }{96}\right)-\frac{815 \nu }{96}+\frac{293}{192}\right]+\chi_a \chi_s \left[\frac{185 \delta {\blue \kappa_{+}}}{96}+\frac{293 \delta }{96} \right.\right.\right.\nonumber\\
        &\left.\left.+ {\blue \kappa_{-}} \left(\frac{185}{96}-\frac{185 \nu }{48}\right)\right]+\chi_s^2 \left[\frac{185 \delta  {\blue \kappa_{-}}}{192}+{\blue \kappa_{+}}
   \left(\frac{185}{192}-\frac{185 \nu }{96}\right)+\frac{229 \nu }{96}+\frac{293}{192}\right]\right\} + y_0^4 \left\{ \chi_a^2 \left[-\frac{89 \delta  {\blue \kappa_{-}}}{192} \right.\right.\nonumber\\
   &\left.+ {\blue \kappa_{+}} \left(\frac{89 \nu }{96}-\frac{89}{192}\right)+\frac{623 \nu }{96}-\frac{293}{192}\right]+\chi_a \chi_s \left[-\frac{89 \delta 
   {\blue \kappa_{+}}}{96}-\frac{293 \delta }{96}+{\blue \kappa_{-}} \left(\frac{89 \nu }{48}-\frac{89}{96}\right)\right]+\chi_s^2 \left[-\frac{89 \delta  {\blue \kappa_{-}}}{192} \right.\nonumber\\
   &\left.\left.\left.+ {\blue \kappa_{+}} \left(\frac{89 \nu
   }{96}-\frac{89}{192}\right)-\frac{37 \nu }{96}-\frac{293}{192}\right] \right\} \right) \,,
   \end{align}
    
   \begin{align}
   \mathcal{E}_{\rm SO,circ}^{\rm 2.5PN} &= y^5 \left[\chi_a \left(8 \delta -\frac{31 \delta  \nu }{9}\right)+\left(\frac{2 \nu ^2}{9}-\frac{121 \nu }{9}+8\right) \chi_s\right] \,,\\
   \mathcal{E}_{\rm SO,ecc}^{\rm 2.5PN} &= e_0^2 \left(\frac{y_0}{y}\right)^{19/3} \left\{ y^5 \left[\chi_a \left(\frac{1855859 \delta }{272160}-\frac{295501 \delta  \nu }{9720}\right)+\left(\frac{4298 \nu ^2}{243}-\frac{1211369 \nu }{68040}+\frac{1855859}{272160}\right) \chi_s\right] \right.\nonumber\\
   & + y_0^5 \left[\chi_a \left(\frac{1279073 \delta }{38880}-\frac{242719 \delta  \nu }{9720}\right)+\left(\frac{4322 \nu ^2}{243}-\frac{146807 \nu }{2430}+\frac{1279073}{38880}\right) \chi_s\right] + y^2 y_0^3 \left[\chi_a \left(\frac{29987 \delta  \nu }{1944} \right.\right.\nonumber\\
   &\left.\left.- \frac{365653 \delta }{54432}\right)+\left(-\frac{10505 \nu ^2}{972}+\frac{547913 \nu }{27216}-\frac{365653}{54432}\right) \chi_s\right] + y^3 y_0^2 \left[\chi_a \left(\frac{59297 \delta  \nu }{1944}-\frac{121819 \delta }{7776}\right)+\left(-\frac{17927 \nu ^2}{972} \right.\right.\nonumber\\
   &\left.\left.\left. + \frac{155423 \nu }{3888}-\frac{121819}{7776}\right) \chi_s\right] \right\} \,,\\
   \mathcal{E}_{\rm SO,ecc}^{\rm 3PN} &= e_0^2 \left(\frac{y_0}{y}\right)^{19/3} \left\{ y^6 \left[\frac{533621 \pi  \delta  \chi_a}{15552}+\left(\frac{533621 \pi }{15552}-\frac{72421 \pi  \nu }{3888}\right) \chi_s\right] + y_0^6 \left[\frac{157043 \pi  \delta  \chi_a}{15552}+\left(\frac{157043 \pi }{15552} \right.\right.\right.\nonumber\\
   & \left.\left.\left.- \frac{37663 \pi  \nu }{3888}\right) \chi_s\right] + y^3 y_0^3 \left[\left(\frac{27521 \pi  \nu }{972}-\frac{86333 \pi }{1944}\right) \chi_s-\frac{86333 \pi  \delta  \chi_a}{1944}\right] \right\}\,,
   \end{align}
   \begin{align}
\mathcal{E}_{\rm SS,circ}^{\rm 3PN} =&  y^6 \left\{\chi_a^2 \left[{\blue \kappa_{-}} \left(\frac{25 \delta  \nu }{12}-\frac{35 \delta }{12}\right)+{\blue \kappa_{+}} \left(-\frac{5 \nu ^2}{6}+\frac{95 \nu }{12}-\frac{35}{12}\right)-\frac{5 \nu ^2}{3}+\frac{10 \nu}{9}+\frac{20}{9}\right]+\chi_a \chi_s \left[{\blue \kappa_{+}} \left(\frac{25 \delta  \nu }{6} \right.\right.\right.\nonumber\\
&\left.\left.- \frac{35 \delta }{6}\right)+\frac{70 \delta  \nu }{9}+\frac{40 \delta }{9}+{\blue \kappa_{-}} \left(-\frac{5 \nu^2}{3}+\frac{95 \nu }{6}-\frac{35}{6}\right)\right]+\chi_s^2 \left[{\blue \kappa_{-}} \left(\frac{25 \delta  \nu }{12}-\frac{35 \delta }{12}\right)+{\blue \kappa_{+}} \left(-\frac{5 \nu ^2}{6}+\frac{95 \nu
}{12}-\frac{35}{12}\right) \right.\nonumber\\
&\left.\left.- \frac{25 \nu ^2}{9}-\frac{20 \nu }{9}+\frac{20}{9}\right]\right\} \,,\\
\mathcal{E}_{\rm SS,ecc}^{\rm 3PN} &= e_0^2 \left(\frac{y_0}{y}\right)^{19/3} \left[ y^6 \left\{\chi_a^2 \left[{\blue \kappa_{-}} \left(\frac{28637 \delta  \nu }{6912}-\frac{106027 \delta }{21504}\right)+{\blue \kappa_{+}} \left(-\frac{47605 \nu ^2}{3456}+\frac{1355161 \nu
}{96768}-\frac{106027}{21504}\right) \right.\right.\right.\nonumber\\
&\left.- \frac{146515 \nu ^2}{3456}+\frac{196667449 \nu }{2612736}-\frac{56676329}{5225472}\right]+\chi_a \chi_s \left[{\blue \kappa_{+}} \left(\frac{28637 \delta  \nu
}{3456}-\frac{106027 \delta }{10752}\right)+\frac{470839 \delta  \nu }{93312}-\frac{56676329 \delta }{2612736} \right.\nonumber\\
&\left.+ {\blue \kappa_{-}} \left(-\frac{47605 \nu ^2}{1728}+\frac{1355161 \nu
}{48384}-\frac{106027}{10752}\right)\right]+\chi_s^2 \left[{\blue \kappa_{-}} \left(\frac{28637 \delta  \nu }{6912}-\frac{106027 \delta }{21504}\right)+{\blue \kappa_{+}} \left(-\frac{47605 \nu ^2}{3456}+\frac{1355161 \nu
}{96768} \right.\right.\nonumber\\
& \left.\left.\left.- \frac{106027}{21504}\right)+\frac{2238659 \nu ^2}{93312}-\frac{10018757 \nu }{373248}-\frac{56676329}{5225472}\right]\right\} + y_0^6 \left\{\chi_a^2 \left[{\blue \kappa_{-}} \left(\frac{12877 \delta  \nu }{2304}-\frac{906103 \delta }{193536}\right) \right.\right.\nonumber\\
&\left.+ {\blue \kappa_{+}} \left(-\frac{24271 \nu ^2}{3456}+\frac{1446937 \nu
}{96768}-\frac{906103}{193536}\right)-\frac{169897 \nu ^2}{3456}+\frac{40961143 \nu }{373248}-\frac{17465819}{746496}\right] \nonumber\\
&+ \chi_a \chi_s \left[{\blue \kappa_{+}} \left(\frac{12877 \delta  \nu }{1152}-\frac{906103
	\delta }{96768}\right)+\frac{5526373 \delta  \nu }{93312}-\frac{17465819 \delta }{373248}+{\blue \kappa_{-}} \left(-\frac{24271 \nu ^2}{1728}+\frac{1446937 \nu }{48384}-\frac{906103}{96768}\right)\right] \nonumber\\
& +\chi_s^2
\left[{\blue \kappa_{-}} \left(\frac{12877 \delta  \nu }{2304}-\frac{906103 \delta }{193536}\right)+{\blue \kappa_{+}} \left(-\frac{24271 \nu ^2}{3456}+\frac{1446937 \nu }{96768}-\frac{906103}{193536}\right)-\frac{433639 \nu
	^2}{93312}+\frac{16075987 \nu }{373248} \right.\nonumber\\
& \left.\left.- \frac{17465819}{746496}\right]\right\} +  y^3 y_0^3 \left\{\chi_a \chi_s \left(\frac{47257 \delta }{1458}-\frac{15421 \delta  \nu }{729}\right)+\left(\frac{5005 \nu ^2}{729}-\frac{15421 \nu }{729}+\frac{47257}{2916}\right) \chi_s^2 \right.\nonumber\\
&\left.+ \left(\frac{47257}{2916}-\frac{47257 \nu }{729}\right) \chi_a^2\right\} + y^2 y_0^4 \left\{\chi_a^2 \left[{\blue \kappa_{-}} \left(\frac{207281 \delta }{193536}-\frac{16999 \delta  \nu }{6912}\right)+{\blue \kappa_{+}} \left(\frac{16999 \nu ^2}{3456}-\frac{445267 \nu
}{96768}+\frac{207281}{193536}\right) \right.\right.\nonumber\\
&\left.+ \frac{118993 \nu ^2}{3456}-\frac{319207 \nu }{13824}+\frac{682397}{193536}\right]+\chi_a \chi_s \left[{\blue \kappa_{+}} \left(\frac{207281 \delta }{96768}-\frac{16999 \delta 
	\nu }{3456}\right)-\frac{55963 \delta  \nu }{3456}+\frac{682397 \delta }{96768} \right.\nonumber\\
&\left.+ {\blue \kappa_{-}} \left(\frac{16999 \nu ^2}{1728}-\frac{445267 \nu }{48384}+\frac{207281}{96768}\right)\right]+\chi_s^2 \left[{\blue \kappa_{-}}
\left(\frac{207281 \delta }{193536}-\frac{16999 \delta  \nu }{6912}\right)+{\blue \kappa_{+}} \left(\frac{16999 \nu ^2}{3456}-\frac{445267 \nu }{96768} \right.\right.\nonumber\\
&\left.\left.\left.+ \frac{207281}{193536}\right)-\frac{7067 \nu ^2}{3456}-\frac{697309 \nu
}{96768}+\frac{682397}{193536}\right]\right\} + y^4 y_0^2 \left\{ \chi_a^2 \left[{\blue \kappa_{-}} \left(\frac{524105 \delta }{193536}-\frac{36445 \delta  \nu }{6912}\right)+{\blue \kappa_{+}} \left(\frac{36445 \nu ^2}{3456} \right.\right.\right.\nonumber\\
&\left.\left.- \frac{1034335 \nu }{96768}+\frac{524105}{193536}\right)+\frac{160555
	\nu ^2}{3456}-\frac{3116989 \nu }{96768}+\frac{830069}{193536}\right]+\chi_a \chi_s \left[{\blue \kappa_{+}} \left(\frac{524105 \delta }{96768}-\frac{36445 \delta  \nu }{3456}\right) \right.\nonumber\\
&\left.- \frac{57721 \delta  \nu
}{3456}+\frac{830069 \delta }{96768}+{\blue \kappa_{-}} \left(\frac{36445 \nu ^2}{1728}-\frac{1034335 \nu }{48384}+\frac{524105}{96768}\right)\right]+\chi_s^2 \left[{\blue \kappa_{-}} \left(\frac{524105 \delta
}{193536}-\frac{36445 \delta  \nu }{6912}\right) \right.\nonumber\\
&\left.\left.\left.+ {\blue \kappa_{+}} \left(\frac{36445 \nu ^2}{3456}-\frac{1034335 \nu }{96768}+\frac{524105}{193536}\right)-\frac{45113 \nu ^2}{3456}-\frac{159337 \nu
}{96768}+\frac{830069}{193536}\right] \right\} \right]\,.
    \end{align}
\end{subequations}
\end{widetext}
{\blue where $\kappa_{-} = 2 \kappa_a$ and $\kappa_{+} = 2 \kappa_s$. Note that, $\kappa_{+,-}$ are un-normalized symmetric and asymmetric combinations of $\kappa_{1,2}$ and are used sometimes as an alternate representation to the one involving $\kappa_{s,a}$; see for instance Ref.~\citep{bohe:2015ana}. We express inputs and results in terms of $\kappa_{s,a}$ everywhere else (except in this section) in this paper.}\\

Similarly, the energy flux can be written as follows
\begin{widetext}
\begin{equation}\label{eq:flux-structure}
    \mathcal{F}\left(y\right) = \frac{32 \nu^2 y^{10}}{5} \left(\mathcal{F}_{\rm NS}^{\rm Newt} + \mathcal{F}_{\rm SO}^{\rm 1.5PN} + \mathcal{F}_{\rm SS}^{\rm 2PN} + \mathcal{F}_{\rm SO}^{\rm 2.5PN} + \mathcal{F}_{\rm SO}^{\rm 3PN} + \mathcal{F}_{\rm SS}^{\rm 3PN}\right)\,,
\end{equation}
where,
\begin{subequations}
    \begin{align}
        \mathcal{F}_{\rm NS}^{\rm Newt} &= 1 + \frac{37}{24} \left(\frac{y_0}{y}\right)^{19/3} e_0^2 + \mathcal{O}\left(y^2, e_0^4\right)\,, \\
        \mathcal{F}_{\rm SO,circ}^{\rm 1.5PN} &= y^3 \left[\left(3 \nu -\frac{11}{4}\right) \chi_s-\frac{11 \delta  \chi_a}{4}\right]\,, \\
        \mathcal{F}_{\rm SO,ecc}^{\rm 1.5PN} &= \frac{37}{24} \left(\frac{y_0}{y}\right)^{19/3} e_0^2 \left\{ y^3 \left[\left(\frac{7613 \nu }{999}-\frac{41099}{1998}\right) \chi_s-\frac{41099 \delta  \chi_a}{1998}\right] + y_0^3 \left[\left(\frac{55 \nu }{27}-\frac{157}{54}\right) \chi_s-\frac{157 \delta  \chi_a}{54}\right] \right\} \,,\\
        \mathcal{F}_{\rm SS,circ}^{\rm 2PN} &= y^4 \left\{ \chi_a^2 \left[2 \delta  \kappa_a+\kappa_s (2-4 \nu )-4 \nu +\frac{1}{16}\right]+\chi_a \chi_s \left[4 \delta  \kappa_s+\frac{\delta }{8}+\kappa_a (4-8 \nu
   )\right]+\chi_s^2 \left[2 \delta  \kappa_a+\kappa_s (2-4 \nu ) \right.\right.\nonumber\\
   &\left.\left.+ \frac{15 \nu }{4}+\frac{1}{16}\right] \right\}\,, \\
        \mathcal{F}_{\rm SS,ecc}^{\rm 2PN} &= \frac{37}{24} \left(\frac{y_0}{y}\right)^{19/3} e_0^2 \left( y^4 \left\{ \chi_a^2 \left[\frac{99235 \delta  \kappa_a}{7104}+\kappa_s \left(\frac{99235}{7104}-\frac{99235 \nu }{3552}\right)-\frac{79477 \nu }{3552}-\frac{8537}{7104}\right]+\chi_a \chi_s
   \left[\frac{99235 \delta  \kappa_s}{3552} \right.\right.\right.\nonumber\\
   & \left.\left.\left.- \frac{8537 \delta }{3552}+\kappa_a \left(\frac{99235}{3552}-\frac{99235 \nu }{1776}\right)\right]+\chi_s^2 \left[\frac{99235 \delta  \kappa_a}{7104}+\kappa_s \left(\frac{99235}{7104}-\frac{99235 \nu }{3552}\right)+\frac{96551 \nu }{3552}-\frac{8537}{7104}\right] \right\} \right)\,, \\
        \mathcal{F}_{\rm SO,circ}^{\rm 2.5PN} &= y^5 \left[ \chi_a \left(\frac{701 \delta  \nu }{36}-\frac{59 \delta }{16}\right)+\left(-\frac{157 \nu ^2}{9}+\frac{227 \nu }{9}-\frac{59}{16}\right) \chi_s \right]\,,\\
        \mathcal{F}_{\rm SO,ecc}^{\rm 2.5PN} &= \frac{37}{24} \left(\frac{y_0}{y}\right)^{19/3} e_0^2 \left\{ y^5 \left[ \chi_a \left(\frac{26398327 \delta  \nu }{359640}-\frac{892925123 \delta }{10069920}\right)+\left(-\frac{826165 \nu ^2}{17982}+\frac{643447193 \nu }{2517480}-\frac{892925123}{10069920}\right) \chi_s \right] \right.\nonumber\\
        & + y_0^5 \left[\chi_a \left(\frac{242719 \delta  \nu }{9720}-\frac{1279073 \delta }{38880}\right)+\left(-\frac{4322 \nu ^2}{243}+\frac{146807 \nu }{2430}-\frac{1279073}{38880}\right) \chi_s\right] + y^2 y_0^3 \left[ \chi_a \left(\frac{2335375 \delta  \nu }{71928} \right.\right.\nonumber\\
        & \left.\left.- \frac{63891935 \delta }{2013984}\right)+\left(-\frac{818125 \nu ^2}{35964}+\frac{55077775 \nu }{1006992}-\frac{63891935}{2013984}\right) \chi_s \right] + y^3 y_0^2 \left[ \chi_a \left(\frac{8096503 \delta  \nu }{71928}-\frac{116433467 \delta }{2013984}\right) \right.\nonumber\\
        &\left.\left.+ \left(-\frac{1499761 \nu ^2}{35964}+\frac{134918671 \nu }{1006992}-\frac{116433467}{2013984}\right) \chi_s \right] \right\}\,, \\
        \mathcal{F}_{\rm SO,circ}^{\rm 3PN} &= y^6 \left[ \left(\frac{34 \pi  \nu }{3}-\frac{65 \pi }{6}\right) \chi_s-\frac{65 \pi  \delta  \chi_a}{6} \right]\,, \\
        \mathcal{F}_{\rm SO,ecc}^{\rm 3PN} &= \frac{37}{24} \left(\frac{y_0}{y}\right)^{19/3} e_0^2 \left\{ y^6 \left[ \left(\frac{1300057 \pi  \nu }{143856}-\frac{18439049 \pi }{575424}\right) \chi_s-\frac{18439049 \pi  \delta  \chi_a}{575424} \right] + y_0^6 \left[ \left(\frac{37663 \pi  \nu }{3888}-\frac{157043 \pi }{15552}\right) \chi_s \right.\right.\nonumber\\
        &\left.\left.- \frac{157043 \pi  \delta  \chi_a}{15552} \right] + y^3 y_0^3 \left[\left(\frac{2127583 \pi  \nu }{35964}-\frac{9724027 \pi }{71928}\right) \chi_s-\frac{9724027 \pi  \delta  \chi_a}{71928}\right] \right\} \,,\\
        \mathcal{F}_{\rm SS,circ}^{\rm 3PN} &= y^6 \left\{ \chi_a^2 \left[\kappa_a \left(\frac{\delta }{7}-\frac{127 \delta  \nu }{8}\right)+\kappa_s \left(\frac{43 \nu ^2}{2}-\frac{905 \nu }{56}+\frac{1}{7}\right)+\frac{43 \nu ^2}{2}-\frac{25 \nu
   }{63}+\frac{77}{72}\right]+\chi_a \chi_s \left[\kappa_s \left(\frac{2 \delta }{7}-\frac{127 \delta  \nu }{4}\right) \right.\right.\nonumber\\
   & \left.- \frac{475 \delta  \nu }{36}+\frac{77 \delta }{36}+\kappa_a \left(43 \nu
   ^2-\frac{905 \nu }{28}+\frac{2}{7}\right)\right]+\chi_s^2 \left[\kappa_a \left(\frac{\delta }{7}-\frac{127 \delta  \nu }{8}\right)+\kappa_s \left(\frac{43 \nu ^2}{2}-\frac{905 \nu
   }{56}+\frac{1}{7}\right)-\frac{107 \nu ^2}{9} \right.\nonumber\\
   &\left.\left.- \frac{4303 \nu }{252}+\frac{77}{72}\right] \right\}\,, \\
        \mathcal{F}_{\rm SS,ecc}^{\rm 3PN} &= \frac{37}{24} \left(\frac{y_0}{y}\right)^{19/3} e_0^2\left( y^6 \left\{ \chi_a^2 \left[\kappa_a \left(\frac{38752555 \delta }{340992}-\frac{33416267 \delta  \nu }{255744}\right)+\kappa_s \left(\frac{14989123 \nu ^2}{127872}-\frac{183090199 \nu
   }{511488} \right.\right.\right.\right. \nonumber\\
   &\left.\left.+ \frac{38752555}{340992}\right)+\frac{150001 \nu ^2}{3456}-\frac{38341393513 \nu }{96671232}+\frac{11991613877}{193342464}\right]+\chi_a \chi_s \left[\kappa_s \left(\frac{38752555 \delta
   }{170496}-\frac{33416267 \delta  \nu }{127872}\right) \right.\nonumber\\
   & \left.- \frac{609975481 \delta  \nu }{3452544}+\frac{11991613877 \delta }{96671232}+\kappa_a \left(\frac{14989123 \nu ^2}{63936}-\frac{183090199 \nu
   }{255744}+\frac{38752555}{170496}\right)\right]+\chi_s^2 \left[\kappa_a \left(\frac{38752555 \delta }{340992} \right.\right.\nonumber\\
   &\left.\left.\left.- \frac{33416267 \delta  \nu }{255744}\right)+\kappa_s \left(\frac{14989123 \nu
   ^2}{127872}-\frac{183090199 \nu }{511488}+\frac{38752555}{340992}\right)-\frac{113766965 \nu ^2}{3452544}-\frac{388735387 \nu }{13810176} \right.\right. \nonumber\\
   &\left.\left. +\frac{11991613877}{193342464}\right] \right\} + y_0^6 \left\{ \chi_a^2 \left[\kappa_a \left(\frac{906103 \delta }{193536}-\frac{12877 \delta  \nu }{2304}\right)+\kappa_s \left(\frac{24271 \nu ^2}{3456}-\frac{1446937 \nu
   }{96768}+\frac{906103}{193536}\right)+\frac{169897 \nu ^2}{3456}  \right.\right.
   \nonumber\\
   &\left. -\frac{40961143 \nu }{373248} + \frac{17465819}{746496}\right] + \chi_a \chi_s \left[\kappa_s \left(\frac{906103 \delta }{96768}-\frac{12877
   \delta  \nu }{1152}\right)-\frac{5526373 \delta  \nu }{93312}+\frac{17465819 \delta }{373248}+\kappa_a \left(\frac{24271 \nu ^2}{1728}  \right.\right.\nonumber\\
   &\left.\left. -\frac{1446937 \nu }{48384} + \frac{906103}{96768}\right)\right]+\chi_s^2
   \left[\kappa_a \left(\frac{906103 \delta }{193536}-\frac{12877 \delta  \nu }{2304}\right)+\kappa_s \left(\frac{24271 \nu ^2}{3456}-\frac{1446937 \nu }{96768}+\frac{906103}{193536}\right)  \right.\nonumber\\
   &\left.\left. +\frac{433639 \nu
   ^2}{93312} - \frac{16075987 \nu }{373248}+\frac{17465819}{746496}\right] \right\} + y^2 y_0^4 \left\{ \chi_a^2 \left[\kappa_a \left(\frac{36218995 \delta }{7160832}-\frac{1323875 \delta  \nu }{255744}\right)+\kappa_s \left(\frac{1323875 \nu ^2}{127872}  \right.\right.\right.\nonumber\\
   &\left.\left. -\frac{54753245 \nu
   }{3580416} + \frac{36218995}{7160832}\right)+\frac{9267125 \nu ^2}{127872}-\frac{44935745 \nu }{511488}+\frac{119237815}{7160832}\right]+\chi_a \chi_s \left[\kappa_s \left(\frac{36218995 \delta
   }{3580416}  \right. \right.\nonumber\\
   &\left.\left. -\frac{1323875 \delta  \nu }{127872}\right) - \frac{4358375 \delta  \nu }{127872}+\frac{119237815 \delta }{3580416}+\kappa_a \left(\frac{1323875 \nu ^2}{63936}-\frac{54753245 \nu
   }{1790208}+\frac{36218995}{3580416}\right)\right]\nonumber\\
   &\left. +\chi_s^2 \left[\kappa_a \left(\frac{36218995 \delta }{7160832} - \frac{1323875 \delta  \nu }{255744}\right)+\kappa_s \left(\frac{1323875 \nu
   ^2}{127872}-\frac{54753245 \nu }{3580416}+\frac{36218995}{7160832}\right)-\frac{14875 \nu ^2}{3456} \right.\right. \nonumber\\
   &\left.\left. -\frac{45959915 \nu }{3580416}+\frac{119237815}{7160832}\right] \right\} + y^3 y_0^3 \left[ \chi_a \chi_s \left(\frac{6452543 \delta }{53946}-\frac{1727843 \delta  \nu }{26973}\right)+\left(\frac{418715 \nu ^2}{26973}-\frac{1727843 \nu }{26973}  \right.\right.\nonumber\\
   &\left.\left. + \frac{6452543}{107892}\right) \chi_s^2+\left(\frac{6452543}{107892} - \frac{6452543 \nu }{26973}\right) \chi_a^2 \right] + y^4 y_0^2 \left\{ \chi_a^2 \left[\kappa_a \left(\frac{281132755 \delta }{7160832}-\frac{19549295 \delta  \nu }{255744}\right)  \right.\right.\nonumber\\
   &\left. +\kappa_s \left(\frac{19549295 \nu ^2}{127872}-\frac{554822885 \nu
   }{3580416} + \frac{281132755}{7160832}\right)+\frac{15656969 \nu ^2}{127872}-\frac{201613295 \nu }{3580416}-\frac{24185321}{7160832}\right] \nonumber\\
   & +\chi_a \chi_s \left[\kappa_s \left(\frac{281132755 \delta
   }{3580416}-\frac{19549295 \delta  \nu }{127872}\right) + \frac{1681789 \delta  \nu }{127872}-\frac{24185321 \delta }{3580416}+\kappa_a \left(\frac{19549295 \nu ^2}{63936}-\frac{554822885 \nu
   }{1790208} \right.\right.\nonumber\\
   &\left.\left.\left. +\frac{281132755}{3580416}\right)\right]+\chi_s^2 \left[\kappa_a \left(\frac{281132755 \delta }{7160832} - \frac{19549295 \delta  \nu }{255744}\right)+\kappa_s \left(\frac{19549295 \nu
   ^2}{127872}-\frac{554822885 \nu }{3580416}+\frac{281132755}{7160832}\right) \right.\right.\nonumber\\
   & \left.\left.\left. -\frac{19020547 \nu ^2}{127872}+\frac{42439147 \nu }{511488}-\frac{24185321}{7160832}\right] \right\} \right)\,.
    \end{align}
\end{subequations}
\end{widetext}
We also give here $\frac{dE}{dy}$ for the convenience of the reader
\begin{widetext}
\begin{equation}\label{eq:dedy-structure}
    \frac{dE}{dy} = -y\left(\mathcal{G}_{\rm NS}^{\rm Newt} + \mathcal{G}_{\rm SO}^{\rm 1.5PN} + \mathcal{G}_{\rm SS}^{\rm 2PN} + \mathcal{G}_{\rm SO}^{\rm 2.5PN} + \mathcal{G}_{\rm SO}^{\rm 3PN} + \mathcal{G}_{\rm SS}^{\rm 3PN}\right)\,,
\end{equation}
where,
\begin{subequations}
    \begin{align}
        \mathcal{G}_{\rm NS}^{\rm Newt} &= 1 - e_0^2 \left(\frac{y_0}{y}\right)^{19/3} + \mathcal{O}\left(y^2, e_0^4\right)\,, \\
        \mathcal{G}_{\rm SO,circ}^{\rm 1.5PN} &= y^3 \left[\frac{20 \delta  \chi_a}{3}+\left(\frac{20}{3}-\frac{10 \nu }{3}\right) \chi_s\right] \,,\\
        \mathcal{G}_{\rm SO,ecc}^{\rm 1.5PN} &= -e_0^2 \left(\frac{y_0}{y}\right)^{19/3} \left\{ y^3 \left[ \frac{517 \delta  \chi_a}{54}+\left(\frac{517}{54}-\frac{145 \nu }{27}\right) \chi_s \right] + y_0^3 \left[ \left(\frac{55 \nu }{27}-\frac{157}{54}\right) \chi_s-\frac{157 \delta  \chi_a}{54} \right] \right\} \,,
    \end{align}

    \begin{align}
        \mathcal{G}_{\rm SS,circ}^{\rm 2PN} &= y^4 \left\{\chi_a^2 \left[-\frac{3 \delta  {\blue \kappa_{-}}}{2}+{\blue \kappa_{+}} \left(3 \nu -\frac{3}{2}\right)+6 \nu \right]+\chi_a \chi_s \left[{\blue \kappa_{-}} (6 \nu -3)-3 \delta  {\blue \kappa_{+}}\right]+\chi_s^2 \left[-\frac{3 \delta  {\blue \kappa_{-}}}{2}+{\blue \kappa_{+}} \left(3 \nu -\frac{3}{2}\right)-6 \nu \right]\right\}\,,\\
        \mathcal{G}_{\rm SS,ecc}^{\rm 2PN} &= -e_0^2 \left(\frac{y_0}{y}\right)^{19/3} \left( y^4 \left\{ \chi_a^2 \left[-\frac{377 \delta  {\blue \kappa_{-}}}{192}+{\blue \kappa_{+}} \left(\frac{377 \nu }{96}-\frac{377}{192}\right)+\frac{1199 \nu }{96}-\frac{293}{192}\right]+\chi_a \chi_s \left[-\frac{377 \delta 
   {\blue \kappa_{+}}}{96}-\frac{293 \delta }{96} \right.\right.\right. \nonumber\\
   & \left.\left.+ {\blue \kappa_{-}} \left(\frac{377 \nu }{48}-\frac{377}{96}\right)\right]+\chi_s^2 \left[-\frac{377 \delta  {\blue \kappa_{-}}}{192}+{\blue \kappa_{+}} \left(\frac{377 \nu
   }{96}-\frac{377}{192}\right)-\frac{613 \nu }{96}-\frac{293}{192}\right] \right\} + y_0^4 \left\{ \chi_a^2 \left[\frac{89 \delta  {\blue \kappa_{-}}}{192} \right.\right. \nonumber\\
   & \left.+ {\blue \kappa_{+}} \left(\frac{89}{192}-\frac{89 \nu }{96}\right)-\frac{623 \nu }{96}+\frac{293}{192}\right]+\chi_a \chi_s \left[\frac{89 \delta 
   {\blue \kappa_{+}}}{96}+\frac{293 \delta }{96}+{\blue \kappa_{-}} \left(\frac{89}{96}-\frac{89 \nu }{48}\right)\right]+\chi_s^2 \left[\frac{89 \delta  {\blue \kappa_{-}}}{192}+{\blue \kappa_{+}} \left(\frac{89}{192} \right.\right.\nonumber\\
   &\left.\left.\left.\left.- \frac{89
   \nu }{96}\right)+\frac{37 \nu }{96}+\frac{293}{192}\right] \right\} \right) \,,
   \end{align}

   \begin{align}
        \mathcal{G}_{\rm SO,circ}^{\rm 2.5PN} &= y^5 \left[\chi_a \left(28 \delta -\frac{217 \delta  \nu }{18}\right)+\left(\frac{7 \nu ^2}{9}-\frac{847 \nu }{18}+28\right) \chi_s\right]\,,\\
        \mathcal{G}_{\rm SO,ecc}^{\rm 2.5PN} &= -e_0^2 \left(\frac{y_0}{y}\right)^{19/3} \left\{ y^5 \left[ \chi_a \left(\frac{733051 \delta  \nu }{9720}-\frac{16313459 \delta }{272160}\right)+\left(-\frac{10655 \nu ^2}{243}+\frac{3030787 \nu }{34020}-\frac{16313459}{272160}\right) \chi_s \right] \right.\nonumber\\
        &+ y_0^5 \left[ 
\chi_a \left(\frac{242719 \delta  \nu }{9720}-\frac{1279073 \delta }{38880}\right)+\left(-\frac{4322 \nu ^2}{243}+\frac{146807 \nu }{2430}-\frac{1279073}{38880}\right) \chi_s \right] + y^2 y_0^3 \left[ \chi_a \left(\frac{286525 \delta }{54432} \right.\right.\nonumber\\
&\left.\left.- \frac{29045 \delta  \nu }{1944}\right)+\left(\frac{10175 \nu ^2}{972}-\frac{507005 \nu }{27216}+\frac{286525}{54432}\right) \chi_s \right] + y^3 y_0^2 \left[ \chi_a \left(\frac{1464661 \delta }{54432}-\frac{101849 \delta  \nu }{1944}\right)+\left(\frac{28565 \nu ^2}{972} \right.\right.\nonumber\\
&\left.\left.\left.- \frac{1836671 \nu }{27216}+\frac{1464661}{54432}\right) \chi_s \right] \right\} \,,\\
        \mathcal{G}_{\rm SO,circ}^{\rm 3PN} &= 0 \,,\\
        \mathcal{G}_{\rm SO,ecc}^{\rm 3PN} &= -e_0^2 \left(\frac{y_0}{y}\right)^{19/3} \left\{ y^6 \left[ \left(\frac{113137 \pi  \nu }{3888}-\frac{859349 \pi }{15552}\right) \chi_s-\frac{859349 \pi  \delta  \chi_a}{15552} \right] + y_0^6 \left[ \left(\frac{37663 \pi  \nu }{3888}-\frac{157043 \pi }{15552}\right) \chi_s \right.\right.\nonumber\\
        &\left.\left.- \frac{157043 \pi  \delta  \chi_a}{15552} \right] + y^3 y_0^3 \left[ \frac{127049 \pi  \delta  \chi_a}{1944}+\left(\frac{127049 \pi }{1944}-\frac{9425 \pi  \nu }{243}\right) \chi_s \right] \right\} \,,
    \end{align}
    \begin{align}
        \mathcal{G}_{\rm SS,circ}^{\rm 3PN} &= y^6 \left\{\chi_a^2 \left[{\blue \kappa_{-}} \left(\frac{25 \delta  \nu }{3}-\frac{35 \delta }{3}\right)+{\blue \kappa_{+}} \left(-\frac{10 \nu ^2}{3}+\frac{95 \nu }{3}-\frac{35}{3}\right)-\frac{20 \nu ^2}{3}+\frac{40 \nu
   }{9}+\frac{80}{9}\right]+\chi_a \chi_s \left[{\blue \kappa_{+}} \left(\frac{50 \delta  \nu }{3} \right.\right.\right.\nonumber\\
   &\left.\left.- \frac{70 \delta }{3}\right)+\frac{280 \delta  \nu }{9}+\frac{160 \delta }{9}+{\blue \kappa_{-}} \left(-\frac{20 \nu
   ^2}{3}+\frac{190 \nu }{3}-\frac{70}{3}\right)\right]+\chi_s^2 \left[{\blue \kappa_{-}} \left(\frac{25 \delta  \nu }{3}-\frac{35 \delta }{3}\right)+{\blue \kappa_{+}} \left(-\frac{10 \nu ^2}{3}+\frac{95 \nu
   }{3} \right.\right.\nonumber\\
   &\left.\left.\left.- \frac{35}{3}\right)-\frac{100 \nu ^2}{9}-\frac{80 \nu }{9}+\frac{80}{9}\right]\right\} \,,\\
        \mathcal{G}_{\rm SS,ecc}^{\rm 3PN} &= -e_0^2 \left(\frac{y_0}{y}\right)^{19/3} \left( y^6 \left\{ \chi_a^2 \left[{\blue \kappa_{-}} \left(\frac{1613401 \delta }{64512}-\frac{107399 \delta  \nu }{6912}\right)+{\blue \kappa_{+}} \left(\frac{140191 \nu ^2}{3456}-\frac{6343789 \nu
   }{96768}+\frac{1613401}{64512}\right) \right.\right.\right.\nonumber\\
   &\left.+ \frac{329017 \nu ^2}{3456}-\frac{847050901 \nu }{2612736}+\frac{273707297}{5225472}\right]+\chi_a \chi_s \left[{\blue \kappa_{+}} \left(\frac{1613401 \delta
   }{32256}-\frac{107399 \delta  \nu }{3456}\right)-\frac{4587853 \delta  \nu }{93312}+\frac{273707297 \delta }{2612736} \right.\nonumber\\
   &\left.+ {\blue \kappa_{-}} \left(\frac{140191 \nu ^2}{1728}-\frac{6343789 \nu
   }{48384}+\frac{1613401}{32256}\right)\right]+\chi_s^2 \left[{\blue \kappa_{-}} \left(\frac{1613401 \delta }{64512}-\frac{107399 \delta  \nu }{6912}\right)+{\blue \kappa_{+}} \left(\frac{140191 \nu ^2}{3456}-\frac{6343789 \nu
   }{96768} \right.\right.\nonumber\\
   &\left.\left.\left.+ \frac{1613401}{64512}\right)-\frac{7005401 \nu ^2}{93312}+\frac{171176423 \nu }{2612736}+\frac{273707297}{5225472}\right] \right\} + y_0^6 \left\{ \chi_a^2 \left[{\blue \kappa_{-}} \left(\frac{906103 \delta }{193536}-\frac{12877 \delta  \nu }{2304}\right) \right.\right.\nonumber\\
   &\left.+ {\blue \kappa_{+}} \left(\frac{24271 \nu ^2}{3456}-\frac{1446937 \nu
   }{96768}+\frac{906103}{193536}\right)+\frac{169897 \nu ^2}{3456}-\frac{40961143 \nu }{373248}+\frac{17465819}{746496}\right]+\chi_a \chi_s \left[{\blue \kappa_{+}} \left(\frac{906103 \delta }{96768} - \right.\right.\nonumber\\
   &\left.\left.\frac{12877
   \delta  \nu }{1152}\right)-\frac{5526373 \delta  \nu }{93312}+\frac{17465819 \delta }{373248}+{\blue \kappa_{-}} \left(\frac{24271 \nu ^2}{1728}-\frac{1446937 \nu }{48384}+\frac{906103}{96768}\right)\right]+\chi_s^2
   \left[{\blue \kappa_{-}} \left(\frac{906103 \delta }{193536} \right.\right.\nonumber\\
   &\left.\left.\left.- \frac{12877 \delta  \nu }{2304}\right)+{\blue \kappa_{+}} \left(\frac{24271 \nu ^2}{3456}-\frac{1446937 \nu }{96768}+\frac{906103}{193536}\right)+\frac{433639 \nu
   ^2}{93312}-\frac{16075987 \nu }{373248}+\frac{17465819}{746496}\right] \right\} \nonumber\\
   &+ y^2 y_0^4 \left\{ \chi_a^2 \left[{\blue \kappa_{-}} \left(\frac{16465 \delta  \nu }{6912}-\frac{162425 \delta }{193536}\right)+{\blue \kappa_{+}} \left(-\frac{16465 \nu ^2}{3456}+\frac{392935 \nu
   }{96768}-\frac{162425}{193536}\right)-\frac{115255 \nu ^2}{3456}+\frac{270835 \nu }{13824} \right.\right.\nonumber\\
   &\left.- \frac{534725}{193536}\right]+\chi_a \chi_s \left[{\blue \kappa_{+}} \left(\frac{16465 \delta  \nu }{3456}-\frac{162425
   \delta }{96768}\right)+\frac{54205 \delta  \nu }{3456}-\frac{534725 \delta }{96768}+{\blue \kappa_{-}} \left(-\frac{16465 \nu ^2}{1728}+\frac{392935 \nu }{48384} \right.\right.\nonumber\\
   &\left.\left.- \frac{162425}{96768}\right)\right]+\chi_s^2
   \left[{\blue \kappa_{-}} \left(\frac{16465 \delta  \nu }{6912}-\frac{162425 \delta }{193536}\right)+{\blue \kappa_{+}} \left(-\frac{16465 \nu ^2}{3456}+\frac{392935 \nu }{96768}-\frac{162425}{193536}\right)+\frac{6845 \nu
   ^2}{3456} \right.\nonumber\\
   &\left.\left.+ \frac{691345 \nu }{96768}-\frac{534725}{193536}\right] \right\} + y^3 y_0^3 \left[ \chi_a \chi_s \left(\frac{25600 \delta  \nu }{729}-\frac{81169 \delta }{1458}\right)+\left(-\frac{7975 \nu ^2}{729}+\frac{25600 \nu }{729}-\frac{81169}{2916}\right) \chi_s \right.\nonumber\\
   &\left.+ \left(\frac{81169 \nu
   }{729}-\frac{81169}{2916}\right) \chi_a^2 \right] + y^4 y_0^2 \left\{ \chi_a^2 \left[{\blue \kappa_{-}} \left(\frac{74269 \delta  \nu }{6912}-\frac{1068041 \delta }{193536}\right)+{\blue \kappa_{+}} \left(-\frac{74269 \nu ^2}{3456}+\frac{2107807 \nu
   }{96768} \right.\right.\right.\nonumber\\
   &\left.\left.- \frac{1068041}{193536}\right)-\frac{236203 \nu ^2}{3456}+\frac{4204861 \nu }{96768}-\frac{830069}{193536}\right]+\chi_a \chi_s \left[{\blue \kappa_{+}} \left(\frac{74269 \delta  \nu }{3456}-\frac{1068041
   \delta }{96768}\right)+\frac{57721 \delta  \nu }{3456} \right.\nonumber\\
   &\left. -\frac{830069 \delta }{96768} + {\blue \kappa_{-}} \left(-\frac{74269 \nu ^2}{1728}+\frac{2107807 \nu }{48384}-\frac{1068041}{96768}\right)\right]+\chi_s^2
   \left[{\blue \kappa_{-}} \left(\frac{74269 \delta  \nu }{6912}-\frac{1068041 \delta }{193536}\right)+{\blue \kappa_{+}} \left(-\frac{74269 \nu ^2}{3456} \right.\right.\nonumber\\
   &\left.\left.\left.\left.+ \frac{2107807 \nu }{96768}-\frac{1068041}{193536}\right)+\frac{120761 \nu
   ^2}{3456}-\frac{928535 \nu }{96768}-\frac{830069}{193536}\right] \right\} \right)\,.
    \end{align}
\end{subequations}
\end{widetext}

\subsection{\TTthree{}}
{\blue The \TTthree{} phase (ignoring all higher order circular and non-spinning, eccentric corrections) is as follows,}
\begin{widetext}
\begin{equation}
\begin{split}
    F =  &\frac{\theta^3}{8 M \pi} \biggl\{1 
    -\frac{471}{344}e_0^2 \biggl(\frac{\theta_0}{\theta}\biggr)^{19/3} \biggl[1 
    +F_{\rm SO,ecc} 
    + F_{\rm SS,ecc}  \biggr] \biggr\}\,,
\end{split}
\end{equation}
where,
\begin{subequations}
\begin{align}
    F^{\rm 1.5PN}_{\rm SO,ecc} &= \theta _0^3 \biggl[\frac{4871 \delta  \chi
   _A}{4320}+ \frac{4871}{4320} \biggl(1-\frac{3232 \nu }{4871}\biggr)\chi_s\biggr]+\theta ^3
   \biggl[-\frac{1893215 \delta  \chi_a}{2306016}-\frac{1893215}{2306016} \biggl(1-\frac{739666 \nu
   }{1893215}\biggr)\chi_s\biggr] \,, \\
    F^{\rm 2PN}_{\rm SS,ecc} &=\theta _0^4 \biggl\{\biggl[\frac{7}{3072} \biggl(1+\frac{2328 \nu }{7}\biggr)-\frac{97 \delta  \kappa_a}{256}-\frac{97}{256} (1-2 \nu ) \kappa_s\biggr]
   \chi_a^2+\biggl[-\frac{97}{128} (1-2 \nu ) \kappa_a+\delta  \biggl(\frac{7}{1536}-\frac{97 \kappa_s}{128}\biggr)\biggr] \chi_a \chi_s\nonumber \\
   &+\biggl[\frac{7}{3072} \biggl(1-\frac{2356 \nu }{7}\biggr)-\frac{97 \delta  \kappa_a}{256}-\frac{97}{256} (1-2 \nu ) \kappa_s\biggr] \chi_s^2\biggr\}+\theta ^4 \biggl\{\biggl[-\frac{11179}{3737856} \biggl(1+\frac{2102946 \nu }{11179}\biggr)+\frac{350491 \delta  \kappa_a}{1245952}\nonumber \\
   &+\frac{350491}{1245952} (1-2 \nu ) \kappa_s\biggr] \chi_a^2+\biggl[\frac{350491}{622976} (1-2 \nu ) \kappa_a+\delta 
   \biggl(-\frac{11179}{1868928}+\frac{350491 \kappa_s}{622976}\biggr)\biggr] \chi_a \chi_s\nonumber \\
   &+\biggl[-\frac{11179}{3737856} \biggl(1-\frac{2147662 \nu
   }{11179}\biggr)+\frac{350491 \delta  \kappa_a}{1245952}+\frac{350491}{1245952} (1-2 \nu ) \kappa_s\biggr] \chi_s^2\biggr\}\,,\\
    F^{\rm 2.5PN}_{\rm SO,ecc} &= \theta _0^5 \biggl[
   \frac{69146501}{13063680} \biggl(1-\frac{79387834 \nu }{69146501}+\frac{7548240 \nu
   ^2}{69146501}\biggr)\chi_s+ \frac{69146501}{13063680} \biggl(1-\frac{19488462 \nu
   }{69146501}\biggr)\delta  \chi_a\biggr]\nonumber \\
   &+\theta ^3 \theta _0^2 \biggl[
   -\frac{8415340675}{7969591296} \biggl(1-\frac{444 \nu }{889}\biggr)\delta  \chi_a-\frac{8415340675}{7969591296} \biggl(1-\frac{1498150534 \nu }{1683068135}+\frac{328411704 \nu
   ^2}{1683068135}\biggr)\chi_s\biggr]\nonumber \\
   &-\theta ^2 \theta _0^3 \biggl[
   \frac{37248834131}{303547668480} \biggl(1-\frac{17585652 \nu }{7647061}\biggr)\delta  \chi_a+\frac{37248834131}{303547668480} \biggl(1-\frac{110375012044 \nu }{37248834131}+\frac{56836827264 \nu
   ^2}{37248834131}\biggr)\chi_s\biggr]\nonumber \\
   &+\theta ^5 \biggl[
   -\frac{7261740747155}{3612217254912} \biggl(1-\frac{80220876078 \nu }{1452348149431}\biggr)\delta  \chi_a-\frac{7261740747155}{3612217254912} \biggl(1-\frac{17211071085917 \nu
   }{14523481494310} \nonumber\\
   &-\frac{521561712378 \nu ^2}{36308703735775}\biggr)\chi_s\biggr]\,,\\
    F^{\rm 3PN}_{\rm SO,ecc} &= \theta _0^6 \biggl[-\frac{22358051 \pi 
   \delta  \chi_a}{24883200}-\frac{22358051 \pi }{24883200} \biggl(1-\frac{16608892 \nu
   }{22358051}\biggr)\chi_s\biggr]+\theta ^3 \theta _0^3 \biggl[-\frac{1482249323 \pi  \delta  \chi_a}{13282652160}\nonumber \\
   &-\frac{1482249323 \pi }{13282652160} \biggl(1-\frac{920485108 \nu
   }{1482249323}\biggr)\chi_s\biggr]+\theta ^6 \biggl[\frac{55371553 \pi  \delta  \chi_a}{84602880}+
   \frac{55371553 \pi }{84602880} \biggl(1-\frac{29754273364 \nu
   }{43466669105}\biggr)\chi_s\biggr] \,,\\
    F^{\rm 3PN}_{\rm SS,ecc} &= \theta ^2 \theta _0^4 \biggl\{\chi_a
   \chi_s \biggl[-\frac{7647061 \delta
   }{15418294272}\biggl(1-\frac{17585652 \nu }{7647061}\biggr) + \frac{741764917}{8994004992} \biggl(1-\frac{32879774 \nu
   }{7647061}+\frac{35171304 \nu ^2}{7647061}\biggr)\kappa_a\nonumber \\
   &+\frac{741764917 \delta }{8994004992}
   \biggl(1-\frac{17585652 \nu }{7647061}\biggr)\kappa_s \biggr]+\chi_s^2 \biggl[
   \frac{741764917}{17988009984} \biggl(1-\frac{32879774 \nu }{7647061}+\frac{35171304 \nu
   ^2}{7647061}\biggr)\kappa_s\nonumber \\
   &-\frac{7647061}{30836588544} \biggl(1-\frac{18139575280 \nu
   }{53529427}+\frac{5918828016 \nu ^2}{7647061}\biggr)+ \frac{741764917 \delta }{17988009984}
   \biggl(1-\frac{17585652 \nu }{7647061}\biggr)\kappa_a\biggr]\nonumber \\
   &+\chi_a^2 \biggl[\frac{741764917}{17988009984} \biggl(1-\frac{32879774 \nu }{7647061}+\frac{35171304 \nu
   ^2}{7647061}\biggr)\kappa_s+ \frac{741764917 \delta }{17988009984} \biggl(1-\frac{17585652 \nu
   }{7647061}\biggr)\kappa_a\nonumber \\
   &-\frac{7647061}{30836588544} \biggl(1+\frac{17679258444 \nu }{53529427}-\frac{5848485408
   \nu ^2}{7647061}\biggr)\biggr]\biggr\}+\theta ^4 \theta _0^2 \biggl\{\chi_a \chi_s \biggl[
   \frac{1557932495}{2153005056} \biggl(1-\frac{2222 \nu }{889}\nonumber \\
   &+\frac{888 \nu ^2}{889}\biggr) \kappa_a+
   \frac{1557932495 \delta }{2153005056} \biggl(1-\frac{444 \nu }{889}\biggr)\kappa_s-\frac{49690655 \delta
   }{6459015168} \biggl(1-\frac{444 \nu }{889}\biggr)\biggr]+\chi_s^2 \biggl[ \frac{1557932495}{4306010112} \biggl(1-\frac{2222 \nu }{889}\nonumber \\
   &+\frac{888 \nu ^2}{889}\biggr) \kappa_s+
   \frac{1557932495 \delta }{4306010112} \biggl(1-\frac{444 \nu
   }{889}\biggr)\kappa_a-\frac{49690655}{12918030336} \biggl(1-\frac{273462142 \nu }{1419733}+\frac{953561928 \nu
   ^2}{9938131}\biggr)\biggr]\nonumber \\
   &+\chi_a^2 \biggl[\frac{1557932495}{4306010112} \biggl(1-\frac{2222 \nu
   }{889}+\frac{888 \nu ^2}{889}\biggr)\kappa_s+ \frac{1557932495 \delta }{4306010112}
   \biggl(1-\frac{444 \nu }{889}\biggr)\kappa_a-\frac{49690655}{12918030336} \biggl(1\nonumber \\
   &+\frac{266365074 \nu
   }{1419733}-\frac{933708024 \nu ^2}{9938131}\biggr)\biggr]\biggr\}+\theta ^3 \theta _0^3 \biggl\{-\frac{1844370053}{1992397824}(1-4 \nu ) \chi_a^2 -\frac{1844370053}{1992397824}
   \biggl(1-\frac{9721783966 \nu }{9221850265}\nonumber \\
   &+\frac{2390600512 \nu ^2}{9221850265}\biggr) \chi_s^2-\frac{1844370053 \delta }{996198912} \biggl(1-\frac{4860891983 \nu
   }{9221850265}\biggr)\chi_a \chi_s\biggr\}+\theta _0^6 \biggl\{\chi_s^2 \biggl[-\frac{15796363}{6193152}
   \biggl(1-\frac{38091274 \nu }{15796363}\nonumber \\
   &+\frac{3292968 \nu ^2}{15796363}\biggr)\kappa_s - \frac{15796363 \delta }{6193152} \biggl(1-\frac{6498548 \nu
   }{15796363}\biggr)\kappa_a+\frac{29393895511}{16721510400} \biggl(1-\frac{84309055624 \nu
   }{29393895511}+\frac{4422254704 \nu ^2}{29393895511}\biggr)\biggr]\nonumber \\
   &+\chi_a \chi_s \biggl[
   -\frac{15796363}{3096576} \biggl(1-\frac{38091274 \nu }{15796363}+\frac{3292968 \nu
   ^2}{15796363}\biggr)\kappa_a-\frac{15796363 \delta }{3096576} \biggl(1-\frac{6498548 \nu
   }{15796363}\biggr)\kappa_s\nonumber \\
   &+\frac{29393895511 \delta }{8360755200} \biggl(1-\frac{3294795812 \nu
   }{29393895511}\biggr)\biggr]+\chi_a^2 \biggl[-\frac{15796363}{6193152} \biggl(1-\frac{38091274
   \nu }{15796363}+\frac{3292968 \nu ^2}{15796363}\biggr)\kappa_s\nonumber \\
   &-\frac{15796363 \delta }{6193152}
   \biggl(1-\frac{6498548 \nu }{15796363}\biggr)\kappa_a+\frac{29393895511}{16721510400} \biggl(1-\frac{39856118044
   \nu }{29393895511}-\frac{8891013600 \nu ^2}{29393895511}\biggr)\biggr]\biggr\}\nonumber \\
   &+\theta ^6 \biggl\{\chi_a^2
   \biggl[\frac{5865969154439}{4646902579200} \biggl(1-\frac{41973254021326 \nu
   }{17597907463317}-\frac{974298467240 \nu ^2}{5865969154439}\biggr)\kappa_s+ \frac{5865969154439
   \delta }{4646902579200} \biggl(1\nonumber \\
   &-\frac{6777439094692 \nu
   }{17597907463317}\biggr)\kappa_a-\frac{1206460449012419}{2559513940623360} \biggl(1+\frac{2979651431961776 \nu
   }{6032302245062095}+\frac{536643595755792 \nu ^2}{1206460449012419}\biggr)\biggr] \nonumber \\
   &+\chi_s^2 \biggl[
   \frac{5865969154439}{4646902579200} \biggl(1-\frac{41973254021326 \nu }{17597907463317}-\frac{974298467240
   \nu ^2}{5865969154439}\biggr)\kappa_s+ \frac{5865969154439 \delta }{4646902579200}
   \biggl(1\nonumber \\
   &-\frac{6777439094692 \nu }{17597907463317}\biggr)\kappa_a-\frac{1206460449012419}{2559513940623360}
   \biggl(1-\frac{3624177344679476 \nu }{1206460449012419}-\frac{1846637721438496 \nu
   ^2}{1206460449012419}\biggr)\biggr] \nonumber \\
   &+\chi_a \chi_s \biggl[\frac{5865969154439 \delta }{2323451289600} \biggl(1-\frac{6777439094692 \nu }{17597907463317}\biggr)\kappa_s+
   \frac{5865969154439}{2323451289600} \biggl(1-\frac{41973254021326 \nu }{17597907463317} \nonumber \\
   &-\frac{974298467240
   \nu ^2}{5865969154439}\biggr)\kappa_a-\frac{1206460449012419 \delta }{1279756970311680}
   \biggl(1+\frac{4493986844406388 \nu }{6032302245062095}\biggr)\biggr]\biggr\}\,.
\end{align}
\end{subequations}
\end{widetext}

This solution is now used on Eq.~\eqref{TaylorT2 phase} to provide the time domain phasing $\phi$ as a function of time, $\langle \phi \rangle (t) = \langle \phi \rangle \left[y=y(t)\right]$. The secular section of the orbital phasing $\langle \phi \rangle$ as a function of $\theta$ {\blue (ignoring all higher order circular and non-spinning, eccentric corrections)} is then given by
\begin{widetext}
\begin{equation}
\begin{split}
    \langle \phi \rangle - \phi_c=  &-\frac{1}{\nu \theta^5} \biggl\{
    1-\frac{7065}{11696}e_0^2 \biggl(\frac{\theta_0}{\theta}\biggr)^{19/3} \biggl[
    1+\phi_{\rm SO,ecc}+ \phi_{\rm SS,ecc} \biggr] \biggr\} \,,
\end{split}
\end{equation}
where,
\begin{subequations}

\begin{align}
    \phi^{\rm 1.5PN}_{\rm SO,ecc} &= \theta _0^3 \biggl[\frac{4871 \delta  \chi_a}{4320}+
   \frac{4871}{4320} \biggl(1-\frac{3232 \nu }{4871}\biggr)\chi_s\biggr]+\theta ^3 \biggl[-\frac{378643
   \delta  \chi_a}{339120}-\frac{378643}{339120} \biggl(1-\frac{739666 \nu
   }{1893215}\biggr)\chi_s\biggr] \,, \\
    \phi^{\rm 2PN}_{\rm SS,ecc} &= \theta _0^4 \biggl\{\biggl[\frac{7}{3072} \biggl(1+\frac{2328
   \nu }{7}\biggr)-\frac{97 \delta  \kappa_a}{256}-\frac{97}{256} (1-2 \nu )\kappa_s\biggr] \chi_a^2+\biggl[-\frac{97}{128} (1-2 \nu ) \kappa_a+\delta  \biggl(\frac{7}{1536}-\frac{97 \kappa
   _S}{128}\biggr)\biggr] \chi_a \chi_s\nonumber \\
   &+\biggl[\frac{7}{3072} \biggl(1-\frac{2356 \nu }{7}\biggr)-\frac{97 \delta 
   \kappa_a}{256}-\frac{97}{256} (1-2 \nu ) \kappa_s\biggr] \chi_s^2\biggr\}+\theta ^4
   \biggl\{\biggl[-\frac{190043}{41116416} \biggl(1+\frac{2102946 \nu }{11179}\biggr)+\frac{5958347 \delta  \kappa_a}{13705472}\nonumber \\
   &+\frac{5958347}{13705472} (1-2 \nu ) \kappa_s\biggr] \chi_a^2+\biggl[\frac{5958347}{6852736} (1-2 \nu ) \kappa_a+\delta 
   \biggl(-\frac{190043}{20558208}+\frac{5958347 \kappa_s}{6852736}\biggr)\biggr] \chi_a \chi_s\nonumber \\
   &+\biggl[-\frac{190043}{41116416} \biggl(1-\frac{2147662 \nu }{11179}\biggr)+\frac{5958347 \delta  \kappa_a}{13705472}+\frac{5958347}{13705472} (1-2 \nu ) \kappa_s\biggr] \chi_s^2\biggr\}\,,\\
    \phi^{\rm 2.5PN}_{\rm SO,ecc} &= \theta _0^5 \biggl[ \frac{69146501}{13063680} \biggl(1-\frac{79387834 \nu
   }{69146501}+\frac{7548240 \nu ^2}{69146501}\biggr)\chi_s+ \frac{69146501}{13063680}
   \biggl(1-\frac{19488462 \nu }{69146501}\biggr)\delta  \chi_a\biggr]\nonumber \\
   &+\theta ^3 \theta _0^2 \biggl[
  -\frac{336613627}{234399744} \biggl(1-\frac{444 \nu }{889}\biggr)\delta  \chi_a-\frac{336613627}{234399744} \biggl(1-\frac{1498150534 \nu }{1683068135}+\frac{328411704 \nu
   ^2}{1683068135}\biggr)\chi_s\biggr]\nonumber \\
   &+\theta ^5 \biggl[ \frac{5033550410711}{1009295997696}
   \biggl(1+\frac{14362504401 \nu }{117059311877}\biggr)\delta  \chi_a+ \frac{5033550410711}{1009295997696}
   \biggl(1-\frac{346680975625 \nu }{468237247508}\nonumber \\
   &-\frac{111702828657 \nu
   ^2}{585296559385}\biggr)\chi_s\biggr]+\theta ^2 \theta _0^3 \biggl[
   -\frac{633230180227}{4249667358720} \biggl(1-\frac{110375012044 \nu }{37248834131}+\frac{56836827264 \nu
   ^2}{37248834131}\biggr)\chi_s\nonumber \\
   &-\frac{633230180227}{4249667358720} \biggl(1-\frac{17585652
   \nu }{7647061}\biggr)\delta  \chi_a\biggr]\,,\\
    \phi^{\rm 3PN}_{\rm SO,ecc} &= \theta _0^6 \biggl[-\frac{22358051 \pi 
   \delta  \chi_a}{24883200}-\frac{22358051 \pi }{24883200} \biggl(1-\frac{16608892 \nu
   }{22358051}\biggr)\chi_s\biggr]+\theta ^6 \biggl[\frac{55371553 \pi  \delta  \chi_a}{39813120}\nonumber \\
   &+
   \frac{55371553 \pi }{39813120} \biggl(1-\frac{29754273364 \nu }{43466669105}\biggr)\chi_s\biggr]-\theta
   ^3 \theta _0^3 \biggl[\frac{1482249323 \pi  \delta  \chi_a}{9766656000}+\frac{1482249323 \pi
   }{9766656000} \biggl(1-\frac{920485108 \nu }{1482249323}\biggr)\chi_s\biggr]\,,\\
    \phi^{\rm 3PN}_{\rm SS,ecc} &= \theta ^2 \theta _0^4 \biggl\{\chi_a \chi_s
   \biggl[\frac{12610003589}{125916069888} \biggl(1-\frac{32879774 \nu }{7647061}+\frac{35171304 \nu ^2}{7647061}\biggr)\kappa_a+
   \frac{12610003589 \delta }{125916069888} \biggl(1-\frac{17585652 \nu }{7647061}\biggr)\kappa_s\nonumber \\
   &-\frac{130000037 \delta }{215856119808} \biggl(1-\frac{17585652
   \nu }{7647061}\biggr)\biggr]+\chi_s^2 \biggl[\frac{12610003589}{251832139776} \biggl(1-\frac{32879774 \nu }{7647061}+\frac{35171304 \nu
   ^2}{7647061}\biggr)\kappa_s\nonumber \\
   &-\frac{130000037}{431712239616} \biggl(1-\frac{18139575280 \nu }{53529427}+\frac{5918828016 \nu ^2}{7647061}\biggr)+
   \frac{12610003589 \delta }{251832139776} \biggl(1-\frac{17585652 \nu }{7647061}\biggr)\kappa_a\biggr]\nonumber \\
   &+\chi_a^2 \biggl[
   \frac{12610003589}{251832139776} \biggl(1-\frac{32879774 \nu }{7647061}+\frac{35171304 \nu ^2}{7647061}\biggr)\kappa_s+ \frac{12610003589
   \delta }{251832139776} \biggl(1-\frac{17585652 \nu }{7647061}\biggr)\kappa_a\nonumber \\
   &-\frac{130000037}{431712239616} \biggl(1+\frac{17679258444 \nu
   }{53529427}-\frac{5848485408 \nu ^2}{7647061}\biggr)\biggr]\biggr\}+\theta ^4 \theta _0^2 \biggl\{\chi_a \chi_s \biggl[
   \frac{26484852415}{23683055616} \biggl(1-\frac{2222 \nu }{889}\nonumber \\
   &+\frac{888 \nu ^2}{889}\biggr) \kappa_a+ \frac{26484852415 \delta }{23683055616}
   \biggl(1-\frac{444 \nu }{889}\biggr)\kappa_s-\frac{844741135 \delta }{71049166848} \biggl(1-\frac{444 \nu }{889}\biggr)\biggr]+\chi_s^2 \biggl[
   \frac{26484852415}{47366111232} \biggl(1-\frac{2222 \nu }{889}\nonumber \\
   &+\frac{888 \nu ^2}{889}\biggr)\kappa_s+ \frac{26484852415 \delta }{47366111232}
   \biggl(1-\frac{444 \nu }{889}\biggr)\kappa_a-\frac{844741135}{142098333696} \biggl(1-\frac{273462142 \nu }{1419733}+\frac{953561928 \nu
   ^2}{9938131}\biggr)\biggr]\nonumber \\
   &+\chi_a^2 \biggl[\frac{26484852415}{47366111232} \biggl(1-\frac{2222 \nu }{889}+\frac{888 \nu
   ^2}{889}\biggr)\kappa_s+ \frac{26484852415 \delta }{47366111232} \biggl(1-\frac{444 \nu }{889}\biggr)\kappa_a-\frac{844741135}{142098333696}
   \biggl(1\nonumber \\
   &+\frac{266365074 \nu }{1419733}-\frac{933708024 \nu ^2}{9938131}\biggr)\biggr]\biggr\}+\theta ^3 \theta _0^3 \biggl[-\frac{1844370053}{1464998400}(1-4 \nu ) \chi_a^2
   -\frac{1844370053}{1464998400} \biggl(1-\frac{9721783966 \nu }{9221850265}\nonumber \\
   &+\frac{2390600512 \nu
   ^2}{9221850265}\biggr)\chi_s^2-\frac{1844370053 \delta }{732499200} \biggl(1-\frac{4860891983 \nu }{9221850265}\biggr)\chi_a \chi_s\biggr]+\theta
   _0^6 \biggl\{\chi_s^2 \biggl[-\frac{15796363}{6193152} \biggl(1-\frac{38091274 \nu }{15796363}\nonumber \\
   &+\frac{3292968 \nu ^2}{15796363}\biggr)\kappa_s-\frac{15796363 \delta }{6193152} \biggl(1-\frac{6498548 \nu }{15796363}\biggr)\kappa
   _A+\frac{29393895511}{16721510400} \biggl(1-\frac{84309055624 \nu
   }{29393895511}+\frac{4422254704 \nu ^2}{29393895511}\biggr)\biggr]\nonumber \\
   &+\chi_a \chi_s \biggl[-\frac{15796363}{3096576} \biggl(1-\frac{38091274 \nu
   }{15796363}+\frac{3292968 \nu ^2}{15796363}\biggr)\kappa_a-\frac{15796363 \delta }{3096576} \biggl(1-\frac{6498548 \nu
   }{15796363}\biggr)\kappa_s\nonumber \\
   &+\frac{29393895511 \delta }{8360755200} \biggl(1-\frac{3294795812 \nu }{29393895511}\biggr)\biggr]+\chi_a^2 \biggl[
   -\frac{15796363}{6193152} \biggl(1-\frac{38091274 \nu }{15796363}+\frac{3292968 \nu ^2}{15796363}\biggr)\kappa_s\nonumber \\
   &-\frac{15796363 \delta
   }{6193152} \biggl(1-\frac{6498548 \nu }{15796363}\biggr)\kappa_a+\frac{29393895511}{16721510400} \biggl(1-\frac{39856118044 \nu }{29393895511}-\frac{8891013600 \nu
   ^2}{29393895511}\biggr)\biggr]\biggr\}\nonumber \\
   &+\theta ^6 \biggl\{\chi_a^2 \biggl[ \frac{99721475625463}{37175220633600} \biggl(1-\frac{41973254021326 \nu
   }{17597907463317}-\frac{974298467240 \nu ^2}{5865969154439}\biggr)\kappa_s+ \frac{99721475625463 \delta }{37175220633600}
   \biggl(1 \nonumber \\
   &-\frac{6777439094692 \nu }{17597907463317}\biggr)\kappa_a-\frac{1206460449012419}{1204477148528640} \biggl(1+\frac{2979651431961776 \nu
   }{6032302245062095}+\frac{536643595755792 \nu ^2}{1206460449012419}\biggr)\biggr]\nonumber \\
   &+\chi_s^2 \biggl[\frac{99721475625463}{37175220633600}
   \biggl(1-\frac{41973254021326 \nu }{17597907463317}-\frac{974298467240 \nu ^2}{5865969154439}\biggr)\kappa_s+ \frac{99721475625463 \delta
   }{37175220633600} \biggl(1\nonumber \\
   &-\frac{6777439094692 \nu }{17597907463317}\biggr)\kappa_a-\frac{1206460449012419}{1204477148528640} \biggl(1-\frac{3624177344679476 \nu
   }{1206460449012419}-\frac{1846637721438496 \nu ^2}{1206460449012419}\biggr)\biggr]\nonumber \\
   &+\chi_a \chi_s \biggl[ \frac{99721475625463}{18587610316800}
   \biggl(1-\frac{41973254021326 \nu }{17597907463317}-\frac{974298467240 \nu ^2}{5865969154439}\biggr)\kappa_a\nonumber \\
   &+ \frac{99721475625463 \delta
   }{18587610316800} \biggl(1-\frac{6777439094692 \nu }{17597907463317}\biggr)\kappa_s-\frac{1206460449012419 \delta }{602238574264320}
   \biggl(1+\frac{4493986844406388 \nu }{6032302245062095}\biggr)\biggr]\biggr\}\,.
\end{align}
\end{subequations}
\end{widetext}
\subsection{\TTfour{}}
{\blue The \TTfour{} phase (ignoring all higher order circular and non-spinning, eccentric corrections) is as follows,}
\begin{widetext}
\begin{equation}
\begin{split}
    \frac{dy}{dt} =  & \frac{32 y^9 \nu}{5 M} \biggl\{ 
    1-\frac{5}{8}e_0^2 \biggl(\frac{y_0}{y}\biggr)^{19/3} \biggl[  
    1+\rho_{\rm SO,ecc} + \rho_{\rm SS,ecc} \biggr]  
    +\cdots+ \mathcal{O}(e_0^8)\biggr\}\,,
\end{split}
\end{equation}
where,
\begin{subequations}

\begin{align}
    \rho^{\rm 1.5PN}_{\rm SO,ecc} &=y^3 \biggl[\frac{4691 \delta  \chi
   _A}{270}+ \frac{4691}{270} \biggl(1+\frac{782 \nu }{4691}\biggr)\chi_s\biggr]+y_0^3
   \biggl[-\frac{157 \delta  \chi_a}{54}-\frac{157}{54} \biggl(1-\frac{110 \nu
   }{157}\biggr)\chi_s\biggr] \,,\\
    \rho^{\rm 2PN}_{\rm SS,ecc} &=y_0^4 \biggl\{\chi_a \chi_s \biggl[\delta  \biggl(\frac{13}{48}+\frac{89 \kappa_s}{24}\biggr)+ \frac{89}{24} (1-2 \nu )\kappa_a\biggr]+\chi_a^2
   \biggl[\frac{13}{96}+\frac{89 \delta  \kappa_a}{48}-\frac{89 \nu }{24}+ \frac{89}{48} (1-2 \nu )\kappa_s\biggr]\nonumber \\
   &+\chi_s^2
   \biggl[\frac{13}{96}+\frac{89 \delta  \kappa_a}{48}+\frac{19 \nu }{6}+ \frac{89}{48} (1-2 \nu )\kappa_s\biggr]\biggr\}+y^4 \biggl\{\chi_a \chi_s
   \biggl[\delta  \biggl(-\frac{137}{240}-\frac{2749 \kappa_s}{120}\biggr)-\frac{2749}{120} (1-2 \nu )\kappa_a\biggr]\nonumber \\
   &-\chi_s^2
   \biggl[\frac{137}{480}+\frac{2749 \delta  \kappa_a}{240}+\frac{653 \nu }{30}+\frac{2749}{240} (1-2 \nu )\kappa_s\biggr]+\chi_a^2
   \biggl[-\frac{137}{480}-\frac{2749 \delta  \kappa_a}{240}+\frac{2749 \nu }{120}-\frac{2749}{240} (1-2 \nu )\kappa_s\biggr]\biggr\}\,,\\
    \rho^{\rm 2.5PN}_{\rm SO,ecc} &=-y_0^5 \biggl[
   \frac{1279073}{38880} \biggl(1-\frac{2348912 \nu }{1279073}+\frac{691520 \nu
   ^2}{1279073}\biggr)\chi_s+ \frac{1279073}{38880} \biggl(1-\frac{970876 \nu
   }{1279073}\biggr)\delta  \chi_a\biggr]\nonumber \\
   &+y^2 y_0^3 \biggl[ \frac{5917801}{54432}
   \biggl(1-\frac{28308 \nu }{37693}\biggr)\delta  \chi_a+ \frac{5917801}{54432} \biggl(1-\frac{8590586 \nu
   }{5917801}+\frac{3113880 \nu ^2}{5917801}\biggr)\chi_s\biggr]\nonumber \\
   &+y^3 y_0^2 \biggl[\frac{13289603}{272160}  \biggl(1-\frac{5516
   \nu }{2833}\biggr)\delta \chi_a + \frac{13289603}{272160} \biggl(1-\frac{23660150
   \nu }{13289603}-\frac{4313512 \nu ^2}{13289603}\biggr)\chi_s\biggr]\nonumber \\
   &+y^5 \biggl[\frac{105998407}{272160}  \biggl(1-\frac{84843948
   \nu }{105998407}\biggr)\delta \chi_a + \frac{105998407}{272160}
   \biggl(1-\frac{193354904 \nu }{105998407}+\frac{53724048 \nu ^2}{105998407}\biggr)\chi_s\biggr]\,,\\
    \rho^{\rm 3PN}_{\rm SO,ecc} &=y_0^6 \biggl[-\frac{157043 \pi 
   \delta  \chi_a}{15552}-\frac{157043 \pi }{15552} \biggl(1-\frac{150652 \nu
   }{157043}\biggr)\chi_s\biggr]+y^3 y_0^3 \biggl[\frac{654587 \pi  \delta  \chi_a}{4860}\nonumber \\
   &+
   \frac{654587 \pi }{4860} \biggl(1-\frac{75154 \nu }{654587}\biggr)\chi_s\biggr]+y^6 \biggl[\frac{13328783
   \pi  \delta  \chi_a}{77760}+ \frac{13328783 \pi }{77760} \biggl(1-\frac{367804 \nu
   }{1025291}\biggr)\chi_s\biggr]\,, \\
    \rho^{\rm 3PN}_{\rm SS,ecc} &=-y^4 y_0^2 \biggl\{\chi_a \chi_s \biggl[\frac{7787917}{120960} \biggl(1-\frac{11182 \nu }{2833}+\frac{11032 \nu ^2}{2833}\biggr)\kappa_a+
   \frac{7787917 \delta }{120960} \biggl(1-\frac{5516 \nu }{2833}\biggr)\kappa_s+\frac{388121 \delta }{241920} \biggl(1-\frac{5516 \nu
   }{2833}\biggr)\biggr]\nonumber \\
   &+\chi_a^2 \biggl[\frac{7787917}{241920} \biggl(1-\frac{11182 \nu }{2833}+\frac{11032 \nu ^2}{2833}\biggr)\kappa_s+
   \frac{7787917 \delta }{241920} \biggl(1-\frac{5516 \nu }{2833}\biggr)\kappa_a+\frac{388121}{483840} \biggl(1-\frac{31907360 \nu }{388121}\, \nonumber \\
   &+\frac{60653936 \nu
   ^2}{388121}\biggr)\biggr]+\chi_s^2 \biggl[\frac{7787917}{241920} \biggl(1-\frac{11182 \nu }{2833}+\frac{11032 \nu ^2}{2833}\biggr)\kappa_s+ \frac{7787917 \delta }{241920} \biggl(1-\frac{5516 \nu }{2833}\biggr)\kappa
   _A+\frac{388121}{483840} \biggl(1 \, \nonumber \\
   &+\frac{28843492 \nu }{388121}-\frac{57631168
   \nu ^2}{388121}\biggr)\biggr]\biggr\}-y^2 y_0^4 \biggl\{\chi_a \chi_s \biggl[\frac{3354677}{24192} \biggl(1-\frac{103694 \nu }{37693}+\frac{56616
   \nu ^2}{37693}\biggr)\kappa_a+ \frac{3354677 \delta }{24192} \biggl(1\, \nonumber \\
   &-\frac{28308 \nu }{37693}\biggr)\kappa_s+\frac{490009 \delta }{48384}
   \biggl(1-\frac{28308 \nu }{37693}\biggr)\biggr]+\chi_a^2 \biggl[\frac{3354677}{48384} \biggl(1-\frac{103694 \nu }{37693}+\frac{56616 \nu
   ^2}{37693}\biggr)\kappa_s+ \frac{3354677 \delta }{48384} \biggl(1\nonumber \\
   &-\frac{28308 \nu }{37693}\biggr)\kappa_a+\frac{490009}{96768}
   \biggl(1-\frac{13786712 \nu }{490009}+\frac{10077648 \nu ^2}{490009}\biggr)\biggr]+\chi_s^2 \biggl[\frac{3354677}{48384} \biggl(1-\frac{103694 \nu
   }{37693}+\frac{56616 \nu ^2}{37693}\biggr)\kappa_s \nonumber \\
   &+ \frac{3354677 \delta }{48384} \biggl(1-\frac{28308 \nu
   }{37693}\biggr)\kappa_a+\frac{490009}{96768} \biggl(1+\frac{11090668 \nu }{490009}-\frac{8605632 \nu ^2}{490009}\biggr)\biggr]\biggr\}+y_0^6 \biggl\{\chi_a^2
   \biggl[\frac{906103}{48384} \biggl(1-\frac{2893874 \nu }{906103}\nonumber \\
   &+\frac{1359176 \nu ^2}{906103}\biggr)\kappa_s+ \frac{906103 \delta
   }{48384} \biggl(1-\frac{1081668 \nu }{906103}\biggr)\kappa_a+\frac{24433195}{2612736} \biggl(1-\frac{169526104 \nu }{24433195}+\frac{73395504 \nu
   ^2}{24433195}\biggr)\biggr]\nonumber \\
   &+\chi_a \chi_s \biggl[\frac{906103}{24192} \biggl(1-\frac{2893874 \nu }{906103}+\frac{1359176 \nu
   ^2}{906103}\biggr)\kappa_a+ \frac{906103 \delta }{24192} \biggl(1-\frac{1081668 \nu }{906103}\biggr)\kappa_s+\frac{24433195 \delta }{1306368}
   \biggl(1\nonumber \\
   &-\frac{33561668 \nu }{24433195}\biggr)\biggr]+\chi_s^2 \biggl[\frac{906103}{48384} \biggl(1-\frac{2893874 \nu }{906103}+\frac{1359176 \nu
   ^2}{906103}\biggr)\kappa_s+ \frac{906103 \delta }{48384} \biggl(1-\frac{1081668 \nu }{906103}\biggr)\kappa_a\nonumber \\
   &+\frac{24433195}{2612736}
   \biggl(1+\frac{4669988 \nu }{24433195}-\frac{42904736 \nu ^2}{24433195}\biggr)\biggr]\biggr\}+y^6 \biggl\{\chi_a^2 \biggl[-\frac{7161173 \delta
   }{26880} \biggl(1-\frac{2082500 \nu }{2222433}\biggr)\kappa_a-\frac{7161173}{26880} \biggl(1 \nonumber \\
   &-\frac{6527366 \nu }{2222433}+\frac{9540328 \nu
   ^2}{7161173}\biggr)\kappa_s-\frac{599862209}{2612736} \biggl(1-\frac{17949354544 \nu }{2999311045}+\frac{4636599408 \nu ^2}{2999311045}\biggr)\biggr]+\chi_a \chi
   _S \biggl[-\frac{7161173 \delta }{13440} \nonumber \\
   &\times \biggl(1-\frac{2082500 \nu }{2222433}\biggr)\kappa_s-\frac{7161173}{13440}
   \biggl(1-\frac{6527366 \nu }{2222433}+\frac{9540328 \nu ^2}{7161173}\biggr)\kappa_a-\frac{599862209 \delta }{1306368} \biggl(1-\frac{1572186644 \nu
   }{2999311045}\biggr)\biggr]\nonumber \\
   &+\chi_s^2 \biggl[-\frac{7161173 \delta }{26880} \biggl(1-\frac{2082500 \nu }{2222433}\biggr)\kappa_a-\frac{7161173}{26880} \biggl(1-\frac{6527366 \nu }{2222433}+\frac{9540328 \nu ^2}{7161173}\biggr)\kappa_s\nonumber \\
   &-\frac{599862209}{2612736}
   \biggl(1+\frac{2807737076 \nu }{2999311045}-\frac{4131331232 \nu ^2}{2999311045}\biggr)\biggr]\biggr\}-y^3 y_0^3 \biggl[\frac{736487 \delta
   }{7290} \biggl(1-\frac{196618 \nu }{736487}\biggr)\chi_a \chi_s\nonumber \\
   &+\frac{736487}{14580} \biggl(1-\frac{393236 \nu }{736487}-\frac{86020 \nu
   ^2}{736487}\biggr)\chi_s^2+\frac{736487}{14580} (1-4 \nu )\chi_a^2\biggr]\,.
\end{align}
\end{subequations}
\end{widetext}

\section{Number of GW cycles estimates for moderate eccentricities}
\begin{widetext}
\begin{table}[H]
	\centering
    \resizebox{2 \columnwidth}{!}{
	\begin{tabular}{|l||c|c||c|c|c||c|c||c|c|}
		\hline
		Detector & \multicolumn{2}{c||}{LIGO/Virgo} & \multicolumn{3}{c||}{3G}& \multicolumn{2}{c||}{DECIGO} & \multicolumn{2}{c|}{LISA}  \\
		\hline
		Masses ($M_\odot$) & $1.4 + 1.4$ & $ 10 + 10$ &  $1.4 + 1.4$ & $50 + 50$ & $500 + 500$ & $500 + 500$ & $5000 + 5000$ & $10^5 + 10^5$ & $10^7 + 10^7$  \\
		\hline
		PN order &\multicolumn{9}{|c|}{cumulative number of cycles} \\
		\hline
		\hline
		1.5PN (circ) & $51.9072$ & $28.8143$ & $250.7891$ & $48.2115$ & $7.0857$ & $232.0146$ & $48.2115$ & $149.0668$ & $2.9067$ \\
		\hline
		1.5PN (ecc) & $-11.7671$ & $-7.1341$ & $-54.6185$ & $-11.3306$ & $-2.3326$ & $-52.5965$ & $-11.3306$ & $-33.1340$ & $-1.1985$\\
		\hline
		2PN (circ)& $-1.6502$ & $-3.5553$ & $-4.1416$ & $-4.9924$ &  $-1.2597$ & $-11.7745$ & $-4.9924$ & $-9.9381$ & $-0.5927$ \\
		\hline
		2PN (ecc) & $0.2398$ & $0.6293$ & $0.5167$ & 
$0.7940$ & $0.3402$ & $1.7112$ & $0.7940$ & $1.3582$ & $0.2057$ \\
		\hline
		2.5PN (circ) & $6.7599$ & $10.2069$ & $10.8024$ & $12.4962$ & $4.8913$ & $15.2098$ & $12.4962$ & $17.8117$ & $2.6020$ \\
		\hline	
		2.5PN (ecc)& $-0.8572$ & $-1.9210$ & $-0.8574$ & $-1.9272$ & $-1.7074$ & $-1.9287$ & $-1.9272$ & $-1.9290$ & $-1.2040$  \\
		\hline 
		3PN (circ)& $-2.2268$ & $-4.9561$ & $-3.0152$ & $-5.5237$ & $-3.0028$ & $-3.6909$ & $-5.5237$ & $-6.4303$ & $-1.7800$ \\
		\hline	
		3PN (ecc)& $0.0542$ & $0.3004$ & $0.0252$ & $0.2432$ & $0.3434$ & $0.1136$ & $0.2432$ & $0.1433$ & $0.1882$ \\
		\hline
            \textbf{Total} & $42.4599$ & $22.3844$ & $199.5010$ & $37.9709$ & $4.3580$ & $179.0590$ & $37.9709$ & $116.9490$ & $1.1274$ \\
		\hline
        
	\end{tabular} }
      
\end{table}
\vspace{-1em}
\begin{center}
        \captionsetup{format=plain}
        \captionof{table}{\justifying{\blue 
        Same as Table~\ref{table_cycles_freq_bands} except that $e_0$ is fixed to 0.5. Estimates are similar (agree within $\mathcal{O}(1)$ cycle) for the same value of $e_0$ with the resummed \TTtwo{} phase presented in Sec.~\ref{sec:comparison} is used.
        }
        } 
        \label{table_cycles_freq_bands_higher_ecc}
    \end{center}
\end{widetext}
\bibliographystyle{apsrev4-1}
\bibliography{ref-list2}

\begin{thebibliography}{91}%
\makeatletter
\providecommand \@ifxundefined [1]{%
 \@ifx{#1\undefined}
}%
\providecommand \@ifnum [1]{%
 \ifnum #1\expandafter \@firstoftwo
 \else \expandafter \@secondoftwo
 \fi
}%
\providecommand \@ifx [1]{%
 \ifx #1\expandafter \@firstoftwo
 \else \expandafter \@secondoftwo
 \fi
}%
\providecommand \natexlab [1]{#1}%
\providecommand \enquote  [1]{``#1''}%
\providecommand \bibnamefont  [1]{#1}%
\providecommand \bibfnamefont [1]{#1}%
\providecommand \citenamefont [1]{#1}%
\providecommand \href@noop [0]{\@secondoftwo}%
\providecommand \href [0]{\begingroup \@sanitize@url \@href}%
\providecommand \@href[1]{\@@startlink{#1}\@@href}%
\providecommand \@@href[1]{\endgroup#1\@@endlink}%
\providecommand \@sanitize@url [0]{\catcode `\\12\catcode `\$12\catcode
  `\&12\catcode `\#12\catcode `\^12\catcode `\_12\catcode `\%12\relax}%
\providecommand \@@startlink[1]{}%
\providecommand \@@endlink[0]{}%
\providecommand \url  [0]{\begingroup\@sanitize@url \@url }%
\providecommand \@url [1]{\endgroup\@href {#1}{\urlprefix }}%
\providecommand \urlprefix  [0]{URL }%
\providecommand \Eprint [0]{\href }%
\providecommand \doibase [0]{http://dx.doi.org/}%
\providecommand \selectlanguage [0]{\@gobble}%
\providecommand \bibinfo  [0]{\@secondoftwo}%
\providecommand \bibfield  [0]{\@secondoftwo}%
\providecommand \translation [1]{[#1]}%
\providecommand \BibitemOpen [0]{}%
\providecommand \bibitemStop [0]{}%
\providecommand \bibitemNoStop [0]{.\EOS\space}%
\providecommand \EOS [0]{\spacefactor3000\relax}%
\providecommand \BibitemShut  [1]{\csname bibitem#1\endcsname}%
\let\auto@bib@innerbib\@empty
\bibitem [{\citenamefont {Abbott}\ \emph
  {et~al.}(2016{\natexlab{a}})\citenamefont {Abbott} \emph
  {et~al.}}]{LIGOScientific:2016aoc}%
  \BibitemOpen
  \bibfield  {author} {\bibinfo {author} {\bibfnamefont {B.~P.}\ \bibnamefont
  {Abbott}} \emph {et~al.} (\bibinfo {collaboration} {LIGO Scientific,
  Virgo}),\ }\href {\doibase 10.1103/PhysRevLett.116.061102} {\bibfield
  {journal} {\bibinfo  {journal} {Phys. Rev. Lett.}\ }\textbf {\bibinfo
  {volume} {116}},\ \bibinfo {pages} {061102} (\bibinfo {year}
  {2016}{\natexlab{a}})},\ \Eprint {http://arxiv.org/abs/1602.03837}
  {arXiv:1602.03837 [gr-qc]} \BibitemShut {NoStop}%
\bibitem [{\citenamefont {Abbott}\ \emph {et~al.}(2017)\citenamefont {Abbott}
  \emph {et~al.}}]{LIGOScientific:2017vwq}%
  \BibitemOpen
  \bibfield  {author} {\bibinfo {author} {\bibfnamefont {B.~P.}\ \bibnamefont
  {Abbott}} \emph {et~al.} (\bibinfo {collaboration} {LIGO Scientific,
  Virgo}),\ }\href {\doibase 10.1103/PhysRevLett.119.161101} {\bibfield
  {journal} {\bibinfo  {journal} {Phys. Rev. Lett.}\ }\textbf {\bibinfo
  {volume} {119}},\ \bibinfo {pages} {161101} (\bibinfo {year} {2017})},\
  \Eprint {http://arxiv.org/abs/1710.05832} {arXiv:1710.05832 [gr-qc]}
  \BibitemShut {NoStop}%
\bibitem [{\citenamefont {Abbott}\ \emph {et~al.}(2020)\citenamefont {Abbott}
  \emph {et~al.}}]{LIGOScientific:2020aai}%
  \BibitemOpen
  \bibfield  {author} {\bibinfo {author} {\bibfnamefont {B.~P.}\ \bibnamefont
  {Abbott}} \emph {et~al.} (\bibinfo {collaboration} {LIGO Scientific,
  Virgo}),\ }\href {\doibase 10.3847/2041-8213/ab75f5} {\bibfield  {journal}
  {\bibinfo  {journal} {Astrophys. J. Lett.}\ }\textbf {\bibinfo {volume}
  {892}},\ \bibinfo {pages} {L3} (\bibinfo {year} {2020})},\ \Eprint
  {http://arxiv.org/abs/2001.01761} {arXiv:2001.01761 [astro-ph.HE]}
  \BibitemShut {NoStop}%
\bibitem [{\citenamefont {Abbott}\ \emph
  {et~al.}(2021{\natexlab{a}})\citenamefont {Abbott} \emph
  {et~al.}}]{LIGOScientific:2021qlt}%
  \BibitemOpen
  \bibfield  {author} {\bibinfo {author} {\bibfnamefont {R.}~\bibnamefont
  {Abbott}} \emph {et~al.} (\bibinfo {collaboration} {LIGO Scientific, KAGRA,
  VIRGO}),\ }\href {\doibase 10.3847/2041-8213/ac082e} {\bibfield  {journal}
  {\bibinfo  {journal} {Astrophys. J. Lett.}\ }\textbf {\bibinfo {volume}
  {915}},\ \bibinfo {pages} {L5} (\bibinfo {year} {2021}{\natexlab{a}})},\
  \Eprint {http://arxiv.org/abs/2106.15163} {arXiv:2106.15163 [astro-ph.HE]}
  \BibitemShut {NoStop}%
\bibitem [{\citenamefont {Aasi}\ \emph {et~al.}(2015)\citenamefont {Aasi} \emph
  {et~al.}}]{LIGOScientific:2014pky}%
  \BibitemOpen
  \bibfield  {author} {\bibinfo {author} {\bibfnamefont {J.}~\bibnamefont
  {Aasi}} \emph {et~al.} (\bibinfo {collaboration} {LIGO Scientific}),\ }\href
  {\doibase 10.1088/0264-9381/32/7/074001} {\bibfield  {journal} {\bibinfo
  {journal} {Class. Quant. Grav.}\ }\textbf {\bibinfo {volume} {32}},\ \bibinfo
  {pages} {074001} (\bibinfo {year} {2015})},\ \Eprint
  {http://arxiv.org/abs/1411.4547} {arXiv:1411.4547 [gr-qc]} \BibitemShut
  {NoStop}%
\bibitem [{\citenamefont {Acernese}\ \emph {et~al.}(2015)\citenamefont
  {Acernese} \emph {et~al.}}]{VIRGO:2014yos}%
  \BibitemOpen
  \bibfield  {author} {\bibinfo {author} {\bibfnamefont {F.}~\bibnamefont
  {Acernese}} \emph {et~al.} (\bibinfo {collaboration} {VIRGO}),\ }\href
  {\doibase 10.1088/0264-9381/32/2/024001} {\bibfield  {journal} {\bibinfo
  {journal} {Class. Quant. Grav.}\ }\textbf {\bibinfo {volume} {32}},\ \bibinfo
  {pages} {024001} (\bibinfo {year} {2015})},\ \Eprint
  {http://arxiv.org/abs/1408.3978} {arXiv:1408.3978 [gr-qc]} \BibitemShut
  {NoStop}%
\bibitem [{\citenamefont {Abbott}\ \emph {et~al.}(2019)\citenamefont {Abbott}
  \emph {et~al.}}]{LIGOScientific:2018mvr}%
  \BibitemOpen
  \bibfield  {author} {\bibinfo {author} {\bibfnamefont {B.~P.}\ \bibnamefont
  {Abbott}} \emph {et~al.} (\bibinfo {collaboration} {LIGO Scientific,
  Virgo}),\ }\href {\doibase 10.1103/PhysRevX.9.031040} {\bibfield  {journal}
  {\bibinfo  {journal} {Phys. Rev. X}\ }\textbf {\bibinfo {volume} {9}},\
  \bibinfo {pages} {031040} (\bibinfo {year} {2019})},\ \Eprint
  {http://arxiv.org/abs/1811.12907} {arXiv:1811.12907 [astro-ph.HE]}
  \BibitemShut {NoStop}%
\bibitem [{\citenamefont {Abbott}\ \emph
  {et~al.}(2021{\natexlab{b}})\citenamefont {Abbott} \emph
  {et~al.}}]{LIGOScientific:2020ibl}%
  \BibitemOpen
  \bibfield  {author} {\bibinfo {author} {\bibfnamefont {R.}~\bibnamefont
  {Abbott}} \emph {et~al.} (\bibinfo {collaboration} {LIGO Scientific,
  Virgo}),\ }\href {\doibase 10.1103/PhysRevX.11.021053} {\bibfield  {journal}
  {\bibinfo  {journal} {Phys. Rev. X}\ }\textbf {\bibinfo {volume} {11}},\
  \bibinfo {pages} {021053} (\bibinfo {year} {2021}{\natexlab{b}})},\ \Eprint
  {http://arxiv.org/abs/2010.14527} {arXiv:2010.14527 [gr-qc]} \BibitemShut
  {NoStop}%
\bibitem [{\citenamefont {Abbott}\ \emph {et~al.}(2024)\citenamefont {Abbott}
  \emph {et~al.}}]{LIGOScientific:2021usb}%
  \BibitemOpen
  \bibfield  {author} {\bibinfo {author} {\bibfnamefont {R.}~\bibnamefont
  {Abbott}} \emph {et~al.} (\bibinfo {collaboration} {LIGO Scientific,
  VIRGO}),\ }\href {\doibase 10.1103/PhysRevD.109.022001} {\bibfield  {journal}
  {\bibinfo  {journal} {Phys. Rev. D}\ }\textbf {\bibinfo {volume} {109}},\
  \bibinfo {pages} {022001} (\bibinfo {year} {2024})},\ \Eprint
  {http://arxiv.org/abs/2108.01045} {arXiv:2108.01045 [gr-qc]} \BibitemShut
  {NoStop}%
\bibitem [{\citenamefont {Abbott}\ \emph {et~al.}(2023)\citenamefont {Abbott}
  \emph {et~al.}}]{KAGRA:2021vkt}%
  \BibitemOpen
  \bibfield  {author} {\bibinfo {author} {\bibfnamefont {R.}~\bibnamefont
  {Abbott}} \emph {et~al.} (\bibinfo {collaboration} {KAGRA, VIRGO, LIGO
  Scientific}),\ }\href {\doibase 10.1103/PhysRevX.13.041039} {\bibfield
  {journal} {\bibinfo  {journal} {Phys. Rev. X}\ }\textbf {\bibinfo {volume}
  {13}},\ \bibinfo {pages} {041039} (\bibinfo {year} {2023})},\ \Eprint
  {http://arxiv.org/abs/2111.03606} {arXiv:2111.03606 [gr-qc]} \BibitemShut
  {NoStop}%
\bibitem [{\citenamefont {Cutler}\ and\ \citenamefont
  {Flanagan}(1994)}]{Cutler:1994ys}%
  \BibitemOpen
  \bibfield  {author} {\bibinfo {author} {\bibfnamefont {C.}~\bibnamefont
  {Cutler}}\ and\ \bibinfo {author} {\bibfnamefont {E.~E.}\ \bibnamefont
  {Flanagan}},\ }\href {\doibase 10.1103/PhysRevD.49.2658} {\bibfield
  {journal} {\bibinfo  {journal} {Phys. Rev. D}\ }\textbf {\bibinfo {volume}
  {49}},\ \bibinfo {pages} {2658} (\bibinfo {year} {1994})},\ \Eprint
  {http://arxiv.org/abs/gr-qc/9402014} {arXiv:gr-qc/9402014} \BibitemShut
  {NoStop}%
\bibitem [{\citenamefont {Poisson}\ and\ \citenamefont
  {Will}(1995)}]{Poisson:1995ef}%
  \BibitemOpen
  \bibfield  {author} {\bibinfo {author} {\bibfnamefont {E.}~\bibnamefont
  {Poisson}}\ and\ \bibinfo {author} {\bibfnamefont {C.~M.}\ \bibnamefont
  {Will}},\ }\href {\doibase 10.1103/PhysRevD.52.848} {\bibfield  {journal}
  {\bibinfo  {journal} {Phys. Rev. D}\ }\textbf {\bibinfo {volume} {52}},\
  \bibinfo {pages} {848} (\bibinfo {year} {1995})},\ \Eprint
  {http://arxiv.org/abs/gr-qc/9502040} {arXiv:gr-qc/9502040} \BibitemShut
  {NoStop}%
\bibitem [{\citenamefont {Krolak}\ \emph {et~al.}(1995)\citenamefont {Krolak},
  \citenamefont {Kokkotas},\ and\ \citenamefont {Schaefer}}]{Krolak:1995md}%
  \BibitemOpen
  \bibfield  {author} {\bibinfo {author} {\bibfnamefont {A.}~\bibnamefont
  {Krolak}}, \bibinfo {author} {\bibfnamefont {K.~D.}\ \bibnamefont
  {Kokkotas}}, \ and\ \bibinfo {author} {\bibfnamefont {G.}~\bibnamefont
  {Schaefer}},\ }\href {\doibase 10.1103/PhysRevD.52.2089} {\bibfield
  {journal} {\bibinfo  {journal} {Phys. Rev. D}\ }\textbf {\bibinfo {volume}
  {52}},\ \bibinfo {pages} {2089} (\bibinfo {year} {1995})},\ \Eprint
  {http://arxiv.org/abs/gr-qc/9503013} {arXiv:gr-qc/9503013} \BibitemShut
  {NoStop}%
\bibitem [{\citenamefont {Abbott}\ \emph
  {et~al.}(2016{\natexlab{b}})\citenamefont {Abbott} \emph
  {et~al.}}]{LIGOScientific:2016vbw}%
  \BibitemOpen
  \bibfield  {author} {\bibinfo {author} {\bibfnamefont {B.~P.}\ \bibnamefont
  {Abbott}} \emph {et~al.} (\bibinfo {collaboration} {LIGO Scientific,
  Virgo}),\ }\href {\doibase 10.1103/PhysRevD.93.122003} {\bibfield  {journal}
  {\bibinfo  {journal} {Phys. Rev. D}\ }\textbf {\bibinfo {volume} {93}},\
  \bibinfo {pages} {122003} (\bibinfo {year} {2016}{\natexlab{b}})},\ \Eprint
  {http://arxiv.org/abs/1602.03839} {arXiv:1602.03839 [gr-qc]} \BibitemShut
  {NoStop}%
\bibitem [{\citenamefont {Blanchet}(2014)}]{Blanchet:2013haa}%
  \BibitemOpen
  \bibfield  {author} {\bibinfo {author} {\bibfnamefont {L.}~\bibnamefont
  {Blanchet}},\ }\href {\doibase 10.12942/lrr-2014-2} {\bibfield  {journal}
  {\bibinfo  {journal} {Living Rev.Rel.}\ }\textbf {\bibinfo {volume} {17}},\
  \bibinfo {pages} {2} (\bibinfo {year} {2014})},\ \Eprint
  {http://arxiv.org/abs/1310.1528} {arXiv:1310.1528 [gr-qc]} \BibitemShut
  {NoStop}%
\bibitem [{SXS()}]{SXS:catalog}%
  \BibitemOpen
  \href@noop {} {}\bibinfo {howpublished}
  {\url{http://www.black-holes.org/waveforms}}\BibitemShut {NoStop}%
\bibitem [{\citenamefont {Healy}\ and\ \citenamefont
  {Lousto}(2022)}]{Healy:2022wdn}%
  \BibitemOpen
  \bibfield  {author} {\bibinfo {author} {\bibfnamefont {J.}~\bibnamefont
  {Healy}}\ and\ \bibinfo {author} {\bibfnamefont {C.~O.}\ \bibnamefont
  {Lousto}},\ }\href {\doibase 10.1103/PhysRevD.105.124010} {\bibfield
  {journal} {\bibinfo  {journal} {Phys. Rev. D}\ }\textbf {\bibinfo {volume}
  {105}},\ \bibinfo {pages} {124010} (\bibinfo {year} {2022})},\ \Eprint
  {http://arxiv.org/abs/2202.00018} {arXiv:2202.00018 [gr-qc]} \BibitemShut
  {NoStop}%
\bibitem [{\citenamefont {Ferguson}\ \emph {et~al.}(2023)\citenamefont
  {Ferguson} \emph {et~al.}}]{Ferguson:2023vta}%
  \BibitemOpen
  \bibfield  {author} {\bibinfo {author} {\bibfnamefont {D.}~\bibnamefont
  {Ferguson}} \emph {et~al.},\ }\href@noop {} {\  (\bibinfo {year} {2023})},\
  \Eprint {http://arxiv.org/abs/2309.00262} {arXiv:2309.00262 [gr-qc]}
  \BibitemShut {NoStop}%
\bibitem [{\citenamefont {Pound}\ and\ \citenamefont
  {Wardell}(2021)}]{Pound:2021qin}%
  \BibitemOpen
  \bibfield  {author} {\bibinfo {author} {\bibfnamefont {A.}~\bibnamefont
  {Pound}}\ and\ \bibinfo {author} {\bibfnamefont {B.}~\bibnamefont
  {Wardell}},\ }\href {\doibase 10.1007/978-981-15-4702-7\_38-1} {\  (\bibinfo
  {year} {2021}),\ 10.1007/978-981-15-4702-7\_38-1},\ \Eprint
  {http://arxiv.org/abs/2101.04592} {arXiv:2101.04592 [gr-qc]} \BibitemShut
  {NoStop}%
\bibitem [{\citenamefont {Peters}\ and\ \citenamefont
  {Mathews}(1963)}]{Peters:1963ux}%
  \BibitemOpen
  \bibfield  {author} {\bibinfo {author} {\bibfnamefont {P.~C.}\ \bibnamefont
  {Peters}}\ and\ \bibinfo {author} {\bibfnamefont {J.}~\bibnamefont
  {Mathews}},\ }\href {\doibase 10.1103/PhysRev.131.435} {\bibfield  {journal}
  {\bibinfo  {journal} {Phys. Rev.}\ }\textbf {\bibinfo {volume} {131}},\
  \bibinfo {pages} {435} (\bibinfo {year} {1963})}\BibitemShut {NoStop}%
\bibitem [{\citenamefont {Peters}(1964)}]{Peters:1964zz}%
  \BibitemOpen
  \bibfield  {author} {\bibinfo {author} {\bibfnamefont {P.~C.}\ \bibnamefont
  {Peters}},\ }\href {\doibase 10.1103/PhysRev.136.B1224} {\bibfield  {journal}
  {\bibinfo  {journal} {Phys. Rev.}\ }\textbf {\bibinfo {volume} {136}},\
  \bibinfo {pages} {B1224} (\bibinfo {year} {1964})}\BibitemShut {NoStop}%
\bibitem [{\citenamefont {Kozai}(1962)}]{Kozai:1962zz}%
  \BibitemOpen
  \bibfield  {author} {\bibinfo {author} {\bibfnamefont {Y.}~\bibnamefont
  {Kozai}},\ }\href {\doibase 10.1086/108790} {\bibfield  {journal} {\bibinfo
  {journal} {Astron. J.}\ }\textbf {\bibinfo {volume} {67}},\ \bibinfo {pages}
  {591} (\bibinfo {year} {1962})}\BibitemShut {NoStop}%
\bibitem [{\citenamefont {Lidov}(1962)}]{LIDOV1962719}%
  \BibitemOpen
  \bibfield  {author} {\bibinfo {author} {\bibfnamefont {M.}~\bibnamefont
  {Lidov}},\ }\href {\doibase https://doi.org/10.1016/0032-0633(62)90129-0}
  {\bibfield  {journal} {\bibinfo  {journal} {Planetary and Space Science}\
  }\textbf {\bibinfo {volume} {9}},\ \bibinfo {pages} {719} (\bibinfo {year}
  {1962})}\BibitemShut {NoStop}%
\bibitem [{\citenamefont {Samsing}\ \emph {et~al.}(2014)\citenamefont
  {Samsing}, \citenamefont {MacLeod},\ and\ \citenamefont
  {Ramirez-Ruiz}}]{Samsing:2013kua}%
  \BibitemOpen
  \bibfield  {author} {\bibinfo {author} {\bibfnamefont {J.}~\bibnamefont
  {Samsing}}, \bibinfo {author} {\bibfnamefont {M.}~\bibnamefont {MacLeod}}, \
  and\ \bibinfo {author} {\bibfnamefont {E.}~\bibnamefont {Ramirez-Ruiz}},\
  }\href {\doibase 10.1088/0004-637X/784/1/71} {\bibfield  {journal} {\bibinfo
  {journal} {Astrophys. J.}\ }\textbf {\bibinfo {volume} {784}},\ \bibinfo
  {pages} {71} (\bibinfo {year} {2014})},\ \Eprint
  {http://arxiv.org/abs/1308.2964} {arXiv:1308.2964 [astro-ph.HE]} \BibitemShut
  {NoStop}%
\bibitem [{\citenamefont {Rodriguez}\ \emph {et~al.}(2016)\citenamefont
  {Rodriguez}, \citenamefont {Chatterjee},\ and\ \citenamefont
  {Rasio}}]{Rodriguez:2016kxx}%
  \BibitemOpen
  \bibfield  {author} {\bibinfo {author} {\bibfnamefont {C.~L.}\ \bibnamefont
  {Rodriguez}}, \bibinfo {author} {\bibfnamefont {S.}~\bibnamefont
  {Chatterjee}}, \ and\ \bibinfo {author} {\bibfnamefont {F.~A.}\ \bibnamefont
  {Rasio}},\ }\href {\doibase 10.1103/PhysRevD.93.084029} {\bibfield  {journal}
  {\bibinfo  {journal} {Phys. Rev. D}\ }\textbf {\bibinfo {volume} {93}},\
  \bibinfo {pages} {084029} (\bibinfo {year} {2016})},\ \Eprint
  {http://arxiv.org/abs/1602.02444} {arXiv:1602.02444 [astro-ph.HE]}
  \BibitemShut {NoStop}%
\bibitem [{\citenamefont {Antonini}\ \emph {et~al.}(2016)\citenamefont
  {Antonini}, \citenamefont {Chatterjee}, \citenamefont {Rodriguez},
  \citenamefont {Morscher}, \citenamefont {Pattabiraman}, \citenamefont
  {Kalogera},\ and\ \citenamefont {Rasio}}]{Antonini:2015zsa}%
  \BibitemOpen
  \bibfield  {author} {\bibinfo {author} {\bibfnamefont {F.}~\bibnamefont
  {Antonini}}, \bibinfo {author} {\bibfnamefont {S.}~\bibnamefont
  {Chatterjee}}, \bibinfo {author} {\bibfnamefont {C.~L.}\ \bibnamefont
  {Rodriguez}}, \bibinfo {author} {\bibfnamefont {M.}~\bibnamefont {Morscher}},
  \bibinfo {author} {\bibfnamefont {B.}~\bibnamefont {Pattabiraman}}, \bibinfo
  {author} {\bibfnamefont {V.}~\bibnamefont {Kalogera}}, \ and\ \bibinfo
  {author} {\bibfnamefont {F.~A.}\ \bibnamefont {Rasio}},\ }\href {\doibase
  10.3847/0004-637X/816/2/65} {\bibfield  {journal} {\bibinfo  {journal}
  {Astrophys. J.}\ }\textbf {\bibinfo {volume} {816}},\ \bibinfo {pages} {65}
  (\bibinfo {year} {2016})},\ \Eprint {http://arxiv.org/abs/1509.05080}
  {arXiv:1509.05080 [astro-ph.GA]} \BibitemShut {NoStop}%
\bibitem [{\citenamefont {Chomiuk}\ \emph {et~al.}(2013)\citenamefont
  {Chomiuk}, \citenamefont {Strader}, \citenamefont {Maccarone}, \citenamefont
  {Miller-Jones}, \citenamefont {Heinke}, \citenamefont {Noyola}, \citenamefont
  {Seth},\ and\ \citenamefont {Ransom}}]{Chomiuk:2013qya}%
  \BibitemOpen
  \bibfield  {author} {\bibinfo {author} {\bibfnamefont {L.}~\bibnamefont
  {Chomiuk}}, \bibinfo {author} {\bibfnamefont {J.}~\bibnamefont {Strader}},
  \bibinfo {author} {\bibfnamefont {T.~J.}\ \bibnamefont {Maccarone}}, \bibinfo
  {author} {\bibfnamefont {J.~C.~A.}\ \bibnamefont {Miller-Jones}}, \bibinfo
  {author} {\bibfnamefont {C.}~\bibnamefont {Heinke}}, \bibinfo {author}
  {\bibfnamefont {E.}~\bibnamefont {Noyola}}, \bibinfo {author} {\bibfnamefont
  {A.~C.}\ \bibnamefont {Seth}}, \ and\ \bibinfo {author} {\bibfnamefont
  {S.}~\bibnamefont {Ransom}},\ }\href {\doibase 10.1088/0004-637X/777/1/69}
  {\bibfield  {journal} {\bibinfo  {journal} {Astrophys. J.}\ }\textbf
  {\bibinfo {volume} {777}},\ \bibinfo {pages} {69} (\bibinfo {year} {2013})},\
  \Eprint {http://arxiv.org/abs/1306.6624} {arXiv:1306.6624 [astro-ph.HE]}
  \BibitemShut {NoStop}%
\bibitem [{\citenamefont {Strader}\ \emph {et~al.}(2012)\citenamefont
  {Strader}, \citenamefont {Chomiuk}, \citenamefont {Maccarone}, \citenamefont
  {Miller-Jones},\ and\ \citenamefont {Seth}}]{Strader:2012wj}%
  \BibitemOpen
  \bibfield  {author} {\bibinfo {author} {\bibfnamefont {J.}~\bibnamefont
  {Strader}}, \bibinfo {author} {\bibfnamefont {L.}~\bibnamefont {Chomiuk}},
  \bibinfo {author} {\bibfnamefont {T.}~\bibnamefont {Maccarone}}, \bibinfo
  {author} {\bibfnamefont {J.}~\bibnamefont {Miller-Jones}}, \ and\ \bibinfo
  {author} {\bibfnamefont {A.}~\bibnamefont {Seth}},\ }\href {\doibase
  10.1038/nature11490} {\bibfield  {journal} {\bibinfo  {journal} {Nature}\
  }\textbf {\bibinfo {volume} {490}},\ \bibinfo {pages} {71} (\bibinfo {year}
  {2012})},\ \Eprint {http://arxiv.org/abs/1210.0901} {arXiv:1210.0901
  [astro-ph.HE]} \BibitemShut {NoStop}%
\bibitem [{\citenamefont {Osburn}\ \emph {et~al.}(2016)\citenamefont {Osburn},
  \citenamefont {Warburton},\ and\ \citenamefont {Evans}}]{Osburn:2015duj}%
  \BibitemOpen
  \bibfield  {author} {\bibinfo {author} {\bibfnamefont {T.}~\bibnamefont
  {Osburn}}, \bibinfo {author} {\bibfnamefont {N.}~\bibnamefont {Warburton}}, \
  and\ \bibinfo {author} {\bibfnamefont {C.~R.}\ \bibnamefont {Evans}},\ }\href
  {\doibase 10.1103/PhysRevD.93.064024} {\bibfield  {journal} {\bibinfo
  {journal} {Phys. Rev. D}\ }\textbf {\bibinfo {volume} {93}},\ \bibinfo
  {pages} {064024} (\bibinfo {year} {2016})},\ \Eprint
  {http://arxiv.org/abs/1511.01498} {arXiv:1511.01498 [gr-qc]} \BibitemShut
  {NoStop}%
\bibitem [{\citenamefont {VanLandingham}\ \emph {et~al.}(2016)\citenamefont
  {VanLandingham}, \citenamefont {Miller}, \citenamefont {Hamilton},\ and\
  \citenamefont {Richardson}}]{VanLandingham:2016ccd}%
  \BibitemOpen
  \bibfield  {author} {\bibinfo {author} {\bibfnamefont {J.~H.}\ \bibnamefont
  {VanLandingham}}, \bibinfo {author} {\bibfnamefont {M.~C.}\ \bibnamefont
  {Miller}}, \bibinfo {author} {\bibfnamefont {D.~P.}\ \bibnamefont
  {Hamilton}}, \ and\ \bibinfo {author} {\bibfnamefont {D.~C.}\ \bibnamefont
  {Richardson}},\ }\href {\doibase 10.3847/0004-637X/828/2/77} {\bibfield
  {journal} {\bibinfo  {journal} {Astrophys. J.}\ }\textbf {\bibinfo {volume}
  {828}},\ \bibinfo {pages} {77} (\bibinfo {year} {2016})},\ \Eprint
  {http://arxiv.org/abs/1604.04948} {arXiv:1604.04948 [astro-ph.HE]}
  \BibitemShut {NoStop}%
\bibitem [{\citenamefont {Hoang}\ \emph {et~al.}(2018)\citenamefont {Hoang},
  \citenamefont {Naoz}, \citenamefont {Kocsis}, \citenamefont {Rasio},\ and\
  \citenamefont {Dosopoulou}}]{Hoang:2017fvh}%
  \BibitemOpen
  \bibfield  {author} {\bibinfo {author} {\bibfnamefont {B.-M.}\ \bibnamefont
  {Hoang}}, \bibinfo {author} {\bibfnamefont {S.}~\bibnamefont {Naoz}},
  \bibinfo {author} {\bibfnamefont {B.}~\bibnamefont {Kocsis}}, \bibinfo
  {author} {\bibfnamefont {F.~A.}\ \bibnamefont {Rasio}}, \ and\ \bibinfo
  {author} {\bibfnamefont {F.}~\bibnamefont {Dosopoulou}},\ }\href {\doibase
  10.3847/1538-4357/aaafce} {\bibfield  {journal} {\bibinfo  {journal}
  {Astrophys. J.}\ }\textbf {\bibinfo {volume} {856}},\ \bibinfo {pages} {140}
  (\bibinfo {year} {2018})},\ \Eprint {http://arxiv.org/abs/1706.09896}
  {arXiv:1706.09896 [astro-ph.HE]} \BibitemShut {NoStop}%
\bibitem [{\citenamefont {Gond\'an}\ \emph {et~al.}(2018)\citenamefont
  {Gond\'an}, \citenamefont {Kocsis}, \citenamefont {Raffai},\ and\
  \citenamefont {Frei}}]{Gondan:2017hbp}%
  \BibitemOpen
  \bibfield  {author} {\bibinfo {author} {\bibfnamefont {L.}~\bibnamefont
  {Gond\'an}}, \bibinfo {author} {\bibfnamefont {B.}~\bibnamefont {Kocsis}},
  \bibinfo {author} {\bibfnamefont {P.}~\bibnamefont {Raffai}}, \ and\ \bibinfo
  {author} {\bibfnamefont {Z.}~\bibnamefont {Frei}},\ }\href {\doibase
  10.3847/1538-4357/aaad0e} {\bibfield  {journal} {\bibinfo  {journal}
  {Astrophys. J.}\ }\textbf {\bibinfo {volume} {855}},\ \bibinfo {pages} {34}
  (\bibinfo {year} {2018})},\ \Eprint {http://arxiv.org/abs/1705.10781}
  {arXiv:1705.10781 [astro-ph.HE]} \BibitemShut {NoStop}%
\bibitem [{\citenamefont {Gond\'an}\ and\ \citenamefont
  {Kocsis}(2021)}]{Gondan:2020svr}%
  \BibitemOpen
  \bibfield  {author} {\bibinfo {author} {\bibfnamefont {L.}~\bibnamefont
  {Gond\'an}}\ and\ \bibinfo {author} {\bibfnamefont {B.}~\bibnamefont
  {Kocsis}},\ }\href {\doibase 10.1093/mnras/stab1722} {\bibfield  {journal}
  {\bibinfo  {journal} {Mon. Not. Roy. Astron. Soc.}\ }\textbf {\bibinfo
  {volume} {506}},\ \bibinfo {pages} {1665} (\bibinfo {year} {2021})},\ \Eprint
  {http://arxiv.org/abs/2011.02507} {arXiv:2011.02507 [astro-ph.HE]}
  \BibitemShut {NoStop}%
\bibitem [{\citenamefont {Kumamoto}\ \emph {et~al.}(2019)\citenamefont
  {Kumamoto}, \citenamefont {Fujii},\ and\ \citenamefont
  {Tanikawa}}]{Kumamoto:2018gdg}%
  \BibitemOpen
  \bibfield  {author} {\bibinfo {author} {\bibfnamefont {J.}~\bibnamefont
  {Kumamoto}}, \bibinfo {author} {\bibfnamefont {M.~S.}\ \bibnamefont {Fujii}},
  \ and\ \bibinfo {author} {\bibfnamefont {A.}~\bibnamefont {Tanikawa}},\
  }\href {\doibase 10.1093/mnras/stz1068} {\bibfield  {journal} {\bibinfo
  {journal} {Mon. Not. Roy. Astron. Soc.}\ }\textbf {\bibinfo {volume} {486}},\
  \bibinfo {pages} {3942} (\bibinfo {year} {2019})},\ \Eprint
  {http://arxiv.org/abs/1811.06726} {arXiv:1811.06726 [astro-ph.HE]}
  \BibitemShut {NoStop}%
\bibitem [{\citenamefont {Fragione}\ and\ \citenamefont
  {Kocsis}(2019)}]{Fragione:2019hqt}%
  \BibitemOpen
  \bibfield  {author} {\bibinfo {author} {\bibfnamefont {G.}~\bibnamefont
  {Fragione}}\ and\ \bibinfo {author} {\bibfnamefont {B.}~\bibnamefont
  {Kocsis}},\ }\href {\doibase 10.1093/mnras/stz1175} {\bibfield  {journal}
  {\bibinfo  {journal} {Mon. Not. Roy. Astron. Soc.}\ }\textbf {\bibinfo
  {volume} {486}},\ \bibinfo {pages} {4781} (\bibinfo {year} {2019})},\ \Eprint
  {http://arxiv.org/abs/1903.03112} {arXiv:1903.03112 [astro-ph.GA]}
  \BibitemShut {NoStop}%
\bibitem [{\citenamefont {O'Leary}\ \emph {et~al.}(2006)\citenamefont
  {O'Leary}, \citenamefont {Rasio}, \citenamefont {Fregeau}, \citenamefont
  {Ivanova},\ and\ \citenamefont {O'Shaughnessy}}]{OLeary:2005vqo}%
  \BibitemOpen
  \bibfield  {author} {\bibinfo {author} {\bibfnamefont {R.~M.}\ \bibnamefont
  {O'Leary}}, \bibinfo {author} {\bibfnamefont {F.~A.}\ \bibnamefont {Rasio}},
  \bibinfo {author} {\bibfnamefont {J.~M.}\ \bibnamefont {Fregeau}}, \bibinfo
  {author} {\bibfnamefont {N.}~\bibnamefont {Ivanova}}, \ and\ \bibinfo
  {author} {\bibfnamefont {R.~W.}\ \bibnamefont {O'Shaughnessy}},\ }\href
  {\doibase 10.1086/498446} {\bibfield  {journal} {\bibinfo  {journal}
  {Astrophys. J.}\ }\textbf {\bibinfo {volume} {637}},\ \bibinfo {pages} {937}
  (\bibinfo {year} {2006})},\ \Eprint {http://arxiv.org/abs/astro-ph/0508224}
  {arXiv:astro-ph/0508224} \BibitemShut {NoStop}%
\bibitem [{\citenamefont {Allen}\ \emph {et~al.}(2012)\citenamefont {Allen},
  \citenamefont {Anderson}, \citenamefont {Brady}, \citenamefont {Brown},\ and\
  \citenamefont {Creighton}}]{Allen:2005fk}%
  \BibitemOpen
  \bibfield  {author} {\bibinfo {author} {\bibfnamefont {B.}~\bibnamefont
  {Allen}}, \bibinfo {author} {\bibfnamefont {W.~G.}\ \bibnamefont {Anderson}},
  \bibinfo {author} {\bibfnamefont {P.~R.}\ \bibnamefont {Brady}}, \bibinfo
  {author} {\bibfnamefont {D.~A.}\ \bibnamefont {Brown}}, \ and\ \bibinfo
  {author} {\bibfnamefont {J.~D.~E.}\ \bibnamefont {Creighton}},\ }\href
  {\doibase 10.1103/PhysRevD.85.122006} {\bibfield  {journal} {\bibinfo
  {journal} {Phys. Rev. D}\ }\textbf {\bibinfo {volume} {85}},\ \bibinfo
  {pages} {122006} (\bibinfo {year} {2012})},\ \Eprint
  {http://arxiv.org/abs/gr-qc/0509116} {arXiv:gr-qc/0509116} \BibitemShut
  {NoStop}%
\bibitem [{\citenamefont {Bhat}\ \emph {et~al.}(2023)\citenamefont {Bhat},
  \citenamefont {Saini}, \citenamefont {Favata},\ and\ \citenamefont
  {Arun}}]{Bhat:2022amc}%
  \BibitemOpen
  \bibfield  {author} {\bibinfo {author} {\bibfnamefont {S.~A.}\ \bibnamefont
  {Bhat}}, \bibinfo {author} {\bibfnamefont {P.}~\bibnamefont {Saini}},
  \bibinfo {author} {\bibfnamefont {M.}~\bibnamefont {Favata}}, \ and\ \bibinfo
  {author} {\bibfnamefont {K.~G.}\ \bibnamefont {Arun}},\ }\href {\doibase
  10.1103/PhysRevD.107.024009} {\bibfield  {journal} {\bibinfo  {journal}
  {Phys. Rev. D}\ }\textbf {\bibinfo {volume} {107}},\ \bibinfo {pages}
  {024009} (\bibinfo {year} {2023})},\ \Eprint
  {http://arxiv.org/abs/2207.13761} {arXiv:2207.13761 [gr-qc]} \BibitemShut
  {NoStop}%
\bibitem [{\citenamefont {Chattaraj}\ \emph {et~al.}(2022)\citenamefont
  {Chattaraj}, \citenamefont {RoyChowdhury}, \citenamefont {Divyajyoti},
  \citenamefont {Mishra},\ and\ \citenamefont {Gupta}}]{Chattaraj:2022tay}%
  \BibitemOpen
  \bibfield  {author} {\bibinfo {author} {\bibfnamefont {A.}~\bibnamefont
  {Chattaraj}}, \bibinfo {author} {\bibfnamefont {T.}~\bibnamefont
  {RoyChowdhury}}, \bibinfo {author} {\bibnamefont {Divyajyoti}}, \bibinfo
  {author} {\bibfnamefont {C.~K.}\ \bibnamefont {Mishra}}, \ and\ \bibinfo
  {author} {\bibfnamefont {A.}~\bibnamefont {Gupta}},\ }\href {\doibase
  10.1103/PhysRevD.106.124008} {\bibfield  {journal} {\bibinfo  {journal}
  {Phys. Rev. D}\ }\textbf {\bibinfo {volume} {106}},\ \bibinfo {pages}
  {124008} (\bibinfo {year} {2022})},\ \Eprint
  {http://arxiv.org/abs/2204.02377} {arXiv:2204.02377 [gr-qc]} \BibitemShut
  {NoStop}%
\bibitem [{\citenamefont {Favata}(2014)}]{Favata2013}%
  \BibitemOpen
  \bibfield  {author} {\bibinfo {author} {\bibfnamefont {M.}~\bibnamefont
  {Favata}},\ }\href {\doibase 10.1103/PhysRevLett.112.101101} {\bibfield
  {journal} {\bibinfo  {journal} {Phys. Rev. Lett.}\ }\textbf {\bibinfo
  {volume} {112}},\ \bibinfo {pages} {101101} (\bibinfo {year} {2014})},\
  \Eprint {http://arxiv.org/abs/1310.8288} {arXiv:1310.8288 [gr-qc]}
  \BibitemShut {NoStop}%
\bibitem [{\citenamefont {Divyajyoti}\ \emph
  {et~al.}(2024{\natexlab{a}})\citenamefont {Divyajyoti}, \citenamefont
  {Kumar}, \citenamefont {Tibrewal}, \citenamefont {Romero-Shaw},\ and\
  \citenamefont {Mishra}}]{Divyajyoti:2023rht}%
  \BibitemOpen
  \bibfield  {author} {\bibinfo {author} {\bibnamefont {Divyajyoti}}, \bibinfo
  {author} {\bibfnamefont {S.}~\bibnamefont {Kumar}}, \bibinfo {author}
  {\bibfnamefont {S.}~\bibnamefont {Tibrewal}}, \bibinfo {author}
  {\bibfnamefont {I.~M.}\ \bibnamefont {Romero-Shaw}}, \ and\ \bibinfo {author}
  {\bibfnamefont {C.~K.}\ \bibnamefont {Mishra}},\ }\href {\doibase
  10.1103/PhysRevD.109.043037} {\bibfield  {journal} {\bibinfo  {journal}
  {Phys. Rev. D}\ }\textbf {\bibinfo {volume} {109}},\ \bibinfo {pages}
  {043037} (\bibinfo {year} {2024}{\natexlab{a}})},\ \Eprint
  {http://arxiv.org/abs/2309.16638} {arXiv:2309.16638 [gr-qc]} \BibitemShut
  {NoStop}%
\bibitem [{\citenamefont {Phukon}\ \emph {et~al.}(2025)\citenamefont {Phukon},
  \citenamefont {Schmidt},\ and\ \citenamefont {Pratten}}]{phukon:2024amh}%
  \BibitemOpen
  \bibfield  {author} {\bibinfo {author} {\bibfnamefont {K.~S.}\ \bibnamefont
  {Phukon}}, \bibinfo {author} {\bibfnamefont {P.}~\bibnamefont {Schmidt}}, \
  and\ \bibinfo {author} {\bibfnamefont {G.}~\bibnamefont {Pratten}},\ }\href
  {\doibase 10.1103/PhysRevD.111.043040} {\bibfield  {journal} {\bibinfo
  {journal} {Phys. Rev. D}\ }\textbf {\bibinfo {volume} {111}},\ \bibinfo
  {pages} {043040} (\bibinfo {year} {2025})},\ \Eprint
  {http://arxiv.org/abs/2412.06433} {arXiv:2412.06433 [gr-qc]} \BibitemShut
  {NoStop}%
\bibitem [{\citenamefont {Cui}\ \emph {et~al.}(2023)\citenamefont {Cui} \emph
  {et~al.}}]{Cui:2023uyb}%
  \BibitemOpen
  \bibfield  {author} {\bibinfo {author} {\bibfnamefont {Y.}~\bibnamefont
  {Cui}} \emph {et~al.},\ }\href {\doibase 10.1038/s41586-023-06479-6}
  {\bibfield  {journal} {\bibinfo  {journal} {Nature}\ }\textbf {\bibinfo
  {volume} {621}},\ \bibinfo {pages} {711} (\bibinfo {year} {2023})},\ \Eprint
  {http://arxiv.org/abs/2310.09015} {arXiv:2310.09015 [astro-ph.HE]}
  \BibitemShut {NoStop}%
\bibitem [{\citenamefont {Taracchini}\ \emph {et~al.}(2012)\citenamefont
  {Taracchini}, \citenamefont {Pan}, \citenamefont {Buonanno}, \citenamefont
  {Barausse}, \citenamefont {Boyle}, \citenamefont {Chu}, \citenamefont
  {Lovelace}, \citenamefont {Pfeiffer},\ and\ \citenamefont
  {Scheel}}]{Taracchini:2012ig}%
  \BibitemOpen
  \bibfield  {author} {\bibinfo {author} {\bibfnamefont {A.}~\bibnamefont
  {Taracchini}}, \bibinfo {author} {\bibfnamefont {Y.}~\bibnamefont {Pan}},
  \bibinfo {author} {\bibfnamefont {A.}~\bibnamefont {Buonanno}}, \bibinfo
  {author} {\bibfnamefont {E.}~\bibnamefont {Barausse}}, \bibinfo {author}
  {\bibfnamefont {M.}~\bibnamefont {Boyle}}, \bibinfo {author} {\bibfnamefont
  {T.}~\bibnamefont {Chu}}, \bibinfo {author} {\bibfnamefont {G.}~\bibnamefont
  {Lovelace}}, \bibinfo {author} {\bibfnamefont {H.~P.}\ \bibnamefont
  {Pfeiffer}}, \ and\ \bibinfo {author} {\bibfnamefont {M.~A.}\ \bibnamefont
  {Scheel}},\ }\href {\doibase 10.1103/PhysRevD.86.024011} {\bibfield
  {journal} {\bibinfo  {journal} {Phys. Rev. D}\ }\textbf {\bibinfo {volume}
  {86}},\ \bibinfo {pages} {024011} (\bibinfo {year} {2012})},\ \Eprint
  {http://arxiv.org/abs/1202.0790} {arXiv:1202.0790 [gr-qc]} \BibitemShut
  {NoStop}%
\bibitem [{\citenamefont {Ramos-Buades}\ \emph {et~al.}(2020)\citenamefont
  {Ramos-Buades}, \citenamefont {Husa}, \citenamefont {Pratten}, \citenamefont
  {Estell\'es}, \citenamefont {Garc\'\i{}a-Quir\'os}, \citenamefont
  {Mateu-Lucena}, \citenamefont {Colleoni},\ and\ \citenamefont
  {Jaume}}]{Ramos-Buades:2019uvh}%
  \BibitemOpen
  \bibfield  {author} {\bibinfo {author} {\bibfnamefont {A.}~\bibnamefont
  {Ramos-Buades}}, \bibinfo {author} {\bibfnamefont {S.}~\bibnamefont {Husa}},
  \bibinfo {author} {\bibfnamefont {G.}~\bibnamefont {Pratten}}, \bibinfo
  {author} {\bibfnamefont {H.}~\bibnamefont {Estell\'es}}, \bibinfo {author}
  {\bibfnamefont {C.}~\bibnamefont {Garc\'\i{}a-Quir\'os}}, \bibinfo {author}
  {\bibfnamefont {M.}~\bibnamefont {Mateu-Lucena}}, \bibinfo {author}
  {\bibfnamefont {M.}~\bibnamefont {Colleoni}}, \ and\ \bibinfo {author}
  {\bibfnamefont {R.}~\bibnamefont {Jaume}},\ }\href {\doibase
  10.1103/PhysRevD.101.083015} {\bibfield  {journal} {\bibinfo  {journal}
  {Phys. Rev. D}\ }\textbf {\bibinfo {volume} {101}},\ \bibinfo {pages}
  {083015} (\bibinfo {year} {2020})},\ \Eprint
  {http://arxiv.org/abs/1909.11011} {arXiv:1909.11011 [gr-qc]} \BibitemShut
  {NoStop}%
\bibitem [{\citenamefont {O'Shea}\ and\ \citenamefont
  {Kumar}(2023)}]{OShea:2021faf}%
  \BibitemOpen
  \bibfield  {author} {\bibinfo {author} {\bibfnamefont {E.}~\bibnamefont
  {O'Shea}}\ and\ \bibinfo {author} {\bibfnamefont {P.}~\bibnamefont {Kumar}},\
  }\href {\doibase 10.1103/PhysRevD.108.104018} {\bibfield  {journal} {\bibinfo
   {journal} {Phys. Rev. D}\ }\textbf {\bibinfo {volume} {108}},\ \bibinfo
  {pages} {104018} (\bibinfo {year} {2023})},\ \Eprint
  {http://arxiv.org/abs/2107.07981} {arXiv:2107.07981 [astro-ph.HE]}
  \BibitemShut {NoStop}%
\bibitem [{\citenamefont {Klein}\ \emph {et~al.}(2013)\citenamefont {Klein},
  \citenamefont {Cornish},\ and\ \citenamefont {Yunes}}]{Klein:2013qda}%
  \BibitemOpen
  \bibfield  {author} {\bibinfo {author} {\bibfnamefont {A.}~\bibnamefont
  {Klein}}, \bibinfo {author} {\bibfnamefont {N.}~\bibnamefont {Cornish}}, \
  and\ \bibinfo {author} {\bibfnamefont {N.}~\bibnamefont {Yunes}},\ }\href
  {\doibase 10.1103/PhysRevD.88.124015} {\bibfield  {journal} {\bibinfo
  {journal} {Phys. Rev. D}\ }\textbf {\bibinfo {volume} {88}},\ \bibinfo
  {pages} {124015} (\bibinfo {year} {2013})},\ \Eprint
  {http://arxiv.org/abs/1305.1932} {arXiv:1305.1932 [gr-qc]} \BibitemShut
  {NoStop}%
\bibitem [{\citenamefont {Huerta}\ \emph {et~al.}(2014)\citenamefont {Huerta},
  \citenamefont {Kumar}, \citenamefont {McWilliams}, \citenamefont
  {O'Shaughnessy},\ and\ \citenamefont {Yunes}}]{Huerta:2014eca}%
  \BibitemOpen
  \bibfield  {author} {\bibinfo {author} {\bibfnamefont {E.~A.}\ \bibnamefont
  {Huerta}}, \bibinfo {author} {\bibfnamefont {P.}~\bibnamefont {Kumar}},
  \bibinfo {author} {\bibfnamefont {S.~T.}\ \bibnamefont {McWilliams}},
  \bibinfo {author} {\bibfnamefont {R.}~\bibnamefont {O'Shaughnessy}}, \ and\
  \bibinfo {author} {\bibfnamefont {N.}~\bibnamefont {Yunes}},\ }\href
  {\doibase 10.1103/PhysRevD.90.084016} {\bibfield  {journal} {\bibinfo
  {journal} {Phys. Rev. D}\ }\textbf {\bibinfo {volume} {90}},\ \bibinfo
  {pages} {084016} (\bibinfo {year} {2014})},\ \Eprint
  {http://arxiv.org/abs/1408.3406} {arXiv:1408.3406 [gr-qc]} \BibitemShut
  {NoStop}%
\bibitem [{\citenamefont {Moore}\ \emph {et~al.}(2018)\citenamefont {Moore},
  \citenamefont {Robson}, \citenamefont {Loutrel},\ and\ \citenamefont
  {Yunes}}]{Moore:2018kvz}%
  \BibitemOpen
  \bibfield  {author} {\bibinfo {author} {\bibfnamefont {B.}~\bibnamefont
  {Moore}}, \bibinfo {author} {\bibfnamefont {T.}~\bibnamefont {Robson}},
  \bibinfo {author} {\bibfnamefont {N.}~\bibnamefont {Loutrel}}, \ and\
  \bibinfo {author} {\bibfnamefont {N.}~\bibnamefont {Yunes}},\ }\href
  {\doibase 10.1088/1361-6382/aaea00} {\bibfield  {journal} {\bibinfo
  {journal} {Class. Quant. Grav.}\ }\textbf {\bibinfo {volume} {35}},\ \bibinfo
  {pages} {235006} (\bibinfo {year} {2018})},\ \Eprint
  {http://arxiv.org/abs/1807.07163} {arXiv:1807.07163 [gr-qc]} \BibitemShut
  {NoStop}%
\bibitem [{\citenamefont {Klein}\ \emph {et~al.}(2018)\citenamefont {Klein},
  \citenamefont {Boetzel}, \citenamefont {Gopakumar}, \citenamefont {Jetzer},\
  and\ \citenamefont {de~Vittori}}]{Klein:2018ybm}%
  \BibitemOpen
  \bibfield  {author} {\bibinfo {author} {\bibfnamefont {A.}~\bibnamefont
  {Klein}}, \bibinfo {author} {\bibfnamefont {Y.}~\bibnamefont {Boetzel}},
  \bibinfo {author} {\bibfnamefont {A.}~\bibnamefont {Gopakumar}}, \bibinfo
  {author} {\bibfnamefont {P.}~\bibnamefont {Jetzer}}, \ and\ \bibinfo {author}
  {\bibfnamefont {L.}~\bibnamefont {de~Vittori}},\ }\href {\doibase
  10.1103/PhysRevD.98.104043} {\bibfield  {journal} {\bibinfo  {journal} {Phys.
  Rev. D}\ }\textbf {\bibinfo {volume} {98}},\ \bibinfo {pages} {104043}
  (\bibinfo {year} {2018})},\ \Eprint {http://arxiv.org/abs/1801.08542}
  {arXiv:1801.08542 [gr-qc]} \BibitemShut {NoStop}%
\bibitem [{\citenamefont {Tanay}\ \emph {et~al.}(2019)\citenamefont {Tanay},
  \citenamefont {Klein}, \citenamefont {Berti},\ and\ \citenamefont
  {Nishizawa}}]{Tanay:2019knc}%
  \BibitemOpen
  \bibfield  {author} {\bibinfo {author} {\bibfnamefont {S.}~\bibnamefont
  {Tanay}}, \bibinfo {author} {\bibfnamefont {A.}~\bibnamefont {Klein}},
  \bibinfo {author} {\bibfnamefont {E.}~\bibnamefont {Berti}}, \ and\ \bibinfo
  {author} {\bibfnamefont {A.}~\bibnamefont {Nishizawa}},\ }\href {\doibase
  10.1103/PhysRevD.100.064006} {\bibfield  {journal} {\bibinfo  {journal}
  {Phys. Rev. D}\ }\textbf {\bibinfo {volume} {100}},\ \bibinfo {pages}
  {064006} (\bibinfo {year} {2019})},\ \Eprint
  {http://arxiv.org/abs/1905.08811} {arXiv:1905.08811 [gr-qc]} \BibitemShut
  {NoStop}%
\bibitem [{\citenamefont {Liu}\ \emph {et~al.}(2020)\citenamefont {Liu},
  \citenamefont {Cao},\ and\ \citenamefont {Shao}}]{Liu:2019jpg}%
  \BibitemOpen
  \bibfield  {author} {\bibinfo {author} {\bibfnamefont {X.}~\bibnamefont
  {Liu}}, \bibinfo {author} {\bibfnamefont {Z.}~\bibnamefont {Cao}}, \ and\
  \bibinfo {author} {\bibfnamefont {L.}~\bibnamefont {Shao}},\ }\href {\doibase
  10.1103/PhysRevD.101.044049} {\bibfield  {journal} {\bibinfo  {journal}
  {Phys. Rev. D}\ }\textbf {\bibinfo {volume} {101}},\ \bibinfo {pages}
  {044049} (\bibinfo {year} {2020})},\ \Eprint
  {http://arxiv.org/abs/1910.00784} {arXiv:1910.00784 [gr-qc]} \BibitemShut
  {NoStop}%
\bibitem [{\citenamefont {Tiwari}\ and\ \citenamefont
  {Gopakumar}(2020)}]{Tiwari:2020hsu}%
  \BibitemOpen
  \bibfield  {author} {\bibinfo {author} {\bibfnamefont {S.}~\bibnamefont
  {Tiwari}}\ and\ \bibinfo {author} {\bibfnamefont {A.}~\bibnamefont
  {Gopakumar}},\ }\href {\doibase 10.1103/PhysRevD.102.084042} {\bibfield
  {journal} {\bibinfo  {journal} {Phys. Rev. D}\ }\textbf {\bibinfo {volume}
  {102}},\ \bibinfo {pages} {084042} (\bibinfo {year} {2020})},\ \Eprint
  {http://arxiv.org/abs/2009.11333} {arXiv:2009.11333 [gr-qc]} \BibitemShut
  {NoStop}%
\bibitem [{\citenamefont {Klein}(2021)}]{Klein:2021jtd}%
  \BibitemOpen
  \bibfield  {author} {\bibinfo {author} {\bibfnamefont {A.}~\bibnamefont
  {Klein}},\ }\href@noop {} {\  (\bibinfo {year} {2021})},\ \Eprint
  {http://arxiv.org/abs/2106.10291} {arXiv:2106.10291 [gr-qc]} \BibitemShut
  {NoStop}%
\bibitem [{\citenamefont {Paul}\ and\ \citenamefont
  {Mishra}(2023)}]{Paul:2022xfy}%
  \BibitemOpen
  \bibfield  {author} {\bibinfo {author} {\bibfnamefont {K.}~\bibnamefont
  {Paul}}\ and\ \bibinfo {author} {\bibfnamefont {C.~K.}\ \bibnamefont
  {Mishra}},\ }\href {\doibase 10.1103/PhysRevD.108.024023} {\bibfield
  {journal} {\bibinfo  {journal} {Phys. Rev. D}\ }\textbf {\bibinfo {volume}
  {108}},\ \bibinfo {pages} {024023} (\bibinfo {year} {2023})},\ \Eprint
  {http://arxiv.org/abs/2211.04155} {arXiv:2211.04155 [gr-qc]} \BibitemShut
  {NoStop}%
\bibitem [{\citenamefont {Huerta}\ \emph {et~al.}(2017)\citenamefont {Huerta}
  \emph {et~al.}}]{Huerta:2016rwp}%
  \BibitemOpen
  \bibfield  {author} {\bibinfo {author} {\bibfnamefont {E.~A.}\ \bibnamefont
  {Huerta}} \emph {et~al.},\ }\href {\doibase 10.1103/PhysRevD.95.024038}
  {\bibfield  {journal} {\bibinfo  {journal} {Phys. Rev. D}\ }\textbf {\bibinfo
  {volume} {95}},\ \bibinfo {pages} {024038} (\bibinfo {year} {2017})},\
  \Eprint {http://arxiv.org/abs/1609.05933} {arXiv:1609.05933 [gr-qc]}
  \BibitemShut {NoStop}%
\bibitem [{\citenamefont {Hinderer}\ and\ \citenamefont
  {Babak}(2017)}]{Hinderer:2017jcs}%
  \BibitemOpen
  \bibfield  {author} {\bibinfo {author} {\bibfnamefont {T.}~\bibnamefont
  {Hinderer}}\ and\ \bibinfo {author} {\bibfnamefont {S.}~\bibnamefont
  {Babak}},\ }\href {\doibase 10.1103/PhysRevD.96.104048} {\bibfield  {journal}
  {\bibinfo  {journal} {Phys. Rev. D}\ }\textbf {\bibinfo {volume} {96}},\
  \bibinfo {pages} {104048} (\bibinfo {year} {2017})},\ \Eprint
  {http://arxiv.org/abs/1707.08426} {arXiv:1707.08426 [gr-qc]} \BibitemShut
  {NoStop}%
\bibitem [{\citenamefont {Hinder}\ \emph {et~al.}(2018)\citenamefont {Hinder},
  \citenamefont {Kidder},\ and\ \citenamefont {Pfeiffer}}]{Hinder:2017sxy}%
  \BibitemOpen
  \bibfield  {author} {\bibinfo {author} {\bibfnamefont {I.}~\bibnamefont
  {Hinder}}, \bibinfo {author} {\bibfnamefont {L.~E.}\ \bibnamefont {Kidder}},
  \ and\ \bibinfo {author} {\bibfnamefont {H.~P.}\ \bibnamefont {Pfeiffer}},\
  }\href {\doibase 10.1103/PhysRevD.98.044015} {\bibfield  {journal} {\bibinfo
  {journal} {Phys. Rev. D}\ }\textbf {\bibinfo {volume} {98}},\ \bibinfo {eid}
  {044015} (\bibinfo {year} {2018})},\ \Eprint
  {http://arxiv.org/abs/1709.02007} {arXiv:1709.02007 [gr-qc]} \BibitemShut
  {NoStop}%
\bibitem [{\citenamefont {Huerta}\ \emph {et~al.}(2018)\citenamefont {Huerta}
  \emph {et~al.}}]{Huerta:2017kez}%
  \BibitemOpen
  \bibfield  {author} {\bibinfo {author} {\bibfnamefont {E.~A.}\ \bibnamefont
  {Huerta}} \emph {et~al.},\ }\href {\doibase 10.1103/PhysRevD.97.024031}
  {\bibfield  {journal} {\bibinfo  {journal} {Phys. Rev. D}\ }\textbf {\bibinfo
  {volume} {97}},\ \bibinfo {pages} {024031} (\bibinfo {year} {2018})},\
  \Eprint {http://arxiv.org/abs/1711.06276} {arXiv:1711.06276 [gr-qc]}
  \BibitemShut {NoStop}%
\bibitem [{\citenamefont {Chen}\ \emph {et~al.}(2021)\citenamefont {Chen},
  \citenamefont {Huerta}, \citenamefont {Adamo}, \citenamefont {Haas},
  \citenamefont {O'Shea}, \citenamefont {Kumar},\ and\ \citenamefont
  {Moore}}]{Chen:2020lzc}%
  \BibitemOpen
  \bibfield  {author} {\bibinfo {author} {\bibfnamefont {Z.}~\bibnamefont
  {Chen}}, \bibinfo {author} {\bibfnamefont {E.~A.}\ \bibnamefont {Huerta}},
  \bibinfo {author} {\bibfnamefont {J.}~\bibnamefont {Adamo}}, \bibinfo
  {author} {\bibfnamefont {R.}~\bibnamefont {Haas}}, \bibinfo {author}
  {\bibfnamefont {E.}~\bibnamefont {O'Shea}}, \bibinfo {author} {\bibfnamefont
  {P.}~\bibnamefont {Kumar}}, \ and\ \bibinfo {author} {\bibfnamefont
  {C.}~\bibnamefont {Moore}},\ }\href {\doibase 10.1103/PhysRevD.103.084018}
  {\bibfield  {journal} {\bibinfo  {journal} {Phys. Rev. D}\ }\textbf {\bibinfo
  {volume} {103}},\ \bibinfo {pages} {084018} (\bibinfo {year} {2021})},\
  \Eprint {http://arxiv.org/abs/2008.03313} {arXiv:2008.03313 [gr-qc]}
  \BibitemShut {NoStop}%
\bibitem [{\citenamefont {Chiaramello}\ and\ \citenamefont
  {Nagar}(2020)}]{Chiaramello:2020ehz}%
  \BibitemOpen
  \bibfield  {author} {\bibinfo {author} {\bibfnamefont {D.}~\bibnamefont
  {Chiaramello}}\ and\ \bibinfo {author} {\bibfnamefont {A.}~\bibnamefont
  {Nagar}},\ }\href {\doibase 10.1103/PhysRevD.101.101501} {\bibfield
  {journal} {\bibinfo  {journal} {Phys. Rev. D}\ }\textbf {\bibinfo {volume}
  {101}},\ \bibinfo {pages} {101501} (\bibinfo {year} {2020})},\ \Eprint
  {http://arxiv.org/abs/2001.11736} {arXiv:2001.11736 [gr-qc]} \BibitemShut
  {NoStop}%
\bibitem [{\citenamefont {Manna}\ \emph {et~al.}(2024)\citenamefont {Manna},
  \citenamefont {RoyChowdhury},\ and\ \citenamefont {Mishra}}]{Manna:2024ycx}%
  \BibitemOpen
  \bibfield  {author} {\bibinfo {author} {\bibfnamefont {P.}~\bibnamefont
  {Manna}}, \bibinfo {author} {\bibfnamefont {T.}~\bibnamefont {RoyChowdhury}},
  \ and\ \bibinfo {author} {\bibfnamefont {C.~K.}\ \bibnamefont {Mishra}},\
  }\href@noop {} {\  (\bibinfo {year} {2024})},\ \Eprint
  {http://arxiv.org/abs/2409.10672} {arXiv:2409.10672 [gr-qc]} \BibitemShut
  {NoStop}%
\bibitem [{\citenamefont {Paul}\ \emph {et~al.}(2024)\citenamefont {Paul},
  \citenamefont {Maurya}, \citenamefont {Henry}, \citenamefont {Sharma},
  \citenamefont {Satheesh}, \citenamefont {Divyajyoti}, \citenamefont {Kumar},\
  and\ \citenamefont {Mishra}}]{Paul:2024ujx}%
  \BibitemOpen
  \bibfield  {author} {\bibinfo {author} {\bibfnamefont {K.}~\bibnamefont
  {Paul}}, \bibinfo {author} {\bibfnamefont {A.}~\bibnamefont {Maurya}},
  \bibinfo {author} {\bibfnamefont {Q.}~\bibnamefont {Henry}}, \bibinfo
  {author} {\bibfnamefont {K.}~\bibnamefont {Sharma}}, \bibinfo {author}
  {\bibfnamefont {P.}~\bibnamefont {Satheesh}}, \bibinfo {author} {\bibnamefont
  {Divyajyoti}}, \bibinfo {author} {\bibfnamefont {P.}~\bibnamefont {Kumar}}, \
  and\ \bibinfo {author} {\bibfnamefont {C.~K.}\ \bibnamefont {Mishra}},\
  }\href@noop {} {\  (\bibinfo {year} {2024})},\ \Eprint
  {http://arxiv.org/abs/2409.13866} {arXiv:2409.13866 [gr-qc]} \BibitemShut
  {NoStop}%
\bibitem [{\citenamefont {Mishra}\ \emph {et~al.}(2015)\citenamefont {Mishra},
  \citenamefont {Arun},\ and\ \citenamefont {Iyer}}]{Mishra:2015bqa}%
  \BibitemOpen
  \bibfield  {author} {\bibinfo {author} {\bibfnamefont {C.~K.}\ \bibnamefont
  {Mishra}}, \bibinfo {author} {\bibfnamefont {K.~G.}\ \bibnamefont {Arun}}, \
  and\ \bibinfo {author} {\bibfnamefont {B.~R.}\ \bibnamefont {Iyer}},\ }\href
  {\doibase 10.1103/PhysRevD.91.084040} {\bibfield  {journal} {\bibinfo
  {journal} {Phys. Rev. D}\ }\textbf {\bibinfo {volume} {91}},\ \bibinfo
  {pages} {084040} (\bibinfo {year} {2015})},\ \Eprint
  {http://arxiv.org/abs/1501.07096} {arXiv:1501.07096 [gr-qc]} \BibitemShut
  {NoStop}%
\bibitem [{\citenamefont {Arun}\ \emph {et~al.}(2009)\citenamefont {Arun},
  \citenamefont {Buonanno}, \citenamefont {Faye},\ and\ \citenamefont
  {Ochsner}}]{Arun:2008kb}%
  \BibitemOpen
  \bibfield  {author} {\bibinfo {author} {\bibfnamefont {K.~G.}\ \bibnamefont
  {Arun}}, \bibinfo {author} {\bibfnamefont {A.}~\bibnamefont {Buonanno}},
  \bibinfo {author} {\bibfnamefont {G.}~\bibnamefont {Faye}}, \ and\ \bibinfo
  {author} {\bibfnamefont {E.}~\bibnamefont {Ochsner}},\ }\href {\doibase
  10.1103/PhysRevD.79.104023} {\bibfield  {journal} {\bibinfo  {journal} {Phys.
  Rev. D}\ }\textbf {\bibinfo {volume} {79}},\ \bibinfo {pages} {104023}
  (\bibinfo {year} {2009})},\ \bibinfo {note} {[Erratum: Phys.Rev.D 84, 049901
  (2011)]},\ \Eprint {http://arxiv.org/abs/0810.5336} {arXiv:0810.5336 [gr-qc]}
  \BibitemShut {NoStop}%
\bibitem [{\citenamefont {Buonanno}\ \emph {et~al.}(2013)\citenamefont
  {Buonanno}, \citenamefont {Faye},\ and\ \citenamefont
  {Hinderer}}]{Buonanno:2012rv}%
  \BibitemOpen
  \bibfield  {author} {\bibinfo {author} {\bibfnamefont {A.}~\bibnamefont
  {Buonanno}}, \bibinfo {author} {\bibfnamefont {G.}~\bibnamefont {Faye}}, \
  and\ \bibinfo {author} {\bibfnamefont {T.}~\bibnamefont {Hinderer}},\ }\href
  {\doibase 10.1103/PhysRevD.87.044009} {\bibfield  {journal} {\bibinfo
  {journal} {Phys. Rev. D}\ }\textbf {\bibinfo {volume} {87}},\ \bibinfo
  {pages} {044009} (\bibinfo {year} {2013})},\ \Eprint
  {http://arxiv.org/abs/1209.6349} {arXiv:1209.6349 [gr-qc]} \BibitemShut
  {NoStop}%
\bibitem [{\citenamefont {Henry}\ \emph {et~al.}(2022)\citenamefont {Henry},
  \citenamefont {Marsat},\ and\ \citenamefont {Khalil}}]{Henry:2022dzx}%
  \BibitemOpen
  \bibfield  {author} {\bibinfo {author} {\bibfnamefont {Q.}~\bibnamefont
  {Henry}}, \bibinfo {author} {\bibfnamefont {S.}~\bibnamefont {Marsat}}, \
  and\ \bibinfo {author} {\bibfnamefont {M.}~\bibnamefont {Khalil}},\ }\href
  {\doibase 10.1103/PhysRevD.106.124018} {\bibfield  {journal} {\bibinfo
  {journal} {Phys. Rev. D}\ }\textbf {\bibinfo {volume} {106}},\ \bibinfo
  {pages} {124018} (\bibinfo {year} {2022})},\ \Eprint
  {http://arxiv.org/abs/2209.00374} {arXiv:2209.00374 [gr-qc]} \BibitemShut
  {NoStop}%
\bibitem [{\citenamefont {Kidder}\ \emph {et~al.}(1993)\citenamefont {Kidder},
  \citenamefont {Will},\ and\ \citenamefont {Wiseman}}]{Kidder:1992fr}%
  \BibitemOpen
  \bibfield  {author} {\bibinfo {author} {\bibfnamefont {L.~E.}\ \bibnamefont
  {Kidder}}, \bibinfo {author} {\bibfnamefont {C.~M.}\ \bibnamefont {Will}}, \
  and\ \bibinfo {author} {\bibfnamefont {A.~G.}\ \bibnamefont {Wiseman}},\
  }\href {\doibase 10.1103/PhysRevD.47.R4183} {\bibfield  {journal} {\bibinfo
  {journal} {Phys. Rev. D}\ }\textbf {\bibinfo {volume} {47}},\ \bibinfo
  {pages} {R4183} (\bibinfo {year} {1993})},\ \Eprint
  {http://arxiv.org/abs/gr-qc/9211025} {arXiv:gr-qc/9211025} \BibitemShut
  {NoStop}%
\bibitem [{\citenamefont {Kidder}(1995)}]{Kidder:1995zr}%
  \BibitemOpen
  \bibfield  {author} {\bibinfo {author} {\bibfnamefont {L.~E.}\ \bibnamefont
  {Kidder}},\ }\href {\doibase 10.1103/PhysRevD.52.821} {\bibfield  {journal}
  {\bibinfo  {journal} {Phys. Rev. D}\ }\textbf {\bibinfo {volume} {52}},\
  \bibinfo {pages} {821} (\bibinfo {year} {1995})},\ \Eprint
  {http://arxiv.org/abs/gr-qc/9506022} {arXiv:gr-qc/9506022} \BibitemShut
  {NoStop}%
\bibitem [{\citenamefont {Vasuth}\ and\ \citenamefont
  {Majar}(2007)}]{Vasuth:2007gx}%
  \BibitemOpen
  \bibfield  {author} {\bibinfo {author} {\bibfnamefont {M.}~\bibnamefont
  {Vasuth}}\ and\ \bibinfo {author} {\bibfnamefont {J.}~\bibnamefont {Majar}},\
  }\href {\doibase 10.1142/S0217751X07036488} {\bibfield  {journal} {\bibinfo
  {journal} {Int. J. Mod. Phys. A}\ }\textbf {\bibinfo {volume} {22}},\
  \bibinfo {pages} {2405} (\bibinfo {year} {2007})},\ \Eprint
  {http://arxiv.org/abs/0705.3481} {arXiv:0705.3481 [gr-qc]} \BibitemShut
  {NoStop}%
\bibitem [{\citenamefont {Majar}\ and\ \citenamefont
  {Vasuth}(2008)}]{Majar:2008zz}%
  \BibitemOpen
  \bibfield  {author} {\bibinfo {author} {\bibfnamefont {J.}~\bibnamefont
  {Majar}}\ and\ \bibinfo {author} {\bibfnamefont {M.}~\bibnamefont {Vasuth}},\
  }\href {\doibase 10.1103/PhysRevD.77.104005} {\bibfield  {journal} {\bibinfo
  {journal} {Phys. Rev. D}\ }\textbf {\bibinfo {volume} {77}},\ \bibinfo
  {pages} {104005} (\bibinfo {year} {2008})},\ \Eprint
  {http://arxiv.org/abs/0806.2273} {arXiv:0806.2273 [gr-qc]} \BibitemShut
  {NoStop}%
\bibitem [{\citenamefont {Khalil}\ \emph {et~al.}(2021)\citenamefont {Khalil},
  \citenamefont {Buonanno}, \citenamefont {Steinhoff},\ and\ \citenamefont
  {Vines}}]{Khalil:2021txt}%
  \BibitemOpen
  \bibfield  {author} {\bibinfo {author} {\bibfnamefont {M.}~\bibnamefont
  {Khalil}}, \bibinfo {author} {\bibfnamefont {A.}~\bibnamefont {Buonanno}},
  \bibinfo {author} {\bibfnamefont {J.}~\bibnamefont {Steinhoff}}, \ and\
  \bibinfo {author} {\bibfnamefont {J.}~\bibnamefont {Vines}},\ }\href
  {\doibase 10.1103/PhysRevD.104.024046} {\bibfield  {journal} {\bibinfo
  {journal} {Phys. Rev. D}\ }\textbf {\bibinfo {volume} {104}},\ \bibinfo
  {pages} {024046} (\bibinfo {year} {2021})},\ \Eprint
  {http://arxiv.org/abs/2104.11705} {arXiv:2104.11705 [gr-qc]} \BibitemShut
  {NoStop}%
\bibitem [{\citenamefont {Henry}\ and\ \citenamefont
  {Khalil}(2023)}]{Henry:2023tka}%
  \BibitemOpen
  \bibfield  {author} {\bibinfo {author} {\bibfnamefont {Q.}~\bibnamefont
  {Henry}}\ and\ \bibinfo {author} {\bibfnamefont {M.}~\bibnamefont {Khalil}},\
  }\href {\doibase 10.1103/PhysRevD.108.104016} {\bibfield  {journal} {\bibinfo
   {journal} {Phys. Rev. D}\ }\textbf {\bibinfo {volume} {108}},\ \bibinfo
  {pages} {104016} (\bibinfo {year} {2023})},\ \Eprint
  {http://arxiv.org/abs/2308.13606} {arXiv:2308.13606 [gr-qc]} \BibitemShut
  {NoStop}%
\bibitem [{\citenamefont {Blanchet}\ \emph {et~al.}(1995)\citenamefont
  {Blanchet}, \citenamefont {Damour},\ and\ \citenamefont
  {Iyer}}]{Blanchet:1995fg}%
  \BibitemOpen
  \bibfield  {author} {\bibinfo {author} {\bibfnamefont {L.}~\bibnamefont
  {Blanchet}}, \bibinfo {author} {\bibfnamefont {T.}~\bibnamefont {Damour}}, \
  and\ \bibinfo {author} {\bibfnamefont {B.~R.}\ \bibnamefont {Iyer}},\ }\href
  {\doibase 10.1103/PhysRevD.51.5360} {\bibfield  {journal} {\bibinfo
  {journal} {Phys. Rev. D}\ }\textbf {\bibinfo {volume} {51}},\ \bibinfo
  {pages} {5360} (\bibinfo {year} {1995})},\ \bibinfo {note} {[Erratum:
  Phys.Rev.D 54, 1860 (1996)]},\ \Eprint {http://arxiv.org/abs/gr-qc/9501029}
  {arXiv:gr-qc/9501029} \BibitemShut {NoStop}%
\bibitem [{\citenamefont {Blanchet}\ \emph {et~al.}(2004)\citenamefont
  {Blanchet}, \citenamefont {Damour}, \citenamefont {Esposito-Farese},\ and\
  \citenamefont {Iyer}}]{Blanchet:2004ek}%
  \BibitemOpen
  \bibfield  {author} {\bibinfo {author} {\bibfnamefont {L.}~\bibnamefont
  {Blanchet}}, \bibinfo {author} {\bibfnamefont {T.}~\bibnamefont {Damour}},
  \bibinfo {author} {\bibfnamefont {G.}~\bibnamefont {Esposito-Farese}}, \ and\
  \bibinfo {author} {\bibfnamefont {B.~R.}\ \bibnamefont {Iyer}},\ }\href
  {\doibase 10.1103/PhysRevLett.93.091101} {\bibfield  {journal} {\bibinfo
  {journal} {Phys. Rev. Lett.}\ }\textbf {\bibinfo {volume} {93}},\ \bibinfo
  {pages} {091101} (\bibinfo {year} {2004})},\ \Eprint
  {http://arxiv.org/abs/gr-qc/0406012} {arXiv:gr-qc/0406012} \BibitemShut
  {NoStop}%
\bibitem [{\citenamefont {Blanchet}\ \emph {et~al.}(2023)\citenamefont
  {Blanchet}, \citenamefont {Faye}, \citenamefont {Henry}, \citenamefont
  {Larrouturou},\ and\ \citenamefont {Trestini}}]{Blanchet:2023bwj}%
  \BibitemOpen
  \bibfield  {author} {\bibinfo {author} {\bibfnamefont {L.}~\bibnamefont
  {Blanchet}}, \bibinfo {author} {\bibfnamefont {G.}~\bibnamefont {Faye}},
  \bibinfo {author} {\bibfnamefont {Q.}~\bibnamefont {Henry}}, \bibinfo
  {author} {\bibfnamefont {F.}~\bibnamefont {Larrouturou}}, \ and\ \bibinfo
  {author} {\bibfnamefont {D.}~\bibnamefont {Trestini}},\ }\href {\doibase
  10.1103/PhysRevLett.131.121402} {\bibfield  {journal} {\bibinfo  {journal}
  {Phys. Rev. Lett.}\ }\textbf {\bibinfo {volume} {131}},\ \bibinfo {pages}
  {121402} (\bibinfo {year} {2023})},\ \Eprint
  {http://arxiv.org/abs/2304.11185} {arXiv:2304.11185 [gr-qc]} \BibitemShut
  {NoStop}%
\bibitem [{\citenamefont {Moore}\ \emph {et~al.}(2016)\citenamefont {Moore},
  \citenamefont {Favata}, \citenamefont {Arun},\ and\ \citenamefont
  {Mishra}}]{Moore:2016qxz}%
  \BibitemOpen
  \bibfield  {author} {\bibinfo {author} {\bibfnamefont {B.}~\bibnamefont
  {Moore}}, \bibinfo {author} {\bibfnamefont {M.}~\bibnamefont {Favata}},
  \bibinfo {author} {\bibfnamefont {K.~G.}\ \bibnamefont {Arun}}, \ and\
  \bibinfo {author} {\bibfnamefont {C.~K.}\ \bibnamefont {Mishra}},\ }\href
  {\doibase 10.1103/PhysRevD.93.124061} {\bibfield  {journal} {\bibinfo
  {journal} {Phys. Rev. D}\ }\textbf {\bibinfo {volume} {93}},\ \bibinfo
  {pages} {124061} (\bibinfo {year} {2016})},\ \Eprint
  {http://arxiv.org/abs/1605.00304} {arXiv:1605.00304 [gr-qc]} \BibitemShut
  {NoStop}%
\bibitem [{SBP()}]{SBPM:suppl}%
  \BibitemOpen
  \href@noop {} {}\bibinfo {howpublished} {See Supplemental Material at
  \url{https://github.com/kaushikush/Spinning-eccentric-phasing/blob/main/Suppl_SBPM.m}
  for spin effects in the phasing formula of eccentric compact binary inspirals
  till the third post-Newtonian order.}\BibitemShut {Stop}%
\bibitem [{\citenamefont {Shoemaker}(2010)}]{Shoemaker:T0900288-v3}%
  \BibitemOpen
  \bibfield  {author} {\bibinfo {author} {\bibfnamefont {D.~L.~C.}\
  \bibnamefont {Shoemaker}},\ }\href
  {https://dcc.ligo.org/LIGO-T0900288/public} {\  (\bibinfo {year}
  {2010})}\BibitemShut {NoStop}%
\bibitem [{\citenamefont {Dwyer}\ \emph {et~al.}(2015)\citenamefont {Dwyer},
  \citenamefont {Sigg}, \citenamefont {Ballmer}, \citenamefont {Barsotti},
  \citenamefont {Mavalvala},\ and\ \citenamefont {Evans}}]{Dwyer:2014fpa}%
  \BibitemOpen
  \bibfield  {author} {\bibinfo {author} {\bibfnamefont {S.}~\bibnamefont
  {Dwyer}}, \bibinfo {author} {\bibfnamefont {D.}~\bibnamefont {Sigg}},
  \bibinfo {author} {\bibfnamefont {S.~W.}\ \bibnamefont {Ballmer}}, \bibinfo
  {author} {\bibfnamefont {L.}~\bibnamefont {Barsotti}}, \bibinfo {author}
  {\bibfnamefont {N.}~\bibnamefont {Mavalvala}}, \ and\ \bibinfo {author}
  {\bibfnamefont {M.}~\bibnamefont {Evans}},\ }\href {\doibase
  10.1103/PhysRevD.91.082001} {\bibfield  {journal} {\bibinfo  {journal} {Phys.
  Rev. D}\ }\textbf {\bibinfo {volume} {91}},\ \bibinfo {pages} {082001}
  (\bibinfo {year} {2015})},\ \Eprint {http://arxiv.org/abs/arXiv:1410.0612}
  {arXiv:arXiv:1410.0612 [astro-ph.IM]} \BibitemShut {NoStop}%
\bibitem [{\citenamefont {Punturo}\ \emph {et~al.}(2010)\citenamefont {Punturo}
  \emph {et~al.}}]{Punturo:2010zz}%
  \BibitemOpen
  \bibfield  {author} {\bibinfo {author} {\bibfnamefont {M.}~\bibnamefont
  {Punturo}} \emph {et~al.},\ }\href {\doibase 10.1088/0264-9381/27/19/194002}
  {\bibfield  {journal} {\bibinfo  {journal} {Class. Quant. Grav.}\ }\textbf
  {\bibinfo {volume} {27}},\ \bibinfo {pages} {194002} (\bibinfo {year}
  {2010})}\BibitemShut {NoStop}%
\bibitem [{\citenamefont {Sato}\ \emph {et~al.}(2017)\citenamefont {Sato} \emph
  {et~al.}}]{Sato_2017}%
  \BibitemOpen
  \bibfield  {author} {\bibinfo {author} {\bibfnamefont {S.}~\bibnamefont
  {Sato}} \emph {et~al.},\ }\href {\doibase 10.1088/1742-6596/840/1/012010}
  {\bibfield  {journal} {\bibinfo  {journal} {Journal of Physics: Conference
  Series}\ }\textbf {\bibinfo {volume} {840}},\ \bibinfo {pages} {012010}
  (\bibinfo {year} {2017})}\BibitemShut {NoStop}%
\bibitem [{\citenamefont {Amaro-Seoane}\ \emph {et~al.}(2017)\citenamefont
  {Amaro-Seoane}, \citenamefont {Audley}, \citenamefont {Babak}, \citenamefont
  {Baker}, \citenamefont {Barausse}, \citenamefont {Bender}, \citenamefont
  {Berti}, \citenamefont {Binetruy}, \citenamefont {Born}, \citenamefont
  {Bortoluzzi} \emph {et~al.}}]{amaro2017laser}%
  \BibitemOpen
  \bibfield  {author} {\bibinfo {author} {\bibfnamefont {P.}~\bibnamefont
  {Amaro-Seoane}}, \bibinfo {author} {\bibfnamefont {H.}~\bibnamefont
  {Audley}}, \bibinfo {author} {\bibfnamefont {S.}~\bibnamefont {Babak}},
  \bibinfo {author} {\bibfnamefont {J.}~\bibnamefont {Baker}}, \bibinfo
  {author} {\bibfnamefont {E.}~\bibnamefont {Barausse}}, \bibinfo {author}
  {\bibfnamefont {P.}~\bibnamefont {Bender}}, \bibinfo {author} {\bibfnamefont
  {E.}~\bibnamefont {Berti}}, \bibinfo {author} {\bibfnamefont
  {P.}~\bibnamefont {Binetruy}}, \bibinfo {author} {\bibfnamefont
  {M.}~\bibnamefont {Born}}, \bibinfo {author} {\bibfnamefont {D.}~\bibnamefont
  {Bortoluzzi}},  \emph {et~al.},\ }\href@noop {} {\bibfield  {journal}
  {\bibinfo  {journal} {arXiv preprint arXiv:1702.00786}\ } (\bibinfo {year}
  {2017})}\BibitemShut {NoStop}%
\bibitem [{\citenamefont {Van Den~Broeck}\ and\ \citenamefont
  {Sengupta}(2007)}]{VanDenBroeck:2006qu}%
  \BibitemOpen
  \bibfield  {author} {\bibinfo {author} {\bibfnamefont {C.}~\bibnamefont {Van
  Den~Broeck}}\ and\ \bibinfo {author} {\bibfnamefont {A.~S.}\ \bibnamefont
  {Sengupta}},\ }\href {\doibase 10.1088/0264-9381/24/1/009} {\bibfield
  {journal} {\bibinfo  {journal} {Class. Quant. Grav.}\ }\textbf {\bibinfo
  {volume} {24}},\ \bibinfo {pages} {155} (\bibinfo {year} {2007})},\ \Eprint
  {http://arxiv.org/abs/gr-qc/0607092} {arXiv:gr-qc/0607092} \BibitemShut
  {NoStop}%
\bibitem [{\citenamefont {{Van Den Broeck}}\ and\ \citenamefont
  {{Sengupta}}(2007)}]{VanDenBroeck:2006ar}%
  \BibitemOpen
  \bibfield  {author} {\bibinfo {author} {\bibfnamefont {C.}~\bibnamefont {{Van
  Den Broeck}}}\ and\ \bibinfo {author} {\bibfnamefont {A.~S.}\ \bibnamefont
  {{Sengupta}}},\ }\href {\doibase 10.1088/0264-9381/24/5/005} {\bibfield
  {journal} {\bibinfo  {journal} {Class.~Quantum Grav.}\ }\textbf {\bibinfo
  {volume} {24}},\ \bibinfo {pages} {1089} (\bibinfo {year} {2007})},\ \Eprint
  {http://arxiv.org/abs/gr-qc/0610126} {arXiv:gr-qc/0610126} \BibitemShut
  {NoStop}%
\bibitem [{\citenamefont {{Buonanno}}\ \emph {et~al.}(2009)\citenamefont
  {{Buonanno}}, \citenamefont {{Iyer}}, \citenamefont {{Ochsner}},
  \citenamefont {{Pan}},\ and\ \citenamefont
  {{Sathyaprakash}}}]{buonanno-iyer-oshsner-pan-sathya-templatecomparison-PRD2009}%
  \BibitemOpen
  \bibfield  {author} {\bibinfo {author} {\bibfnamefont {A.}~\bibnamefont
  {{Buonanno}}}, \bibinfo {author} {\bibfnamefont {B.~R.}\ \bibnamefont
  {{Iyer}}}, \bibinfo {author} {\bibfnamefont {E.}~\bibnamefont {{Ochsner}}},
  \bibinfo {author} {\bibfnamefont {Y.}~\bibnamefont {{Pan}}}, \ and\ \bibinfo
  {author} {\bibfnamefont {B.~S.}\ \bibnamefont {{Sathyaprakash}}},\ }\href
  {\doibase 10.1103/PhysRevD.80.084043} {\bibfield  {journal} {\bibinfo
  {journal} {Phys.~Rev.~D}\ }\textbf {\bibinfo {volume} {80}},\ \bibinfo
  {pages} {084043} (\bibinfo {year} {2009})},\ \Eprint
  {http://arxiv.org/abs/arXiv:0907.0700 [gr-qc]} {arXiv:0907.0700 [gr-qc]}
  \BibitemShut {NoStop}%
\bibitem [{\citenamefont {Boetzel}\ \emph {et~al.}(2017)\citenamefont
  {Boetzel}, \citenamefont {Susobhanan}, \citenamefont {Gopakumar},
  \citenamefont {Klein},\ and\ \citenamefont {Jetzer}}]{Boetzel:2017zza}%
  \BibitemOpen
  \bibfield  {author} {\bibinfo {author} {\bibfnamefont {Y.}~\bibnamefont
  {Boetzel}}, \bibinfo {author} {\bibfnamefont {A.}~\bibnamefont {Susobhanan}},
  \bibinfo {author} {\bibfnamefont {A.}~\bibnamefont {Gopakumar}}, \bibinfo
  {author} {\bibfnamefont {A.}~\bibnamefont {Klein}}, \ and\ \bibinfo {author}
  {\bibfnamefont {P.}~\bibnamefont {Jetzer}},\ }\href {\doibase
  10.1103/PhysRevD.96.044011} {\bibfield  {journal} {\bibinfo  {journal} {Phys.
  Rev. D}\ }\textbf {\bibinfo {volume} {96}},\ \bibinfo {pages} {044011}
  (\bibinfo {year} {2017})},\ \Eprint {http://arxiv.org/abs/1707.02088}
  {arXiv:1707.02088 [gr-qc]} \BibitemShut {NoStop}%
\bibitem [{\citenamefont {Krishnendu}\ \emph {et~al.}(2017)\citenamefont
  {Krishnendu}, \citenamefont {Arun},\ and\ \citenamefont
  {Mishra}}]{Krishnendu:2017shb}%
  \BibitemOpen
  \bibfield  {author} {\bibinfo {author} {\bibfnamefont {N.~V.}\ \bibnamefont
  {Krishnendu}}, \bibinfo {author} {\bibfnamefont {K.~G.}\ \bibnamefont
  {Arun}}, \ and\ \bibinfo {author} {\bibfnamefont {C.~K.}\ \bibnamefont
  {Mishra}},\ }\href {\doibase 10.1103/PhysRevLett.119.091101} {\bibfield
  {journal} {\bibinfo  {journal} {Phys. Rev. Lett.}\ }\textbf {\bibinfo
  {volume} {119}},\ \bibinfo {pages} {091101} (\bibinfo {year} {2017})},\
  \Eprint {http://arxiv.org/abs/1701.06318} {arXiv:1701.06318 [gr-qc]}
  \BibitemShut {NoStop}%
\bibitem [{\citenamefont {Divyajyoti}\ \emph
  {et~al.}(2024{\natexlab{b}})\citenamefont {Divyajyoti}, \citenamefont
  {Krishnendu}, \citenamefont {Saleem}, \citenamefont {Colleoni}, \citenamefont
  {Vijaykumar}, \citenamefont {Arun},\ and\ \citenamefont
  {Mishra}}]{Divyajyoti:2023izl}%
  \BibitemOpen
  \bibfield  {author} {\bibinfo {author} {\bibnamefont {Divyajyoti}}, \bibinfo
  {author} {\bibfnamefont {N.~V.}\ \bibnamefont {Krishnendu}}, \bibinfo
  {author} {\bibfnamefont {M.}~\bibnamefont {Saleem}}, \bibinfo {author}
  {\bibfnamefont {M.}~\bibnamefont {Colleoni}}, \bibinfo {author}
  {\bibfnamefont {A.}~\bibnamefont {Vijaykumar}}, \bibinfo {author}
  {\bibfnamefont {K.~G.}\ \bibnamefont {Arun}}, \ and\ \bibinfo {author}
  {\bibfnamefont {C.~K.}\ \bibnamefont {Mishra}},\ }\href {\doibase
  10.1103/PhysRevD.109.023016} {\bibfield  {journal} {\bibinfo  {journal}
  {Phys. Rev. D}\ }\textbf {\bibinfo {volume} {109}},\ \bibinfo {pages}
  {023016} (\bibinfo {year} {2024}{\natexlab{b}})},\ \Eprint
  {http://arxiv.org/abs/2311.05506} {arXiv:2311.05506 [gr-qc]} \BibitemShut
  {NoStop}%
\bibitem [{\citenamefont {P\"urrer}(2014)}]{Purrer:2014fza}%
  \BibitemOpen
  \bibfield  {author} {\bibinfo {author} {\bibfnamefont {M.}~\bibnamefont
  {P\"urrer}},\ }\href {\doibase 10.1088/0264-9381/31/19/195010} {\bibfield
  {journal} {\bibinfo  {journal} {Class. Quant. Grav.}\ }\textbf {\bibinfo
  {volume} {31}},\ \bibinfo {pages} {195010} (\bibinfo {year} {2014})},\
  \Eprint {http://arxiv.org/abs/1402.4146} {arXiv:1402.4146 [gr-qc]}
  \BibitemShut {NoStop}%
\bibitem [{\citenamefont {Boh\'e}\ \emph {et~al.}(2015)\citenamefont {Boh\'e},
  \citenamefont {Faye}, \citenamefont {Marsat},\ and\ \citenamefont
  {Porter}}]{bohe:2015ana}%
  \BibitemOpen
  \bibfield  {author} {\bibinfo {author} {\bibfnamefont {A.}~\bibnamefont
  {Boh\'e}}, \bibinfo {author} {\bibfnamefont {G.}~\bibnamefont {Faye}},
  \bibinfo {author} {\bibfnamefont {S.}~\bibnamefont {Marsat}}, \ and\ \bibinfo
  {author} {\bibfnamefont {E.~K.}\ \bibnamefont {Porter}},\ }\href {\doibase
  10.1088/0264-9381/32/19/195010} {\bibfield  {journal} {\bibinfo  {journal}
  {Class. Quant. Grav.}\ }\textbf {\bibinfo {volume} {32}},\ \bibinfo {pages}
  {195010} (\bibinfo {year} {2015})},\ \Eprint
  {http://arxiv.org/abs/1501.01529} {arXiv:1501.01529 [gr-qc]} \BibitemShut
  {NoStop}%
\end{thebibliography}%
\end{document}